\documentclass[copyright,creativecommons]{eptcs}

\usepackage{madoko}

\providecommand{\event}{MSFP 2014}
              \providecommand{\publicationstatus}{To appear in EPTCS'14.}
\begin{document}
\newcommand{\tfun}[3]{#1\,\rightarrow\,#2\;#3}
\newcommand{\ec}[1]{\mathit{#1}}
\newcommand{\empt}{\langle\rangle}
\newcommand{\row}[1]{\langle#1\rangle}
\newcommand{\ext}[2]{\row{#1\,|\,#2}}
\newcommand{\e}{\epsilon}
\renewcommand{\t}{\tau}
\renewcommand{\exp}{e}
\newcommand{\p}{p}
\newcommand{\s}{\sigma}

\newcommand{\textkw}[1]{\mathsf{#1}}
\newcommand{\labs}[1]{\textkw{labels}(#1)}
\newcommand{\tail}[1]{\textkw{tail}(#1)}

\newcommand{\infer}[4]{#1\vdash\,#2:#3\;|\;#4}
\newcommand{\infers}[4]{#1\vdash_s\,#2:#3\;|\;#4}
\newcommand{\inferi}[4]{#1\vdash_i\,#2:#3\;|\;#4}

\newcommand{\keyword}[1]{\mathkw{#1}}  
\newcommand{\dom}[1]{\textkw{dom}(#1)}
\newcommand{\rdom}[1]{\textkw{rdom}(#1)}
\newcommand{\codom}[1]{\textkw{codom}(#1)}
\newcommand{\ftv}[1]{\textkw{ftv}(#1)}
\newcommand{\fv}[1]{\textkw{fv}(#1)}
\newcommand{\frv}[1]{\textkw{frv}(#1)}

\newcommand{\len}[1]{|\hspace{-.6ex}#1\hspace{-0.9ex}|} 
\newcommand{\wt}[1]{\,\lceil#1\rceil} 
\renewcommand{\min}[1]{\textsf{min}\!\wt{#1}}
\newcommand{\bind}[3]{#1 \leftarrow #2;\,#3}

\renewcommand{\P}{{P}}
\newcommand{\Q}{{Q}}

\newcommand{\bag}[1]{\{\!\!\{#1\}\!\!\}}
\newcommand{\set}[1]{\{#1\}}

\newcommand{\ffont}[1]{\textsf{#1}}

\newcommand{\bindop}{\triangleright} 
\newcommand{\opbind}[1]{\,\bindop_{#1}}
\newcommand{\unit}[1]{\ffont{unit}_{#1}}

\newcommand{\fx}{\ffont{x}}
\newcommand{\fy}{\ffont{y}}
\newcommand{\fm}{\ffont{m}}
\newcommand{\ff}{\ffont{f}}
\newcommand{\fe}{\ffont{e}}
\newcommand{\fh}{\ffont{h}}
\newcommand{\ftp}[1]{\llbracket #1 \rrbracket}
\newcommand{\ev}{\ffont{f}}

\newcommand{\stack}[2]{\stackrel{\displaystyle #1}{#2}}

\newcommand{\inl}{\textsf{inl}}
\newcommand{\inr}{\textsf{inr}}
\newcommand{\inductive}{\keyword{inductive}}
\newcommand{\eff}{\textit{eff}}
\newcommand{\heap}{\textit{heap}}

\newcommand{\sem}[1]{\llbracket #1\rrbracket}
\newcommand{\lift}{\ffont{lift}}
\newcommand{\map}[1]{\ffont{map}_{#1}}
\newcommand{\elift}{\ffont{extend}}

\newcommand{\rx}{r}
\newcommand{\hp}[2]{\keyword{hp}\,#1.\,#2}
\newcommand{\fix}{\keyword{fix}}
\newcommand{\new}{\keyword{ref}}
\newcommand{\run}{\keyword{run}}
\newcommand{\throw}{\keyword{throw}}
\newcommand{\catch}{\keyword{catch}}
\newcommand{\letb}[2]{\keyword{let}\,#1\,\keyword{in}\,#2}

\newcommand{\context}[2]{#1[#2]}
\newcommand{\R}[1]{\context{R}{#1}}
\newcommand{\X}[1]{\context{X}{#1}}
\newcommand{\N}[1]{\context{N}{#1}}
\newcommand{\E}[1]{\context{E}{#1}}
\newcommand{\disjoint}{\mathop{\not\!\cap}}

\newcommand\HH[1]{\context{H}{#1}}
\newcommand\HHX[1]{\context{H'}{#1}}

\newcommand\tst[1]{\ec{st}\langle #1\rangle}
\newcommand\sth{\tst{h}}
\newcommand{\exn}{\ec{exn}}
\renewcommand{\div}{\ec{div}}

\newcommand{\kk}{\kappa}
\newcommand{\ke}{\mathsf{e}}
\newcommand{\kl}{\mathsf{k}}
\newcommand{\kh}{\mathsf{h}}
\newcommand{\tref}[2]{\mathit{ref}\langle#1,#2\rangle}

\newcommand{\extdiv}{\ext{\ec{div}}{\e}}

\newcommand{\inference}[2]{\dfrac{#1}{#2}}

\newcommand{\sub}{\theta}
\newcommand{\subempty}{[\,]}
\newcommand{\fresh}[1]{\mathsf{fresh}(#1)}
\newcommand{\unify}[3]{#1 \,\mathop{\sim}\, #2 \;:#3}
\newcommand{\solve}[1]{\mathsf{solve}(#1)}
\newcommand{\gen}[2]{\mathsf{gen}(#1,#2)}

\newcommand{\ontop}[2]{\genfrac{}{}{0pt}{0}{\strut #1}{\strut #2}}
\newcommand{\ontopthree}[3]{\ontop{#1}{\ontop{#2}{#3}}}

\newcommand{\runify}[4]{#1 \simeq #2\;|\;#3 : #4}
\newcommand{\tl}[1]{\mathsf{tl}(#1)}
\newcommand{\ls}{\overline{l}}
\newcommand{\typeof}[1]{\mathsf{typeof}(#1)}
\newcommand{\RX}[1]{\context{R'}{#1}}

\newcommand{\dlongmapsto}{\,\longmapsto\hspace{-2.8ex}\rightarrow\;}

\newcommand{\red}[2]{\mathcal{R}(#1\,|\,#2)}
\newcommand{\redg}[1]{\mathcal{R}(#1)}

\newcommand{\total}{\langle\rangle}\mdTitle{Koka: Programming with Row-polymorphic Effect Types}

\mdAuthor{Daan Leijen}{Microsoft Research}{daan@microsoft.com}{}
\mdMaketitle{}
\begin{mdDiv}[class={abstract},elem={abstract}]%
\begin{mdP}%
We propose a programming model where effects are treated in a disciplined way,
and where the potential side-effects of a function are apparent in its type
signature. The type and effect of expressions can also be inferred
automatically, and we describe a polymorphic type inference system based on
Hindley-Milner style inference. A novel feature is that we support polymorphic
effects through row-polymorphism using duplicate labels. Moreover, we show
that our effects are not just syntactic labels but have a deep semantic
connection to the program. For example, if an expression can be typed without
an \mdEm{exn} effect, then it will never throw an unhandled exception. Similar to
Haskell{'}s \mdSpan[class={code,math-inline}]{$\mathid{runST}$} we show how we can safely encapsulate stateful operations.
Through the state effect, we can also safely combine state with let-polymorphism
without needing either imperative type variables or a syntactic
value restriction. Finally, our system is implemented fully in a new language
called Koka\mdFootnote[id=back-fn-koka,label={\mdSpan[class={footnote-label}]{1}}]{\begin{mdDiv}[class={footnote},id=fn-koka,label={[1]\{.footnote-label\}},elem={footnote}]%
\begin{mdP}%
\mdSpan[class={footnote-before}]{\mdSup{\mdSpan[class={footnote-label}]{1}.} }Koka means {\textquoteleft}effect{\textquoteright} or {\textquoteleft}effective{\textquoteright} in Japanese. 
\mdA[class={footnote-backref,localref}]{back-fn-koka}{}{\ensuremath{\hookleftarrow}}%
\end{mdP}
\end{mdDiv}%
} and has been used successfully on various small to medium-sized 
sample programs ranging from a Markdown processor to a tier-splitted
chat application. You can try out Koka live at 
\mdA{http://www.rise4fun.com/koka/tutorial}{}{www.rise4fun.com/koka/tutorial}.%
\end{mdP}
\end{mdDiv}%
\mdHxx[id=sec-intro,label={[1]\{.heading-label\}},toc={},caption={Introduction}]{\mdSpan[class={heading-before}]{\mdSpan[class={heading-label}]{1}.{\enspace}}Introduction}\begin{mdP}[class={para-continue}]%
We propose a programming model where effects are a part of the type signature
of a function. Currently, types only tell us something about the input and
output value of a function but say nothing about all \mdEm{other} behaviors; for
example, if the function writes to the console or can throw an exception. In
our system, the squaring function:%
\end{mdP}%
\begin{mdDiv}[class={para-block,para-block,input-mathpre},elem={pre}]%
\begin{mdDiv}[class={math-display}]%
\[\begin{mdMathprearray}
\mathkw{function}\mathspace{1}\mathid{sqr}(\mathid{x}\mathspace{1}:\mathspace{1}\mathid{int})\mathspace{1}\{\ \mathid{x}\mathspace{1}*\mathspace{1}\mathid{x}\mathspace{1}\}
\end{mdMathprearray}\]%
\end{mdDiv}
\end{mdDiv}%
\begin{mdP}[class={para-continue}]%
will get the type:%
\end{mdP}%
\begin{mdDiv}[class={para-block,para-block,input-mathpre},elem={pre}]%
\begin{mdDiv}[class={math-display}]%
\[\begin{mdMathprearray}
\mathid{sqr}\mathspace{1}:\mathspace{1}\mathid{int}\mathspace{1}{\rightarrow}\mathspace{1}\mathid{total}\mathspace{1}\mathid{int}
\end{mdMathprearray}\]%
\end{mdDiv}
\end{mdDiv}%
\begin{mdP}[class={para-continue}]%
signifying that \mdSpan[class={code,math-inline}]{$\mathid{sqr}$} has no side effect at all and behaves as a total
function from integers to integers. If we add a \mdSpan[class={code,math-inline}]{$\mathid{print}$} statement though:%
\end{mdP}%
\begin{mdDiv}[class={para-block,para-block,input-mathpre},elem={pre}]%
\begin{mdDiv}[class={math-display}]%
\[\begin{mdMathprearray}
\mathkw{function}\mathspace{1}\mathid{sqr}(\mathid{x}\mathspace{1}:\mathspace{1}\mathid{int})\mathspace{1}\{\ \mathid{print}(\mathid{x});\mathspace{1}\mathid{x}*\mathid{x}\mathspace{1}\}
\end{mdMathprearray}\]%
\end{mdDiv}
\end{mdDiv}%
\begin{mdP}[class={para-continue}]%
the (inferred) type indicates that \mdSpan[class={code,math-inline}]{$\mathid{sqr}$} has an input-output (\mdSpan[class={code,math-inline}]{$\mathid{io}$}) effect:%
\end{mdP}%
\begin{mdDiv}[class={para-block,para-block,input-mathpre},elem={pre}]%
\begin{mdDiv}[class={math-display}]%
\[\begin{mdMathprearray}
\mathid{sqr}\mathspace{1}:\mathspace{1}\mathid{int}\mathspace{1}{\rightarrow}\mathspace{1}\mathid{io}\mathspace{1}\mathid{int}
\end{mdMathprearray}\]%
\end{mdDiv}
\end{mdDiv}%
\begin{mdP}%
Note that there was no need to change the original function nor to promote the
expression \mdSpan[class={code,math-inline}]{$\mathid{x}*\mathid{x}$} into the \mdSpan[class={code,math-inline}]{$\mathid{io}$} effect. One of our goals is to make
effects convenient for the programmer, so we automatically combine effects. In
particular, this makes it convenient for the programmer to use precise effects
without having to insert coercions.%
\end{mdP}%
\begin{mdP}[class={indent}]%
There have been many proposals for effects systems in the past 
\mdSpan[class={citations},target-element={bibitem}]{[\mdA[class={bibref,localref},target-element={bibitem}]{benton:exceptioneffects}{}{\mdSpan[class={bibitem-label}]{2}}, \mdA[class={bibref,localref},target-element={bibitem}]{gifford:imperative}{}{\mdSpan[class={bibitem-label}]{8}}, \mdA[class={bibref,localref},target-element={bibitem}]{lucassen:polyeffect}{}{\mdSpan[class={bibitem-label}]{24}}, \mdA[class={bibref,localref},target-element={bibitem}]{marino:genericeffects}{}{\mdSpan[class={bibitem-label}]{25}}, \mdA[class={bibref,localref},target-element={bibitem}]{nielson:polyeffect}{}{\mdSpan[class={bibitem-label}]{28}}, \mdA[class={bibref,localref},target-element={bibitem}]{scala:effects}{}{\mdSpan[class={bibitem-label}]{33}}, \mdA[class={bibref,localref},target-element={bibitem}]{talpin:effects}{}{\mdSpan[class={bibitem-label}]{36}}, \mdA[class={bibref,localref},target-element={bibitem}]{leijen:effects-tr}{}{\mdSpan[class={bibitem-label}]{38}}, \mdA[class={bibref,localref},target-element={bibitem}]{wadler:marriage}{}{\mdSpan[class={bibitem-label}]{42}}]}.  However, many
such systems suffer from being syntactical in nature (i.e. effects are just
labels), or by being quite restricted, for example being monomorphic or
applied to a very specific set of effects. Some of the more general systems
suffer from having complicated effect types, especially in a polymorphic
setting that generally requires sub-effect constraints.%
\end{mdP}%
\begin{mdP}[class={indent,para-continue}]%
Our main contribution in this paper is the novel combination of existing
techniques into the design of a practical ML-like language with strong effect
typing. In addition, many known techniques are applied in a novel way: ranging 
from effect types as rows with duplicate labels to the safety of \mdSpan[class={code,math-inline}]{$\mathid{runST}$} in
a strict setting. In particular:%
\end{mdP}%
\begin{mdUl}[class={loose}]%
\begin{mdLi}%
\begin{mdP}%
We describe a novel effect system based on row polymorphism which allows
  \mdEm{duplicated} effects. This simplifies the effect types and provides natural
  types to effect elimination forms, like catching exceptions.%
\end{mdP}
\end{mdLi}%
\begin{mdLi}%
\begin{mdP}%
The effect types are not just syntactic labels but they have a deep
  semantic connection to the program (Section{\mdNbsp}\mdA[class={localref},target-element={h1}]{sec-semeffect}{}{\mdSpan[class={heading-label}]{6}}). For example, we
  can prove that if an expression that can be typed without an \mdEm{exn} effect,
  then it will never throw an unhandled exception; or if an expression can be
  typed without a \mdEm{div} effect, then it always terminates.%
\end{mdP}
\end{mdLi}%
\begin{mdLi}%
\begin{mdP}%
The interaction between polymorphism and mutable state is fraught with
  danger. We show that by modeling state as an effect we can safely combine
  mutability with let-polymorphism without needing either imperative type
  variables, nor a syntactic value restriction.
  Moreover, we can safely encapsulate local state operations and we prove
  that such encapsulation is sound where no references or stateful behavior
  can escape the encapsulation scope.%
\end{mdP}%
\begin{mdP}%
  The interaction between divergence and higher-order mutable state is also
  tricky. Again, we show how explicit heap effects allow us to safely infer
  whether stateful operations may diverge.%
\end{mdP}
\end{mdLi}%
\begin{mdLi}%
\begin{mdP}%
We have an extensive experience with the type system within the Koka
  language. The Koka language fully implements the effect types as described in
  this paper and we have used it successfully in various small to medium
  sized code examples ranging from a fully compliant Markdown text processor
  to a tier-splitted chat application (Section{\mdNbsp}\mdA[class={localref},target-element={h2}]{sec-practice}{}{\mdSpan[class={heading-label}]{2.8}}).%
\end{mdP}
\end{mdLi}
\end{mdUl}%
\mdHxx[id=overview,label={[2]\{.heading-label\}},toc={},caption={Overview}]{\mdSpan[class={heading-before}]{\mdSpan[class={heading-label}]{2}.{\enspace}}Overview}\begin{mdP}%
Types tell us about the behavior of functions. For example, if suppose 
we have the type of a function \mdSpan[class={code,math-inline}]{$\mathid{foo}$} in ML with type \mdSpan[class={code,math-inline}]{$\mathid{int}\mathspace{1}{\rightarrow}\mathspace{1}\mathid{int}$}.
We know that \mdSpan[class={code,math-inline}]{$\mathid{foo}$} is well defined on inputs of type \mdSpan[class={code,math-inline}]{$\mathid{int}$} and returns values
of type \mdSpan[class={code,math-inline}]{$\mathid{int}$}. But that is only part of the story, the type tells us
nothing about all \mdEm{other} behaviors: i.e. if it accesses the file system perhaps, 
or  throws exceptions, or never returns a result at all.%
\end{mdP}%
\begin{mdP}[class={indent}]%
Even {\textquoteleft}pure{\textquoteright} functional languages
like Haskell do not fare much better at this. Suppose our function
has the Haskell type \mdSpan[class={code,math-inline}]{$\mathid{Int}\mathspace{1}{\rightarrow}\mathspace{1}\mathid{Int}$}.
Even though we know now there is no arbitrary side-effect, we still do not
know  whether this function terminates or can throw exceptions. Due to
laziness, we do not even know if the result itself, when demanded, will raise
an exception or diverge; i.e. even a simple transformation like \mdSpan[class={code,math-inline}]{$\mathid{x}*0$} to \mdSpan[class={code,math-inline}]{$0$}
is not possible under Haskell{'}s notion of purity.%
\end{mdP}%
\begin{mdP}[class={indent}]%
In essence, in both ML and
Haskell the types are not precise enough to describe many aspects of the
static behavior of  a program. In the Haskell case, the real type is more like
\mdSpan[class={code,math-inline}]{$(\mathid{Int}_\bot {\rightarrow}\mathspace{1}\mathid{Int}_\bot)_\bot $} while the type signature of the ML program
should really include that any kind of side-effect might happen.%
\end{mdP}%
\begin{mdP}[class={indent}]%
Functional programming has been done wrong! We believe it is essential for
types to include potential behaviors like divergence, exceptions, or
statefulness. Being able to reason about these aspects is crucial in many
domains, including safe parallel execution, optimization, query embedding,
tiersplitting, etc.%
\end{mdP}%
\mdHxxx[id=effect-types,label={[2.1]\{.heading-label\}},toc={},caption={Effect types}]{\mdSpan[class={heading-before}]{\mdSpan[class={heading-label}]{2.1}.{\enspace}}Effect types}\begin{mdP}%
To address the previous problems, we take a fresh look at programming with
side-effects in the context of a new language called Koka{\mdNbsp}\mdSpan[class={citations},target-element={bibitem}]{[\mdA[class={bibref,localref},target-element={bibitem}]{koka}{}{\mdSpan[class={bibitem-label}]{18}}, \mdA[class={bibref,localref},target-element={bibitem}]{leijen:kokatr}{}{\mdSpan[class={bibitem-label}]{19}}]}.\mdBr
Like ML,
Koka has strict semantics where arguments are evaluated before calling a
function. This implies that an expression with type \mdSpan[class={code,math-inline}]{$\mathid{int}$} can really be
modeled semantically as an integer (and not as a delayed computation that can
potentially diverge or raise an exception).%
\end{mdP}%
\begin{mdP}[class={indent}]%
As a consequence, the \mdEm{only point where side effects can occur is during
function application}. We write function types as \mdSpan[class={code,math-inline}]{$(\t_1,\dots ,\t_\mathid{n})\mathspace{1}{\rightarrow}\mathspace{1}\e\
\tau$} to denote that a function takes arguments of type \mdSpan[class={math-inline}]{$\t_1$} to \mdSpan[class={math-inline}]{$\t_n$}, and
returns a value of type \mdSpan[class={code,math-inline}]{$\t$} with a potential side effect \mdSpan[class={code,math-inline}]{$\e$}. As apparent
from the type, functions need to be fully applied and are not curried. This is
to make it immediately apparent where side effects can occur. For example, in
ML, an expression like \mdSpan[class={code,math-inline}]{$\mathid{f}\mathspace{1}\mathid{x}\mathspace{1}\mathid{y}$} can have side effects
at different points depending on the arity of the function \mdSpan[class={code,math-inline}]{$\mathid{f}$}. In our system
this is immediately apparent, as one writes either \mdSpan[class={code,math-inline}]{$\mathid{f}(\mathid{x},\mathid{y})$} or \mdSpan[class={code,math-inline}]{$(\mathid{f}(\mathid{x}))(\mathid{y})$}.%
\end{mdP}%
\mdHxxx[id=basic-effects,label={[2.2]\{.heading-label\}},toc={},caption={Basic effects}]{\mdSpan[class={heading-before}]{\mdSpan[class={heading-label}]{2.2}.{\enspace}}Basic effects}\begin{mdP}[class={para-continue}]%
The effects in our system are extensible, but the basic effects defined in
Koka are \mdSpan[class={code,math-inline}]{$\mathid{total}$}, \mdSpan[class={code,math-inline}]{$\mathid{exn}$}, \mdSpan[class={code,math-inline}]{$\mathid{div}$}, \mdSpan[class={code,math-inline}]{$\mathid{ndet}$},  \mdSpan[class={code,math-inline}]{$\mathid{alloc}{\langle}\mathid{h}{\rangle}$}, \mdSpan[class={code,math-inline}]{$\mathid{read}{\langle}\mathid{h}{\rangle}$}, \mdSpan[class={code,math-inline}]{$\mathid{write}{\langle}\mathid{h}{\rangle}$},
and \mdSpan[class={code,math-inline}]{$\mathid{io}$}. Of course \mdSpan[class={code,math-inline}]{$\mathid{total}$} is not really an effect but signifies the absence
of any effect and is assigned to pure mathematical functions. When a function
can throw an exception, it gets the \mdSpan[class={code,math-inline}]{$\mathid{exn}$} effect. Potential divergence or
non-termination is signified by the \mdSpan[class={code,math-inline}]{$\mathid{div}$} effect. Currently, Koka uses a
simple termination analysis based on inductive data types to assign this
effect to recursive functions. Non-deterministic functions get the \mdSpan[class={code,math-inline}]{$\mathid{ndet}$}
effect. The effects \mdSpan[class={code,math-inline}]{$\mathid{alloc}{\langle}\mathid{h}{\rangle}$}, \mdSpan[class={code,math-inline}]{$\mathid{read}{\langle}\mathid{h}{\rangle}$} and \mdSpan[class={code,math-inline}]{$\mathid{write}{\langle}\mathid{h}{\rangle}$} are used for stateful
functions over a heap \mdSpan[class={code,math-inline}]{$\mathid{h}$}. Finally \mdSpan[class={code,math-inline}]{$\mathid{io}$} is used for functions that do any
input/output operations.
Here are some type signatures of common functions in Koka:%
\end{mdP}%
\begin{mdDiv}[class={para-block,para-block,input-mathpre},elem={pre}]%
\begin{mdDiv}[class={math-display}]%
\[\begin{mdMathprearray}
\mathid{random}\mathspace{1}&:\mathspace{1}()\mathspace{1}{\rightarrow}\mathspace{1}\mathid{ndet}\mathspace{1}\mathid{double}\mathbr{}
\mathid{print}\mathspace{2}&:\mathspace{1}\mathid{string}\mathspace{1}{\rightarrow}\mathspace{1}\mathid{io}\mathspace{1}()\mathbr{}
\mathid{error}\mathspace{2}&:\mathspace{1}\forall\alpha.\mathspace{1}\mathid{string}\mathspace{1}{\rightarrow}\mathspace{1}\mathid{exn}\mathspace{1}\mathid{a}\mathbr{}
(:=)\mathspace{3}&:\mathspace{1}\forall\alpha.\mathspace{1}(\mathid{ref}{\langle}\mathid{h},\mathid{a}{\rangle},\mathspace{1}\mathid{a})\mathspace{1}{\rightarrow}\mathspace{1}\mathid{write}{\langle}\mathid{h}{\rangle}\mathspace{1}()
\end{mdMathprearray}\]%
\end{mdDiv}
\end{mdDiv}%
\begin{mdP}[class={para-continue}]%
Note that we use angled brackets to denote type application as usual in
languages like C\# or Scala. We also use angled brackets to denote a \mdEm{row} of
effects. For example, the program:%
\end{mdP}%
\begin{mdDiv}[class={para-block,para-block,input-mathpre},elem={pre}]%
\begin{mdDiv}[class={math-display}]%
\[\begin{mdMathprearray}
\mathkw{function}\mathspace{1}\mathid{sqr}(\mathspace{1}\mathid{x}\mathspace{1}:\mathspace{1}\mathid{int}\mathspace{1})\mathspace{1}\{\mathspace{1}\mathid{error}({\tt "hi"});\mathspace{1}\mathid{sqr}(\mathid{x});\mathspace{1}\mathid{x}*\mathid{x}\mathspace{1}\}
\end{mdMathprearray}\]%
\end{mdDiv}
\end{mdDiv}%
\begin{mdP}[class={para-continue}]%
will get the type%
\end{mdP}%
\begin{mdDiv}[class={para-block,para-block,input-mathpre},elem={pre}]%
\begin{mdDiv}[class={math-display}]%
\[\begin{mdMathprearray}
\mathid{sqr}\mathspace{1}:\mathspace{1}\mathid{int}\mathspace{1}{\rightarrow}\mathspace{1}{\langle}\mathid{exn},\mathid{div}{\rangle}\mathspace{1}\mathid{int}
\end{mdMathprearray}\]%
\end{mdDiv}
\end{mdDiv}%
\begin{mdP}[class={para-continue}]%
where we combined the two basic effects \mdSpan[class={code,math-inline}]{$\mathid{exn}$} and \mdSpan[class={code,math-inline}]{$\mathid{div}$} into a row of effects
\mdSpan[class={code,math-inline}]{${\langle}\mathid{exn},\mathspace{1}\mathid{div}{\rangle}$}. The combination of the exception and divergence effect
corresponds exactly to Haskell{'}s notion of purity, and we call this effect
\mdSpan[class={code,math-inline}]{$\mathid{pure}$}. Common type aliases are:%
\end{mdP}%
\begin{mdDiv}[class={para-block,para-block,input-mathpre},elem={pre}]%
\begin{mdDiv}[class={math-display}]%
\[\begin{mdMathprearray}
\mathkw{alias}\mathspace{1}\mathid{total}\mathspace{1}&=\mathspace{1}{\langle}{\rangle}\mathbr{}
\mathkw{alias}\mathspace{1}\mathid{pure}\mathspace{2}&=\mathspace{1}{\langle}\mathid{exn},\mathspace{1}\mathid{div}{\rangle}\mathbr{}
\mathkw{alias}\mathspace{1}\mathid{st}{\langle}\mathid{h}{\rangle}\mathspace{1}&=\mathspace{1}{\langle}\mathid{alloc}{\langle}\mathid{h}{\rangle},\mathspace{1}\mathid{read}{\langle}\mathid{h}{\rangle},\mathspace{1}\mathid{write}{\langle}\mathid{h}{\rangle}{\rangle}\mathbr{}
\mathkw{alias}\mathspace{1}\mathid{io}\mathspace{4}&=\mathspace{1}{\langle}\mathid{st}{\langle}\mathid{ioheap}{\rangle},\mathspace{1}\mathid{pure},\mathspace{1}\mathid{ndet}{\rangle}
\end{mdMathprearray}\]%
\end{mdDiv}
\end{mdDiv}%
\begin{mdP}%
This hierarchy is clearly inspired by Haskell{'}s standard monads and we use
this as a starting point for more refined effects which we hope to explore in
Koka. For example, blocking, client/server side effects, reversable
operations, etc.%
\end{mdP}%
\mdHxxx[id=polymorphic-effects,label={[2.3]\{.heading-label\}},toc={},caption={Polymorphic effects}]{\mdSpan[class={heading-before}]{\mdSpan[class={heading-label}]{2.3}.{\enspace}}Polymorphic effects}\begin{mdP}[class={para-continue}]%
Often, the effect of a function is determined by the effects of functions
passed to it. For example, the \mdSpan[class={code,math-inline}]{$\mathid{map}$} function which maps a function over all
elements of a list will have the type:%
\end{mdP}%
\begin{mdDiv}[class={para-block,para-block,input-mathpre},elem={pre}]%
\begin{mdDiv}[class={math-display}]%
\[\begin{mdMathprearray}
\mathid{map}\mathspace{1}:\mathspace{1}\forall\alpha\beta\mu.\mathspace{1}(\mathid{list}{\langle}\alpha{\rangle},\mathspace{1}\alpha {\rightarrow}\mathspace{1}\mu\ \beta)\mathspace{1}{\rightarrow}\mathspace{1}\mu\ \mathid{list}{\langle}\beta{\rangle}
\end{mdMathprearray}\]%
\end{mdDiv}
\end{mdDiv}%
\begin{mdP}[class={para-continue}]%
where the effect of the \mdSpan[class={code,math-inline}]{$\mathid{map}$} function itself is completely determined by the
effect of its argument. In this case, a simple and obvious type is assigned to
\mdSpan[class={code,math-inline}]{$\mathid{map}$}, but one can quickly create more complex examples where the type may not
be obvious at first. Consider the following program:%
\end{mdP}%
\begin{mdDiv}[class={para-block,para-block,input-mathpre},elem={pre}]%
\begin{mdDiv}[class={math-display}]%
\[\begin{mdMathprearray}
\mathkw{function}\mathspace{1}\mathid{foo}(\mathid{f},\mathid{g})\mathspace{1}\{\ \mathid{f}();\mathspace{1}\mathid{g}();\mathspace{1}\mathid{error}({\tt "\mathid{hi}"})\mathspace{1}\}
\end{mdMathprearray}\]%
\end{mdDiv}
\end{mdDiv}%
\begin{mdP}[class={para-continue}]%
Clearly, the effect of \mdSpan[class={code,math-inline}]{$\mathid{foo}$} is a combination of the effects of \mdSpan[class={code,math-inline}]{$\mathid{f}$} and \mdSpan[class={code,math-inline}]{$\mathid{g}$},
and the \mdSpan[class={code,math-inline}]{$\mathid{exn}$} effect. One possible design choice is to have a \mdSpan[class={math-inline}]{$\cup$} operation
on effect types, and write the type of \mdSpan[class={code,math-inline}]{$\mathid{foo}$} as:%
\end{mdP}%
\begin{mdDiv}[class={para-block,para-block,input-mathpre},elem={pre}]%
\begin{mdDiv}[class={math-display}]%
\[\begin{mdMathprearray}
\forall\mu_1\mu_2.\mathspace{1}(()\mathspace{1}{\rightarrow}\mathspace{1}\mu_1\mathspace{1}(),\mathspace{1}()\mathspace{1}{\rightarrow}\mathspace{1}\mu_2\mathspace{1}())\mathspace{1}{\rightarrow}\mathspace{1}(\mu_1\cup \mu_2\cup \mathid{exn})\mathspace{1}()
\end{mdMathprearray}\]%
\end{mdDiv}
\end{mdDiv}%
\begin{mdP}[class={para-continue}]%
Unfortunately, this quickly gets us in trouble during type inference:
unification can lead to constraints of the form \mdSpan[class={math-inline}]{$\mu_1 \cup\ \mu_2 \sim \mu_3
\cup\ \mu_4$} which cannot be solved uniquely and must become part of the type
language. Another design choice is to introduce subtyping over effects and
write the type of \mdSpan[class={code,math-inline}]{$\mathid{foo}$} as:%
\end{mdP}%
\begin{mdDiv}[class={para-block,para-block,input-mathpre},elem={pre}]%
\begin{mdDiv}[class={math-display}]%
\[\begin{mdMathprearray}
\forall\mu_1\mu_2\mu_3.\mathspace{1}(\mu_1\leq\mu_3,\mathspace{1}\mu_2\leq\mu_3,\mathspace{1}{\langle}\mathid{exn}{\rangle}\leq\mu_3)\mathspace{1}\Rightarrow (()\mathspace{1}{\rightarrow}\mathspace{1}\mu_1\mathspace{1}(),\mathspace{1}()\mathspace{1}{\rightarrow}\mathspace{1}\mu_2\mathspace{1}())\mathspace{1}{\rightarrow}\mathspace{1}\mu_3\mathspace{1}()
\end{mdMathprearray}\]%
\end{mdDiv}
\end{mdDiv}%
\begin{mdP}%
This is the choice made in an earlier version of Koka described as a technical
report{\mdNbsp}\mdSpan[class={citations},target-element={bibitem}]{[\mdA[class={bibref,localref},target-element={bibitem}]{leijen:effects-tr}{}{\mdSpan[class={bibitem-label}]{38}}]}. However, in our experience with that system in
practice we felt the constraints often became quite complex and the
combination of polymorphism with subtyping can make type inference
undecidable.%
\end{mdP}%
\begin{mdP}[class={indent,para-continue}]%
The approach we advocate in this paper and which is adopted by Koka is the use
of row-polymorphism on effects. Row polymorphism is well understood and used
for many inference systems for record calculi{\mdNbsp}\mdSpan[class={citations},target-element={bibitem}]{[\mdA[class={bibref,localref},target-element={bibitem}]{gasterjones:trex}{}{\mdSpan[class={bibitem-label}]{7}}, \mdA[class={bibref,localref},target-element={bibitem}]{leijen:scopedlabels}{}{\mdSpan[class={bibitem-label}]{17}}, \mdA[class={bibref,localref},target-element={bibitem}]{lindley:effects}{}{\mdSpan[class={bibitem-label}]{23}}, \mdA[class={bibref,localref},target-element={bibitem}]{remy:mlart}{}{\mdSpan[class={bibitem-label}]{31}}, \mdA[class={bibref,localref},target-element={bibitem}]{sulzmann:records}{}{\mdSpan[class={bibitem-label}]{34}}, \mdA[class={bibref,localref},target-element={bibitem}]{sulzmann:recordsrev}{}{\mdSpan[class={bibitem-label}]{35}}]}. We use the notation \mdSpan[class={code,math-inline}]{${\langle}\mathid{l}\mathspace{1}|\mathspace{1}\mu{\rangle}$} to extend an effect row
\mdSpan[class={math-inline}]{$\mu$} with an effect constant \mdSpan[class={math-inline}]{$l$}. Rows can now have two forms, either a
\mdEm{closed} effect \mdSpan[class={code,math-inline}]{${\langle}\mathid{exn},\mathid{div}{\rangle}$}, or an \mdEm{open} effect ending in an effect variable
\mdSpan[class={math-inline}]{$\ext{\exn, \div}{\mu}$}.  Using an open effect, our system infers the
following type for \mdSpan[class={code,math-inline}]{$\mathid{foo}$}:%
\end{mdP}%
\begin{mdDiv}[class={para-block,para-block,input-mathpre},elem={pre}]%
\begin{mdDiv}[class={math-display}]%
\[\begin{mdMathprearray}
\mathid{foo}\mathspace{1}:\mathspace{1}\forall\mu.\mathspace{1}(()\mathspace{1}{\rightarrow}\mathspace{1}{\langle}\mathid{exn}\mathspace{1}|\mathspace{1}\mu{\rangle}\mathspace{1}(),\mathspace{1}()\mathspace{1}{\rightarrow}\mathspace{1}{\langle}\mathid{exn}\mathspace{1}|\mathspace{1}\mu{\rangle}\mathspace{1}())\mathspace{1}{\rightarrow}\mathspace{1}{\langle}\mathid{exn}\mathspace{1}|\mathspace{1}\mu{\rangle}\mathspace{1}()
\end{mdMathprearray}\]%
\end{mdDiv}
\end{mdDiv}%
\begin{mdP}%
The reader may worry at this point that the row polymorphic type is more
restrictive than the earlier type using subtype constraints: indeed, the row
polymorphic type requires that each function argument now has the same effect
\mdSpan[class={math-inline}]{$\ext{\exn}{\mu}$}. However, in a calling context \mdSpan[class={code,math-inline}]{$\mathid{foo}(\mathid{f},\mathid{g})$} our system ensures
that we always infer a polymorphic open effect for each expression \mdSpan[class={math-inline}]{$f$} and
\mdSpan[class={math-inline}]{$g$}. For example, \mdSpan[class={math-inline}]{$f : \tfun{()}{\ext{\exn}{\mu_1}}{()}$} and \mdSpan[class={math-inline}]{$g :
\tfun{()}{\ext{\div}{\mu_2}}{()}$}. This allows the types \mdSpan[class={math-inline}]{$\ext{\exn}{\mu_1}$}
and \mdSpan[class={math-inline}]{$\ext{\div}{\mu_2}$} to unify into a common type \mdSpan[class={math-inline}]{$\ext{\exn, \div}{\mu_3}$}
such that they can be applied to \mdSpan[class={code,math-inline}]{$\mathid{foo}$}, resulting in an inferred effect
\mdSpan[class={math-inline}]{$\ext{\exn, \div}{\mu_3}$} for \mdSpan[class={code,math-inline}]{$\mathid{foo}(\mathid{f},\mathid{g})$}.%
\end{mdP}%
\mdHxxx[id=duplicate-effects,label={[2.4]\{.heading-label\}},toc={},caption={Duplicate effects}]{\mdSpan[class={heading-before}]{\mdSpan[class={heading-label}]{2.4}.{\enspace}}Duplicate effects}\begin{mdP}%
Our effect rows differ in an important way from the usual approaches in that
effect labels can be duplicated, i.e. \mdSpan[class={math-inline}]{$\row{\exn,\exn} \not\equiv \row{\exn}$}
\mdStrong{(1)}. This was first described by Leijen{\mdNbsp}\mdSpan[class={citations},target-element={bibitem}]{[\mdA[class={bibref,localref},target-element={bibitem}]{leijen:scopedlabels}{}{\mdSpan[class={bibitem-label}]{17}}]}
where this was used to enable scoped labels in record types. Enabling
duplicate labels fits our approach well: first of all, it enables
principal unification without needing extra constraints, and secondly, it
enables us to give precise types to effect elimination forms (like catching
exceptions).%
\end{mdP}%
\begin{mdP}[class={indent}]%
In particular, during unification we may end up with constraints of the form
\mdSpan[class={math-inline}]{$\ext{\exn}{\mu} \,\sim\, \row{\exn}$}. With regular row-polymorphism which are
sets of labels, such constraint can have multiple solutions, namely \mdSpan[class={math-inline}]{$\mu =
\row{}$} or \mdSpan[class={math-inline}]{$\mu = \row{\exn}$}. This was first observed by Wand
\mdSpan[class={citations},target-element={bibitem}]{[\mdA[class={bibref,localref},target-element={bibitem}]{wand:records}{}{\mdSpan[class={bibitem-label}]{43}}]} in the context of records. Usually, this problem is fixed by
either introducing \mdEm{lacks} constraints{\mdNbsp}\mdSpan[class={citations},target-element={bibitem}]{[\mdA[class={bibref,localref},target-element={bibitem}]{gasterjones:trex}{}{\mdSpan[class={bibitem-label}]{7}}]} or polymorphic
presence and absence flags on each label{\mdNbsp}\mdSpan[class={citations},target-element={bibitem}]{[\mdA[class={bibref,localref},target-element={bibitem}]{remy:records}{}{\mdSpan[class={bibitem-label}]{32}}]} (as used by Lindley
and Cheney{\mdNbsp}\mdSpan[class={citations},target-element={bibitem}]{[\mdA[class={bibref,localref},target-element={bibitem}]{lindley:effects}{}{\mdSpan[class={bibitem-label}]{23}}]} for an effect system in the context of database
queries). The problem with both approaches is that they complicate the type
system quite a bit. With \mdEm{lacks} contstraints we need a system of qualified 
types as in Haskell, while with presece and absence flags, we need a kind 
system that tracks for each type variable which labels cannot be present.%
\end{mdP}%
\begin{mdP}[class={indent}]%
With rows allowing duplicate labels, we avoid any additional machinery and
can use straight forward type inference techniques. In our case \mdSpan[class={math-inline}]{$\mu = \row{}$}
is the only solution to the above constraint (due to (1)).%
\end{mdP}%
\begin{mdP}[class={indent,para-continue}]%
Moreover, duplicate labels make it easy to give types to effect elimination forms.
For example, catching effects removes the \mdSpan[class={math-inline}]{$\exn$} effect:%
\end{mdP}%
\begin{mdDiv}[class={para-block,para-block,input-mathpre},elem={pre}]%
\begin{mdDiv}[class={math-display}]%
\[\begin{mdMathprearray}
\mathid{catch}\mathspace{1}:\mathspace{1}\forall\alpha\mu.\mathspace{1}(()\mathspace{1}{\rightarrow}\mathspace{1}{\langle}\mathid{exn}\mathspace{1}|\mathspace{1}\mu{\rangle}\ \alpha,\mathspace{1}\mathid{exception}\mathspace{1}{\rightarrow}\mathspace{1}\mu\ \alpha)\mathspace{1}{\rightarrow}\mathspace{1}\mu\ \alpha
\end{mdMathprearray}\]%
\end{mdDiv}
\end{mdDiv}%
\begin{mdP}[class={para-continue}]%
Here we assume that \mdSpan[class={math-inline}]{$catch$} takes two functions, the action and the exception
handler that takes as an argument the thrown \mdSpan[class={code,math-inline}]{$\mathid{exception}$}. The \mdSpan[class={code,math-inline}]{$\mathid{exn}$}
effect of the action is discarded in the final effect \mdSpan[class={code,math-inline}]{$\mu$} since all
exceptions are caught by the handler.  But of course, the handler can itself
throw an exception and have an \mdSpan[class={code,math-inline}]{$\mathid{exn}$} effect itself. In that case \mdSpan[class={code,math-inline}]{$\mu$} will
unify with a type of the form \mdSpan[class={code,math-inline}]{${\langle}\mathid{exn}|\mu'{\rangle}$} giving action the effect
\mdSpan[class={code,math-inline}]{${\langle}\mathid{exn}|\mathid{exn}|\mu'{\rangle}$} where \mdSpan[class={code,math-inline}]{$\mathid{exn}$} occurs duplicated, which gives us exactly the
right behavior. Note that with \mdEm{lacks} constraints we would not be able to
type  this example  because there would be a \mdSpan[class={code,math-inline}]{$\mathid{exn}\not\in \mu$} constraint. We
can type this example using flags but the type would arguably be more complex
with a polymorphic presence/absence flag \mdSpan[class={code,math-inline}]{$\varphi$} on the \mdSpan[class={code,math-inline}]{$\mathid{exn}$} label in the result
effect, something like:%
\end{mdP}%
\begin{mdDiv}[class={para-block,para-block,input-mathpre},elem={pre}]%
\begin{mdDiv}[class={math-display}]%
\[\begin{mdMathprearray}
\mathid{catch}\mathspace{1}:\mathspace{1}\forall\mu\alpha\varphi.\mathspace{1}(()\mathspace{1}{\rightarrow}\mathspace{1}{\langle}\mathid{exn}_\bullet |\mathspace{1}\mu{\rangle}\ \alpha,\mathspace{1}\mathid{exception}\mathspace{1}{\rightarrow}\mathspace{1}{\langle}\mathid{exn}_\varphi |\mathspace{1}\mu{\rangle}\ \alpha)\mathspace{1}{\rightarrow}\mathspace{1}{\langle}\mathid{exn}_\varphi |\mathspace{1}\mu{\rangle}\ \alpha
\end{mdMathprearray}\]%
\end{mdDiv}
\end{mdDiv}%
\begin{mdP}%
There is one situation where an approach with flags is more expressive
though: with flags one can state specifically that a certain effect must be
absent. This is used for example in the effect system by Lindley and 
Cheney{\mdNbsp}\mdSpan[class={citations},target-element={bibitem}]{[\mdA[class={bibref,localref},target-element={bibitem}]{lindley:effects}{}{\mdSpan[class={bibitem-label}]{23}}]} to enforce that database queries never have the
\mdEm{wild} effect{\mdNbsp}(\mdEm{io}). In our setting we can only enforce absence of an effect
by  explicitly listing a closed row of the allowed effects which is less
modular. In our current experience this has not yet proven to be  a problem in
practice though but we may experiment with this in the future.%
\end{mdP}%
\mdHxxxx[id=injection,label={[2.4.1]\{.heading-label\}},toc={},caption={Injection}]{\mdSpan[class={heading-before}]{\mdSpan[class={heading-label}]{2.4.1}.{\enspace}}Injection}\begin{mdP}[class={para-continue}]%
There are some situations where having duplicate effects is quite different
from other approaches though. Intuitively, we can do a monadic translation 
of a Koka program where effect types get translated to a sequence of monad
transformers. Under such interpretation, duplicate effects would have real
semantic significance. For example, we could provide an \mdEm{injection}
operation for exceptions:%
\end{mdP}%
\begin{mdDiv}[class={para-block,para-block,input-mathpre},elem={pre}]%
\begin{mdDiv}[class={math-display}]%
\[\begin{mdMathprearray}
\mathid{inject}\mathspace{1}:\mathspace{1}\forall\mu\alpha.\mathspace{1}(()\mathspace{1}{\rightarrow}\mathspace{1}{\langle}\mathid{exn}|\mu{\rangle}\mathspace{1}\alpha)\mathspace{1}{\rightarrow}\mathspace{1}(()\mathspace{1}{\rightarrow}\mathspace{1}{\langle}\mathid{exn}|\mathid{exn}|\mu{\rangle}\mathspace{1}\alpha)
\end{mdMathprearray}\]%
\end{mdDiv}
\end{mdDiv}%
\begin{mdP}%
Semantically, it would inject an extra exception layer, such that a \mdSpan[class={code,math-inline}]{$\mathid{catch}$} 
operation would only catch exceptions raised from the outer \mdSpan[class={code,math-inline}]{$\mathid{exn}$}, while passing the
inner injected exceptions through. Internally, one can implement this by maintaining
level numbers on thrown exceptions {\textendash} increasing them on an \mdSpan[class={code,math-inline}]{$\mathid{inject}$} and decreasing
them on a \mdSpan[class={code,math-inline}]{$\mathid{catch}$} (and only catching level 0 exceptions).%
\end{mdP}%
\begin{mdP}[class={indent,para-continue}]%
Now, suppose we have some library code whose exceptions we do not want to handle,
but we do want to handle the exceptions in our own code. In that case, we could
write something like:%
\end{mdP}%
\begin{mdDiv}[class={para-block,para-block,input-mathpre},elem={pre}]%
\begin{mdDiv}[class={math-display}]%
\[\begin{mdMathprearray}
\mathid{catch}(\mathspace{1}\mathkw{function}()\mathspace{1}\{\mathspace{1}\mathbr{}
\mathindent{4}...\mathspace{1}\textrm{my code}\mathspace{1}...\mathbr{}
\mathindent{4}\mathid{x}\mathspace{1}=\mathspace{1}\mathid{inject}(\mathspace{1}...\mathspace{1}\textrm{library code}\mathspace{1}...)()\mathbr{}
\mathindent{4}...\mathspace{1}\textrm{my code}\mathspace{1}...\mathspace{1}\mathbr{}
\mathindent{4}\mathid{inject}(\mathspace{1}...\mathrm{more library code}\mathspace{1}...)()\mathbr{}
\},\mathspace{1}\mathid{handler}\mathspace{1})
\end{mdMathprearray}\]%
\end{mdDiv}
\end{mdDiv}%
\begin{mdP}%
In the example, all exceptions in {\textquoteleft}my code{\textquoteright} are caught, while exceptions raised
in the library code are only caught by an outer exception handler.
For this article though, we will not further formalize \mdSpan[class={code,math-inline}]{$\mathid{inject}$} but only
describe the core calculus.%
\end{mdP}%
\mdHxxx[id=sec-state,label={[2.5]\{.heading-label\}},toc={},caption={Heap effects}]{\mdSpan[class={heading-before}]{\mdSpan[class={heading-label}]{2.5}.{\enspace}}Heap effects}\begin{mdP}[class={para-continue}]%
One of the most useful side-effects is of course mutable state.
Here is an example where we give a linear version of the fibonacci function 
using imperative updates:%
\end{mdP}%
\begin{mdDiv}[class={para-block,para-block,input-mathpre},elem={pre}]%
\begin{mdDiv}[class={math-display}]%
\[\begin{mdMathprearray}
\mathkw{function}\mathspace{1}\mathid{fib}(\mathspace{1}\mathid{n}\mathspace{1}:\mathid{int}\mathspace{1})\mathspace{1}\{\mathbr{}
\mathindent{2}\mathkw{val}\mathspace{1}\mathid{x}\mathspace{1}=\mathspace{1}\mathid{ref}(0);\mathspace{1}\mathkw{val}\mathspace{1}\mathid{y}\mathspace{1}=\mathspace{1}\mathid{ref}(1)\mathbr{}
\mathindent{2}\mathkw{repeat}(\mathid{n})\mathspace{1}\{\mathbr{}
\mathindent{4}\mathkw{val}\mathspace{1}\mathid{y}_{0}=\mathspace{1}!\mathid{y};\mathspace{1}\mathid{y}\mathspace{1}:=\mathspace{1}\ !\mathid{x}+!\mathid{y};\mathspace{1}\mathid{x}\mathspace{1}:=\mathspace{1}\mathid{y}_{0}\mathbr{}
\mathindent{2}\}\mathbr{}
\mathindent{2}!\mathid{x}\mathbr{}
\}
\end{mdMathprearray}\]%
\end{mdDiv}
\end{mdDiv}%
\begin{mdP}[class={para-continue}]%
Here \mdSpan[class={code,math-inline}]{$\mathid{x}$} and \mdSpan[class={code,math-inline}]{$\mathid{y}$} are bound to freshly allocated references of type
\mdSpan[class={code,math-inline}]{$\mathid{ref}{\langle}\mathid{h},\mathid{int}{\rangle}$}. The operator \mdSpan[class={code,math-inline}]{$(!)$} dereferences a reference while the operator
\mdSpan[class={code,math-inline}]{$(:=)$} is used for assignment to references.
Due to the reading and writing of \mdSpan[class={code,math-inline}]{$\mathid{x}$} and \mdSpan[class={code,math-inline}]{$\mathid{y}$} of type \mdSpan[class={code,math-inline}]{$\mathid{ref}{\langle}\mathid{h},\mathid{int}{\rangle}$}, the effect
inferred for the body of the function is \mdSpan[class={code,math-inline}]{$\mathid{st}{\langle}\mathid{h}{\rangle}$} for some heap \mdSpan[class={code,math-inline}]{$\mathid{h}$}:%
\end{mdP}%
\begin{mdDiv}[class={para-block,para-block,input-mathpre},elem={pre}]%
\begin{mdDiv}[class={math-display}]%
\[\begin{mdMathprearray}
\mathid{fib}\mathspace{1}:\mathspace{1}\forall \mathid{h}.\mathspace{1}\mathid{int}\mathspace{1}{\rightarrow}\mathspace{1}\mathid{st}{\langle}\mathid{h}{\rangle}\mathspace{1}\mathid{int}
\end{mdMathprearray}\]%
\end{mdDiv}
\end{mdDiv}%
\begin{mdP}[class={para-continue}]%
However, we can of course consider the function \mdSpan[class={code,math-inline}]{$\mathid{fib}$} to be total: for any
input, it always returns the same output since the heap \mdSpan[class={code,math-inline}]{$\mathid{h}$} cannot be modified
or observed from outside this function. In particular, we can safely remove
the effect \mdSpan[class={code,math-inline}]{$\mathid{st}{\langle}\mathid{h}{\rangle}$} whenever the function is polymorphic in the heap \mdSpan[class={code,math-inline}]{$\mathid{h}$} and
where \mdSpan[class={code,math-inline}]{$\mathid{h}$} is not among the free type variables of argument types or result
type. This notion corresponds directly to the use of the higher-ranked \mdSpan[class={code,math-inline}]{$\mathid{runST}$}
function in Haskell{\mdNbsp}\mdSpan[class={citations},target-element={bibitem}]{[\mdA[class={bibref,localref},target-element={bibitem}]{stateinhaskell}{}{\mdSpan[class={bibitem-label}]{30}}]} (which we will call just \mdSpan[class={code,math-inline}]{$\mathid{run}$}):%
\end{mdP}%
\begin{mdDiv}[class={para-block,para-block,input-mathpre},elem={pre}]%
\begin{mdDiv}[class={math-display}]%
\[\begin{mdMathprearray}
\mathid{run}\mathspace{1}:\mathspace{1}\forall\mu\alpha.\mathspace{1}(\forall \mathid{h}.\mathspace{1}()\mathspace{1}{\rightarrow}\mathspace{1}{\langle}\mathid{st}{\langle}\mathid{h}{\rangle}\mathspace{1}|\mathspace{1}\mu{\rangle}\mathspace{1}\alpha)\mathspace{1}{\rightarrow}\mathspace{1}\mu\ \alpha
\end{mdMathprearray}\]%
\end{mdDiv}
\end{mdDiv}%
\begin{mdP}[class={para-continue}]%
Koka will automatically insert a \mdSpan[class={code,math-inline}]{$\mathid{run}$} wrapper at generalization points if it
can be applied, and infers a total type for the above fibonacci function:%
\end{mdP}%
\begin{mdDiv}[class={para-block,para-block,input-mathpre},elem={pre}]%
\begin{mdDiv}[class={math-display}]%
\[\begin{mdMathprearray}
\mathid{fib}\mathspace{1}:\mathspace{1}\mathid{int}\mathspace{1}{\rightarrow}\mathspace{1}\mathid{total}\mathspace{1}\mathid{int}
\end{mdMathprearray}\]%
\end{mdDiv}
\end{mdDiv}%
\begin{mdP}%
Again, using row polymorphism is quite natural to express in the type of \mdSpan[class={code,math-inline}]{$\mathid{run}$}
where the \mdSpan[class={code,math-inline}]{$\mathid{st}{\langle}\mathid{h}{\rangle}$} effect can be dismissed.%
\end{mdP}%
\begin{mdP}[class={indent}]%
One complex example from a type inference perspective where we applied Koka,
is the Garsia-Wachs algorithm as described by Filli{\^{a}}tre
\mdSpan[class={citations},target-element={bibitem}]{[\mdA[class={bibref,localref},target-element={bibitem}]{filliatre:garsiawachs}{}{\mdSpan[class={bibitem-label}]{6}}]}. The given algorithm was originally written
in ML and uses updateable references in the leaf nodes of the trees to achieve
efficiency comparable to the reference C implementation. However,
Filli{\^{a}}tre remarks that these side effects are local and not observable
to any caller. We implemented Filli{\^{a}}tre{'}s algorithm in Koka and our
system correctly inferred that the state effect can be discarded and assigned
a pure effect to the Garsia-Wachs algorithm{\mdNbsp}\mdSpan[class={citations},target-element={bibitem}]{[\mdA[class={bibref,localref},target-element={bibitem}]{koka}{}{\mdSpan[class={bibitem-label}]{18}}]}.%
\end{mdP}%
\mdHxxx[id=heap-safety,label={[2.6]\{.heading-label\}},toc={},sec-heapsafe={},caption={Heap safety}]{\mdSpan[class={heading-before}]{\mdSpan[class={heading-label}]{2.6}.{\enspace}}Heap safety}\begin{mdP}[class={para-continue}]%
Combining polymorphism and imperative state is fraught with difficulty and
requires great care. In particular, \mdEm{let}-polymorphism may lead to unsoundness
if references can be given a polymorphic type. A classical example from ML is:%
\end{mdP}%
\begin{mdDiv}[class={para-block,para-block,input-mathpre},elem={pre}]%
\begin{mdDiv}[class={math-display}]%
\[\begin{mdMathprearray}
\mathkw{let}\mathspace{1}\mathid{r}\mathspace{1}=\mathspace{1}\mathid{ref}\mathspace{1}[\mathspace{1}]\mathspace{1}\mathkw{in}\mathspace{1}(\mathid{r}\mathspace{1}:=\mathspace{1}[\mathid{true}],\mathspace{1}!\mathid{r}+1)\mathspace{1}
\end{mdMathprearray}\]%
\end{mdDiv}
\end{mdDiv}%
\begin{mdP}%
Here, we \mdSpan[class={code,math-inline}]{$\mathkw{let}$} bind \mdSpan[class={math-inline}]{$r$} to a reference with type  \mdSpan[class={code,math-inline}]{$\forall\alpha.\mathspace{1}\mathid{ref}{\langle}\mathid{list}{\langle}\alpha{\rangle}{\rangle}$}. 
The problem is that this type can instantiate later to
both a reference to an integer list and a boolean list. Intuitively, the
problem is that the first binding of \mdSpan[class={math-inline}]{$r$} generalized over type variables that
are actually free in the heap. The ML language considered many solutions to
prevent this from happening, ranging from imperative type variables{\mdNbsp}\mdSpan[class={citations},target-element={bibitem}]{[\mdA[class={bibref,localref},target-element={bibitem}]{tofte:refs}{}{\mdSpan[class={bibitem-label}]{39}}]} 
to the current syntactic value restriction, where only value
expressions can be generalized.%
\end{mdP}%
\begin{mdP}[class={indent}]%
In our system, no such tricks are necessary. Using the effect types, we
restrict generalization to expressions that are total, and we reject the ML
example since we will not generalize over the type of \mdSpan[class={math-inline}]{$r$} since it has an
\mdSpan[class={code,math-inline}]{$\mathid{alloc}{\langle}\mathid{h}{\rangle}$} effect. We prove in Section{\mdNbsp}\mdA[class={localref},target-element={h2}]{sec-subject}{}{\mdSpan[class={heading-label}]{5.1}} that our approach is
semantically sound.  In contrast to the value restriction, we can still
generalize over  any expression that is not stateful regardless of its
syntactic form.%
\end{mdP}%
\begin{mdP}[class={indent,para-continue}]%
The addition of \mdSpan[class={code,math-inline}]{$\mathid{run}$} adds further requirements where we must ensure that
encapsulated stateful computations truly behave like a pure function and do
not {\textquoteleft}leak{\textquoteright} the state. For example, it would be unsound to let a reference
escape its encapsulation:%
\end{mdP}%
\begin{mdDiv}[class={para-block,para-block,input-mathpre},elem={pre}]%
\begin{mdDiv}[class={math-display}]%
\[\begin{mdMathprearray}
\mathid{run}(\mathspace{1}\mathkw{function}()\{\ \mathid{ref}(1)\mathspace{1}\}\mathspace{1})
\end{mdMathprearray}\]%
\end{mdDiv}
\end{mdDiv}%
\begin{mdP}%
or to encapsulate a computation where its effects can still be observed.%
\end{mdP}%
\begin{mdP}[class={indent}]%
We prove in Section{\mdNbsp}\mdA[class={localref},target-element={h2}]{sec-faulty}{}{\mdSpan[class={heading-label}]{5.2}} that well-typed terms never exhibit such
behavior. To our knowledge we are the first to prove this formally for a
strict language in combination with exceptions and divergence. A similar
result is by Launchbury and Sabry{\mdNbsp}\mdSpan[class={citations},target-element={bibitem}]{[\mdA[class={bibref,localref},target-element={bibitem}]{launchbury:monadstate}{}{\mdSpan[class={bibitem-label}]{16}}]}  where they
prove heap safety of the Haskell{'}s ST monad in the context of a lazy store
with lazy evaluation.%
\end{mdP}%
\mdHxxx[id=sec-divergence,label={[2.7]\{.heading-label\}},toc={},caption={Divergence}]{\mdSpan[class={heading-before}]{\mdSpan[class={heading-label}]{2.7}.{\enspace}}Divergence}\begin{mdP}%
Koka uses a simple termination checker (based on{\mdNbsp}\mdSpan[class={citations},target-element={bibitem}]{[\mdA[class={bibref,localref},target-element={bibitem}]{abel:termination}{}{\mdSpan[class={bibitem-label}]{1}}]}) to
assign the  divergence effect to potentially non-terminating functions. To do
this safely, Koka has three kinds of data types, inductive, co-inductive, and
arbitrary recursive data types.  In particular, we restrict (co)inductive
data types such that the type itself cannot occur in a negative position.  Any
function that matches on an arbitrary recursive data type is assumed to be
potentially divergent since one can encode the Y combinator using such data
type and write a non-terminating function that is not syntactically recursive.%
\end{mdP}%
\begin{mdP}[class={indent}]%
Recursively defined functions should of course include the divergence effect
in general. However, if the termination checker finds that each recursive call
decreases the size of an inductive data type (or is productive for a co-
inductive data type), then we do not assign the divergent effect.  The current
analysis is quite limited and syntactically fragile but seems to work well
enough in practice (Section{\mdNbsp}\mdA[class={localref},target-element={h2}]{sec-practice}{}{\mdSpan[class={heading-label}]{2.8}}). For our purpose, we prefer a
predictable analysis with clear rules.%
\end{mdP}%
\begin{mdP}[class={indent,para-continue}]%
However, in combination with higher-order mutable state, we can still define
functions that are not syntactically recursive, but fail to terminate.
Consider {\textquoteleft}Landin's knot{\textquoteright}:%
\end{mdP}%
\begin{mdDiv}[class={para-block,para-block,input-mathpre},elem={pre}]%
\begin{mdDiv}[class={math-display}]%
\[\begin{mdMathprearray}
\mathkw{function}\mathspace{1}\mathid{diverge}()\mathspace{1}\{\mathbr{}
\mathindent{2}\mathkw{val}\mathspace{1}\mathid{r}\mathspace{1}:=\mathspace{1}\mathid{ref}(\mathid{id})\mathspace{10}\mathbr{}
\mathindent{2}\mathkw{function}\mathspace{1}\mathid{foo}()\mathspace{1}\{\ (!\mathid{r})()\mathspace{1}\}\mathspace{3}\mathbr{}
\mathindent{2}\mathid{r}\mathspace{1}:=\mathspace{1}\mathid{foo}\mathspace{12}\mathbr{}
\mathindent{2}\mathid{foo}()\mathspace{15}\mathbr{}
\}
\end{mdMathprearray}\]%
\end{mdDiv}
\end{mdDiv}%
\begin{mdP}%
In this function, we first create a reference \mdSpan[class={code,math-inline}]{$\mathid{r}$} initialized with the
identify function. Next we define a local function \mdSpan[class={code,math-inline}]{$\mathid{foo}$} which calls the
function in \mdSpan[class={code,math-inline}]{$\mathid{r}$}. Then we assign \mdSpan[class={code,math-inline}]{$\mathid{foo}$} itself to \mdSpan[class={code,math-inline}]{$\mathid{r}$} and call \mdSpan[class={code,math-inline}]{$\mathid{foo}$}, which will
now never terminate even though there is no syntactic recursion.%
\end{mdP}%
\begin{mdP}[class={indent}]%
How can we infer in general that \mdSpan[class={code,math-inline}]{$\mathid{diverge}$} must include the \mdSpan[class={code,math-inline}]{$\mathid{div}$} effect?
It turns out that in essence reading from the heap may result in divergence. A
conservative approach would be to assign the \mdSpan[class={code,math-inline}]{$\mathid{div}$} effect to the type of read
\mdSpan[class={code,math-inline}]{$(!)$}. For simplicity, this is what we will do in the formal development.%
\end{mdP}%
\begin{mdP}[class={indent}]%
But in Koka we use a more sophisticated approach. In order to cause
divergence, we actually need to read a function
from the heap which accesses the heap itself. Fortunately, our effect system
makes this behavior already apparent in the inferred types! {\textendash} in our example,
the effect of \mdSpan[class={code,math-inline}]{$\mathid{foo}$} contains \mdSpan[class={code,math-inline}]{$\mathid{read}{\langle}\mathid{h}{\rangle}$}, which is being stored in a reference
in the same heap of type \mdSpan[class={code,math-inline}]{$\mathid{ref}{\langle}\mathid{h},()\mathspace{1}{\rightarrow}\mathspace{1}\mathid{read}{\langle}\mathid{h}{\rangle}\mathspace{1}(){\rangle}$}.%
\end{mdP}%
\begin{mdP}[class={indent}]%
The trick is now that we generate a type constraint \mdSpan[class={code,math-inline}]{$\mathid{hdiv}{\langle}\mathid{h},\tau,\e{\rangle}$} for
every heap read that keeps track of heap type \mdSpan[class={code,math-inline}]{$\mathid{h}$}, the type of the value that
was read \mdSpan[class={code,math-inline}]{$\tau$} and the current effect \mdSpan[class={code,math-inline}]{$\e$}. The constraint \mdSpan[class={code,math-inline}]{$\mathid{hdiv}{\langle}\mathid{h},\tau,\e{\rangle}$}
expresses that if \mdSpan[class={code,math-inline}]{$\mathid{h}\in\ftv\tau$} then the effect \mdSpan[class={code,math-inline}]{$\e$} must include divergence.
In particular, this constraint is fine-grained enough that any reading of a
non-function type, or non-stateful functions will never cause divergence (and
we can dismiss the constraint)  The drawback is that if \mdSpan[class={math-inline}]{$\tau$} is polymorphic
at generalization time, we need to keep the constraint around (as we cannot
decide at that point whether \mdSpan[class={code,math-inline}]{$\mathid{h}$} will ever be in \mdSpan[class={math-inline}]{$\ftv\tau$}), which in turn
means we need to use a system of qualified types{\mdNbsp}\mdSpan[class={citations},target-element={bibitem}]{[\mdA[class={bibref,localref},target-element={bibitem}]{jones:qualifiedtypes}{}{\mdSpan[class={bibitem-label}]{12}}]}.
Currently this is not fully implemented yet in Koka, and if at generalization
time we cannot guarantee \mdSpan[class={math-inline}]{$\tau$} will never contain a reference to the heap
\mdSpan[class={code,math-inline}]{$\mathid{h}$},  we conservatively assume that the function may diverge.%
\end{mdP}%
\mdHxxx[id=sec-practice,label={[2.8]\{.heading-label\}},toc={},caption={Koka in practice}]{\mdSpan[class={heading-before}]{\mdSpan[class={heading-label}]{2.8}.{\enspace}}Koka in practice}\begin{mdP}%
When designing a new type system it is always a question how well it will work
in practice: does it infer the types you expect? Do the types become too
complicated? Is the termination checker strong enough? etc. We have
implemented the effect inference and various extensions in the Koka language
which is freely available on the web{\mdNbsp}\mdSpan[class={citations},target-element={bibitem}]{[\mdA[class={bibref,localref},target-element={bibitem}]{koka}{}{\mdSpan[class={bibitem-label}]{18}}]}. The Koka system currently has a
JavaScript backend and can generate code that runs in NodeJS or a web page.
We have written many small to medium sized samples to see how well the
system works in practice.%
\end{mdP}%
\mdHxxxxx[id=sec-markdown,label={[2.8.0.1]\{.heading-label\}},caption={Markdown}]{Markdown}\begin{mdP}%
One application is a fully compliant Markdown text processor. This program
consists of three phases where it first parses block elements, performs block
analysis, collects link definitions, numbers sections, and finally
renders the inline elements in each block. The program passes the full
Markdown test suite. In fact, this article has been written itself 
as a Madoko document{\mdNbsp}\mdSpan[class={citations},target-element={bibitem}]{[\mdA[class={bibref,localref},target-element={bibitem}]{madoko}{}{\mdSpan[class={bibitem-label}]{20}}]}.%
\end{mdP}%
\begin{mdP}[class={indent,para-continue}]%
Remarkably, almost all functions are inferred to be
\mdSpan[class={code,math-inline}]{$\mathid{total}$}, and only a handful of driver functions perform side effects, like
reading input files. For efficiency though, many  internal functions use local
state. For example, when rendering all inline elements in a block, we use a
local mutable string builder (of type \mdSpan[class={code,math-inline}]{$\mathid{builder}{\langle}\mathid{h}{\rangle}$}) to build the result string
in constant time (actual Koka code):%
\end{mdP}%
\begin{mdDiv}[class={para-block,para-block,input-mathpre},elem={pre}]%
\begin{mdDiv}[class={math-display}]%
\[\begin{mdMathprearray}
\mathkw{function}\mathspace{1}\mathid{formatInline}(\mathspace{1}\mathid{ctx}:\mathid{inlineCtx},\mathspace{1}\mathid{txt}:\mathid{string})\mathspace{1}:\mathid{string}\mathspace{1}\{\mathbr{}
\mathindent{2}\mathid{formatAcc}(\mathid{ctx},\mathspace{1}\mathid{builder}(),\mathspace{1}\mathid{txt})\mathbr{}
\}\mathbr{}
\mathbr{}
\mathkw{function}\mathspace{1}\mathid{formatAcc}(\mathspace{1}\mathid{ctx}:\mathid{inlineCtx},\mathspace{1}\mathid{acc}:\mathid{builder}{\langle}\mathid{h}{\rangle},\mathspace{1}\mathid{txt}:\mathid{string}\mathspace{1})\mathspace{1}:\mathspace{1}\mathid{st}{\langle}\mathid{h}{\rangle}\mathspace{1}\mathid{string}\mathspace{1}\{\mathbr{}
\mathindent{2}\mathkw{if}\mathspace{1}(\mathid{txt}=="")\mathspace{1}\mathkw{return}\mathspace{1}\mathid{acc}.\mathid{string}\mathbr{}
\mathindent{2}\mathkw{val}\mathspace{1}(\mathid{s},\mathid{next})\mathspace{1}=\mathspace{1}\mathid{matchRules}(\mathid{ctx}.\mathid{grammar},\mathspace{1}\mathid{ctx},\mathspace{1}\mathid{txt},\mathspace{1}\mathid{id})\mathbr{}
\mathindent{2}\mathid{formatAcc}(\mathid{ctx},\mathspace{1}\mathid{acc}.\mathid{append}(\mathid{s}),\mathspace{1}\mathid{txt}.\mathid{substr}_{1}(\mathid{next}))\mathspace{2}\mathbr{}
\}
\end{mdMathprearray}\]%
\end{mdDiv}
\end{mdDiv}%
\begin{mdP}%
Note how \mdSpan[class={code,math-inline}]{$\mathid{formatAcc}$} is stateful due to the calls to the \mdSpan[class={code,math-inline}]{$\mathid{append}$} and \mdSpan[class={code,math-inline}]{$\mathid{string}$}
methods of the string builder \mdSpan[class={code,math-inline}]{$\mathid{acc}$}, but the outer function \mdSpan[class={code,math-inline}]{$\mathid{formatInline}$} is
inferred to be \mdSpan[class={code,math-inline}]{$\mathid{total}$} since Koka can automatically apply the \mdSpan[class={code,math-inline}]{$\mathid{run}$} function
and encapsulate the state: indeed it is not observable if we use a mutable
string builder internally or not. This pattern also occurs for example in the
block analysis phase where we use  a mutable hashmap to build the dictionary
of link definitions.%
\end{mdP}%
\mdHxxxxx[id=sec-tiersplit,label={[2.8.0.2]\{.heading-label\}},caption={Safe tier-splitting}]{Safe tier-splitting}\begin{mdP}%
Most of the HTML5 DOM and the Node API{'}s are available in Koka which allows us
to write more substantial programs and evaluate the effect inference system in
practice. We use the effect
\mdSpan[class={code,math-inline}]{$\mathid{dom}$} for functions that may have any side effect through a DOM call.%
\end{mdP}%
\begin{mdP}[class={indent}]%
On the web, many programs are split in a server and client part communicating
with each other. It is advantageous to
write both the client and server part as one program since that 
enables us to share one common type definition for the data they
exchange. Also, their
interaction will be more apparent, and they can share common functionality, like
date parsing, to ensure that both parts behave consistently.%
\end{mdP}%
\begin{mdP}[class={indent}]%
Safely splitting a program into a server and client part is difficult though.
For example, the client code may call a library function that itself calls a
function that can only be run on the server (like writing to a log file), or
the other way  around.  Moreover, if the client and server part both access a
shared global variable (or both call a library function that uses  an internal
global variable) then we cannot split this code anymore.%
\end{mdP}%
\begin{mdP}[class={indent,para-continue}]%
The Koka effect types tackle both problems and enable fully safe tier
splitting. Our main tier splitting function has the following
(simplified) type signature:%
\end{mdP}%
\begin{mdDiv}[class={para-block,para-block,input-mathpre},elem={pre}]%
\begin{mdDiv}[class={math-display}]%
\[\begin{mdMathprearray}
\mathkw{function}\mathspace{1}\mathid{tiersplit}(\mathspace{2}\mathbr{}
\mathindent{1}\mathid{serverPart}:\mathspace{1}()\mathspace{1}{\rightarrow}\mathspace{1}\mathid{server}\mathspace{1}((\alpha {\rightarrow}\mathspace{1}\mathid{server}\mathspace{1}())\mathspace{1}{\rightarrow}\mathspace{1}\mathid{server}\mathspace{1}(\beta {\rightarrow}\mathspace{1}\mathid{server}\mathspace{1}())\mathspace{1}),\mathspace{1}\mathbr{}
\mathindent{1}\mathid{clientPart}:\mathspace{1}(\beta {\rightarrow}\mathspace{1}\mathid{client}\mathspace{1}())\mathspace{1}{\rightarrow}\mathspace{1}\mathid{client}\mathspace{1}(\alpha {\rightarrow}\mathspace{1}\mathid{client}\mathspace{1}())\mathspace{2}\mathbr{}
)\mathspace{1}:\mathspace{1}\mathid{io}\mathspace{1}()
\end{mdMathprearray}\]%
\end{mdDiv}
\end{mdDiv}%
\begin{mdP}[class={para-continue}]%
where the \mdSpan[class={code,math-inline}]{$\mathid{server}$} and \mdSpan[class={code,math-inline}]{$\mathid{client}$} effects are defined as:%
\end{mdP}%
\begin{mdDiv}[class={para-block,para-block,input-mathpre},elem={pre}]%
\begin{mdDiv}[class={math-display}]%
\[\begin{mdMathprearray}
\mathkw{alias}\mathspace{1}\mathid{server}\mathspace{1}=\mathspace{1}\mathid{io}\mathbr{}
\mathkw{alias}\mathspace{1}\mathid{client}\mathspace{1}=\mathspace{1}{\langle}\mathid{dom},\mathid{div}{\rangle}
\end{mdMathprearray}\]%
\end{mdDiv}
\end{mdDiv}%
\begin{mdP}%
The \mdSpan[class={code,math-inline}]{$\mathid{tiersplit}$} function takes a server and client function and sets up a
socket connection. 
On the server it will call the server part function which
can initialize. Now, both the client and server part can be called for each
fresh connection where \mdSpan[class={code,math-inline}]{$\mathid{tiersplit}$} supplies a \mdEm{send} function that takes a
message of type \mdSpan[class={code,math-inline}]{$\alpha$} for client messages, and \mdSpan[class={code,math-inline}]{$\beta$} for server messages.
Both the client and server part return a fresh {\textquoteleft}connection{\textquoteright} function that
handles incoming messages from the server or client respectively. Note how
this type guarantees that messages sent to the client, and messages handled by
the client, are both of type \mdSpan[class={code,math-inline}]{$\alpha$}, while for the server messages they will
be \mdSpan[class={code,math-inline}]{$\beta$}.
Furthermore, because the effect types for \mdSpan[class={code,math-inline}]{$\mathid{server}$} and \mdSpan[class={code,math-inline}]{$\mathid{client}$} are closed,
the client and server part are only able to call functions available for
the client or server respectively.%
\end{mdP}%
\begin{mdP}[class={indent}]%
Finally, the Koka effect system also prevents accidental sharing of global
state by the client and server part. Both the client and server can use state
that is contained in their handler. In that case the \mdSpan[class={code,math-inline}]{$\mathid{st}{\langle}\mathid{h}{\rangle}$} effect will be
inferred, and discarded because \mdSpan[class={code,math-inline}]{$\mathid{h}$} will generalize. However, if either
function tries to access a shared variable in an outer scope, then the \mdSpan[class={code,math-inline}]{$\mathid{h}$}
will \mdEm{not} generalize (because the  variable will have type \mdSpan[class={code,math-inline}]{$\mathid{ref}{\langle}\mathid{h},\mathid{a}{\rangle}$} and
therefore \mdSpan[class={code,math-inline}]{$\mathid{h}$} is not free in the environment), in which case the inferred
\mdSpan[class={code,math-inline}]{$\mathid{st}{\langle}\mathid{h}{\rangle}$} effect cannot be removed. Again, this will lead to a unification
failure and the program will be statically rejected.%
\end{mdP}%
\mdHxx[id=sec-types,label={[3]\{.heading-label\}},toc={},caption={The type system}]{\mdSpan[class={heading-before}]{\mdSpan[class={heading-label}]{3}.{\enspace}}The type system}\begin{mdDiv}[class={figure,align-center},id=fig-types,label={[1]\{.figure-label\}},elem={figure},toc={tof},toc-line={[1]\{.figure-label\}. Syntax of types and kinds. An extra restriction is that effect constants cannot be type variables, i.e. \${\textbackslash}alpha{\textasciicircum}{\textbackslash}kl\$ is illegal.},page-align={top},caption={Syntax of types and kinds. An extra restriction is that effect constants cannot be type variables, i.e. \${\textbackslash}alpha{\textasciicircum}{\textbackslash}kl\$ is illegal.}]%
\begin{mdDiv}[class={div},elem={div},text-align={left},margin-left={4em}]%
\begin{mdTable}[class={madoko,block}]{5}{lllll}

\mdTd[text-align={left},display={table-cell}]{kinds}&\mdTd[display={table-cell}]{\mdSpan[class={math-inline}]{$\kk$}}&\mdTd[display={table-cell}]{\mdSpan[class={code,math-inline}]{$::=$}}&\mdTd[text-align={left},display={table-cell}]{\mdSpan[class={code,math-inline}]{$*$} {\textbar} \mdSpan[class={code,math-inline}]{$\ke$} {\textbar} \mdSpan[class={code,math-inline}]{$\kl$} {\textbar} \mdSpan[class={code,math-inline}]{$\kh$}}&\mdTd[display={table-cell}]{values, effect rows, effect constants, heaps}\\
\mdTd[text-align={left},display={table-cell}]{}&\mdTd[display={table-cell}]{}&\mdTd[display={table-cell}]{{\textbar}}&\mdTd[text-align={left},display={table-cell}]{\mdSpan[class={code,math-inline}]{$(\kk_1,...,\kk_\mathid{n})\mathspace{1}{\rightarrow}\mathspace{1}\kk$}}&\mdTd[display={table-cell}]{type constructor}\\
\mdTd[text-align={left},display={table-cell}]{{\mdNbsp}}&\mdTd[display={table-cell}]{}&\mdTd[display={table-cell}]{}&\mdTd[text-align={left},display={table-cell}]{}&\mdTd[display={table-cell}]{}\\
\mdTd[text-align={left},display={table-cell}]{types}&\mdTd[display={table-cell}]{\mdSpan[class={code,math-inline}]{$\t^\mathid{k}$}}&\mdTd[display={table-cell}]{\mdSpan[class={code,math-inline}]{$::=$}}&\mdTd[text-align={left},display={table-cell}]{\mdSpan[class={code,math-inline}]{$\alpha^\mathid{k}$}}&\mdTd[display={table-cell}]{type variable (using \mdSpan[class={code,math-inline}]{$\mu$} for effects,\mdSpan[class={code,math-inline}]{$\xi$} for heaps)}\\
\mdTd[text-align={left},display={table-cell}]{}&\mdTd[display={table-cell}]{}&\mdTd[display={table-cell}]{{\textbar}}&\mdTd[text-align={left},display={table-cell}]{\mdSpan[class={code,math-inline}]{$\mathid{c}^\kk$}}&\mdTd[display={table-cell}]{type constant}\\
\mdTd[text-align={left},display={table-cell}]{}&\mdTd[display={table-cell}]{}&\mdTd[display={table-cell}]{{\textbar}}&\mdTd[text-align={left},display={table-cell}]{\mdSpan[class={code,math-inline}]{$\mathid{c}^{\kk_0}{\langle}\t_1^{\kk_1},...,\t_\mathid{n}^{\kk_\mathid{n}}{\rangle}$}}&\mdTd[display={table-cell}]{\mdSpan[class={code,math-inline}]{${\scriptstyle \kk_0\mathspace{1}=\mathspace{1}(\kk_1,...,\kk_\mathid{n})\mathspace{1}{\rightarrow}\mathspace{1}\kk}$}}\\
\mdTd[text-align={left},display={table-cell}]{{\mdNbsp}}&\mdTd[display={table-cell}]{}&\mdTd[display={table-cell}]{}&\mdTd[text-align={left},display={table-cell}]{}&\mdTd[display={table-cell}]{}\\
\mdTd[text-align={left},display={table-cell}]{schemes}&\mdTd[display={table-cell}]{\mdSpan[class={code,math-inline}]{$\sigma$}}&\mdTd[display={table-cell}]{\mdSpan[class={code,math-inline}]{$::=$}}&\mdTd[text-align={left},display={table-cell}]{\mdSpan[class={code,math-inline}]{$\forall\alpha_1...\alpha_\mathid{n}.\mathspace{1}\t^{*}$}}&\mdTd[display={table-cell}]{}\\
\end{mdTable}
\begin{mdTable}[class={madoko,block}]{5}{lllll}

\mdTd[text-align={left},display={table-cell}]{constants}&\mdTd[display={table-cell}]{\mdSpan[class={code,math-inline}]{$()$}}&\mdTd[text-align={left},display={table-cell}]{\mdSpan[class={code,math-inline}]{$::$}}&\mdTd[display={table-cell}]{\mdSpan[class={code,math-inline}]{$*$}}&\mdTd[display={table-cell}]{unit type}\\
\mdTd[text-align={left},display={table-cell}]{}&\mdTd[display={table-cell}]{\mdSpan[class={code,math-inline}]{$(\_\mathspace{1}{\rightarrow}\mathspace{1}\_\,\_)$}}&\mdTd[text-align={left},display={table-cell}]{\mdSpan[class={code,math-inline}]{$::$}}&\mdTd[display={table-cell}]{\mdSpan[class={code,math-inline}]{$(*,\ke,*)\mathspace{1}{\rightarrow}\mathspace{1}*\mathspace{1}$}}&\mdTd[display={table-cell}]{functions}\\
\mdTd[text-align={left},display={table-cell}]{}&\mdTd[display={table-cell}]{\mdSpan[class={code,math-inline}]{${\langle}{\rangle}$}}&\mdTd[text-align={left},display={table-cell}]{\mdSpan[class={code,math-inline}]{$::$}}&\mdTd[display={table-cell}]{\mdSpan[class={code,math-inline}]{$\ke$}}&\mdTd[display={table-cell}]{empty effect}\\
\mdTd[text-align={left},display={table-cell}]{}&\mdTd[display={table-cell}]{\mdSpan[class={code,math-inline}]{${\langle}\_$}{\textbar}\mdSpan[class={code,math-inline}]{$\_{\rangle}$}}&\mdTd[text-align={left},display={table-cell}]{\mdSpan[class={code,math-inline}]{$::$}}&\mdTd[display={table-cell}]{\mdSpan[class={code,math-inline}]{$(\kl,\ke)\mathspace{1}{\rightarrow}\mathspace{1}\ke$}}&\mdTd[display={table-cell}]{effect extension}\\
\mdTd[text-align={left},display={table-cell}]{}&\mdTd[display={table-cell}]{\mdSpan[class={math-inline}]{$\ec{ref}$}}&\mdTd[text-align={left},display={table-cell}]{\mdSpan[class={code,math-inline}]{$::$}}&\mdTd[display={table-cell}]{\mdSpan[class={code,math-inline}]{$(\kh,*)\mathspace{1}{\rightarrow}\mathspace{1}*$}}&\mdTd[display={table-cell}]{references}\\
\mdTd[text-align={left},display={table-cell}]{}&\mdTd[display={table-cell}]{\mdSpan[class={math-inline}]{$\ec{exn}$},\mdSpan[class={math-inline}]{$\ec{div}$}}&\mdTd[text-align={left},display={table-cell}]{\mdSpan[class={code,math-inline}]{$::$}}&\mdTd[display={table-cell}]{\mdSpan[class={code,math-inline}]{$\kl$}}&\mdTd[display={table-cell}]{partial, divergent}\\
\mdTd[text-align={left},display={table-cell}]{}&\mdTd[display={table-cell}]{\mdSpan[class={math-inline}]{$\ec{st}$}}&\mdTd[text-align={left},display={table-cell}]{\mdSpan[class={code,math-inline}]{$::$}}&\mdTd[display={table-cell}]{\mdSpan[class={code,math-inline}]{$\kh {\rightarrow}\mathspace{1}\kl$}}&\mdTd[display={table-cell}]{stateful}\\
\end{mdTable}
\begin{mdTable}[class={madoko,block}]{4}{llll}

\mdTd[text-align={left},display={table-cell}]{syntactic sugar}&\mdTd[display={table-cell}]{\mdSpan[class={code,math-inline}]{$\t_1\mathspace{1}{\rightarrow}\mathspace{1}\t_2$}}&\multicolumn{1}{c}{\mdTd[text-align={center},display={table-cell}]{=}}&\mdTd[display={table-cell}]{\mdSpan[class={code,math-inline}]{$\t_1\mathspace{1}{\rightarrow}\mathspace{1}{\langle}{\rangle}\;\t_2$}}\\
\mdTd[text-align={left},display={table-cell}]{}&\mdTd[display={table-cell}]{\mdSpan[class={code,math-inline}]{${\langle}\mathid{l}_1,...,\mathid{l}_\mathid{n}\mathspace{1}$}{\textbar}\mdSpan[class={code,math-inline}]{$\mathindent{1}\e{\rangle}$}}&\multicolumn{1}{c}{\mdTd[text-align={center},display={table-cell}]{=}}&\mdTd[display={table-cell}]{\mdSpan[class={code,math-inline}]{${\langle}\mathid{l}_1\mathspace{1}...\mathspace{1}{\langle}\mathid{l}_\mathid{n}\mathspace{1}$}{\textbar}\mdSpan[class={code,math-inline}]{$\mathindent{1}\e{\rangle}\mathspace{1}...{\rangle}$}}\\
\mdTd[text-align={left},display={table-cell}]{}&\mdTd[display={table-cell}]{\mdSpan[class={code,math-inline}]{${\langle}\mathid{l}_1,...,\mathid{l}_\mathid{n}{\rangle}$}}&\multicolumn{1}{c}{\mdTd[text-align={center},display={table-cell}]{=}}&\mdTd[display={table-cell}]{\mdSpan[class={code,math-inline}]{${\langle}\mathid{l}_1,...,\mathid{l}_\mathid{n}\mathspace{1}$}{\textbar}\mdSpan[class={code,math-inline}]{$\mathindent{1}\empt{\rangle}$}}\\
\end{mdTable}%
\end{mdDiv}%
\mdHr[class={figureline,madoko}]{}\mdSpan[class={figure-caption}]{\mdSpan[class={caption-before}]{\mdStrong{Figure{\mdNbsp}\mdSpan[class={figure-label}]{1}.} }Syntax of types and kinds. An extra restriction is that effect constants cannot be type variables, i.e. \mdSpan[class={math-inline}]{$\alpha^\kl$} is illegal.}%
\end{mdDiv}%
\begin{mdP}%
In this section we are going to give a formal definition of our polymorphic
effect system for a small core-calculus that captures the essence of Koka. We
call this \mdSpan[class={math-inline}]{$\lambda^k$}.  Figure{\mdNbsp}\mdA[class={localref},target-element={figure}]{fig-types}{}{\mdSpan[class={figure-label}]{1}} defines the syntax of types.  The
well-formedness of types \mdSpan[class={math-inline}]{$\tau$} is guaranteed by a simple kind system. We put
the kind \mdSpan[class={math-inline}]{$\kk$} of a type \mdSpan[class={math-inline}]{$\t$} in superscript, as \mdSpan[class={math-inline}]{$\t^\kk$}. We have the usual
kind \mdSpan[class={code,math-inline}]{$*$} and \mdSpan[class={code,math-inline}]{${\rightarrow}$}, but also kinds for effect rows (\mdSpan[class={math-inline}]{$\ke$}), effect constants
(\mdSpan[class={math-inline}]{$\kl$}), and heaps (\mdSpan[class={math-inline}]{$\kh$}). Often the kind of a type is immediately apparent
or not relevant, and most of the time we will not denote the kind to reduce
clutter, and just write plain types \mdSpan[class={math-inline}]{$\t$}. For clarity, we are using \mdSpan[class={math-inline}]{$\alpha$}
for regular type variables, \mdSpan[class={math-inline}]{$\mu$} for effect type variables, and \mdSpan[class={math-inline}]{$\xi$} for
heap type variables.%
\end{mdP}%
\begin{mdP}[class={indent}]%
Effect types are defined as a row of effect labels \mdSpan[class={math-inline}]{$l$}. Such effect row is
either empty \mdSpan[class={math-inline}]{$\empt$}, a polymorphic effect variable \mdSpan[class={math-inline}]{$\mu$}, or an extension of
an effect row \mdSpan[class={math-inline}]{$\e$} with an effect constant \mdSpan[class={math-inline}]{$l$}, written as \mdSpan[class={code,math-inline}]{${\langle}\mathid{l}|\e{\rangle}$}.  The
effect constants can be anything that is interesting to our language. For our
purposes we will restrict the constants to exceptions (\mdSpan[class={math-inline}]{$\ec{exn}$}), divergence
(\mdSpan[class={math-inline}]{$\ec{div}$}), and heap operations (\mdSpan[class={math-inline}]{$\ec{st}$}). It is no problem to generalize
this to the more fine-grained hierarchy of Koka but this simplifies the
presentation and proofs. The kind system ensures that an effect is always
either a \mdEm{closed effect} of the form \mdSpan[class={code,math-inline}]{${\langle}\mathid{l}_1,...,\mathid{l}_\mathid{n}{\rangle}$}, or an \mdEm{open effect} of
the form \mdSpan[class={code,math-inline}]{${\langle}\mathid{l}_1,...,\mathid{l}_\mathid{n}\mathspace{1}|\mathspace{1}\mu{\rangle}$}.%
\end{mdP}%
\begin{mdDiv}[class={figure,align-center},id=fig-eqeffect,label={[2]\{.figure-label\}},elem={figure},toc={tof},toc-line={[2]\{.figure-label\}. Effect equivalence.},page-align={top},caption={Effect equivalence.}]%
\begin{mdP}%
\mdSpan[class={rulename},font-variant={small-caps},font-size={small}]{(eq-refl)} \mdSpan[class={code,math-inline}]{$\e \equiv \e$} {\mdNbsp}{\mdNbsp}
\mdSpan[class={rulename},font-variant={small-caps},font-size={small}]{(eq-trans)} \mdSpan[class={code,math-inline}]{$\dfrac{\e_1\mathspace{1}\equiv \e_2\mathspace{1}\quad \e_2\mathspace{1}\equiv \e_3}{\e_1\mathspace{1}\equiv \e_3}$} \mdBr
{\strut}\mdBr
\mdSpan[class={rulename},font-variant={small-caps},font-size={small}]{(eq-head)}  \mdSpan[class={code,math-inline}]{$\dfrac{\e_1\mathspace{1}\equiv \e_2}{\ext{\mathid{l}}{\e_1}\mathspace{1}\equiv \ext{\mathid{l}}{\e_2}}$}  {\mdNbsp}{\mdNbsp}
\mdSpan[class={rulename},font-variant={small-caps},font-size={small}]{(eq-swap)} \mdSpan[class={code,math-inline}]{$\dfrac{\mathid{l}_1\mathspace{1}\not\equiv \mathid{l}_2}{\ext{\mathid{l}_1}{\ext{\mathid{l}_2}{\e}}\mathspace{1}\equiv \ext{\mathid{l}_2}{\ext{\mathid{l}_1}{\e}}}$} \mdBr
{\strut}\mdBr
\mdSpan[class={rulename},font-variant={small-caps},font-size={small}]{(uneq-lab)} \mdSpan[class={code,math-inline}]{$\dfrac{\mathid{c}\mathspace{1}\neq \mathid{c}'}{\mathid{c}{\langle}\t_1,...,\t_\mathid{n}{\rangle}\mathspace{1}\not\equiv \mathid{c}'{\langle}\t'_1,...,\t'_\mathid{n}{\rangle}}$}%
\end{mdP}%
\mdHr[class={figureline,madoko}]{}\mdSpan[class={figure-caption}]{\mdSpan[class={caption-before}]{\mdStrong{Figure{\mdNbsp}\mdSpan[class={figure-label}]{2}.} }Effect equivalence.}%
\end{mdDiv}%
\begin{mdP}[class={indent}]%
Figure{\mdNbsp}\mdA[class={localref},target-element={figure}]{fig-eqeffect}{}{\mdSpan[class={figure-label}]{2}} defines an equivalence relation \mdSpan[class={math-inline}]{$(\equiv)$} between effect
types where we consider effects equivalent
regardless of the order of the effect constants. In contrast to many record
calculi{\mdNbsp}\mdSpan[class={citations},target-element={bibitem}]{[\mdA[class={bibref,localref},target-element={bibitem}]{gasterjones:trex}{}{\mdSpan[class={bibitem-label}]{7}}, \mdA[class={bibref,localref},target-element={bibitem}]{remy:records}{}{\mdSpan[class={bibitem-label}]{32}}, \mdA[class={bibref,localref},target-element={bibitem}]{sulzmann:records}{}{\mdSpan[class={bibitem-label}]{34}}]}
effect rows \mdEm{do} allow duplicate labels where an effect \mdSpan[class={code,math-inline}]{${\langle}\mathid{exn},\mathid{exn}{\rangle}$} is allowed
(and not equal to the effect \mdSpan[class={code,math-inline}]{${\langle}\mathid{exn}{\rangle}$}). The definition of effect equality is
essentially the same as for scoped labels{\mdNbsp}\mdSpan[class={citations},target-element={bibitem}]{[\mdA[class={bibref,localref},target-element={bibitem}]{leijen:scopedlabels}{}{\mdSpan[class={bibitem-label}]{17}}]} where we
ignore the type components. Note that for rule \mdSpan[class={rulename},font-variant={small-caps},font-size={small}]{(eq-swap)}
we use the rule \mdSpan[class={rulename},font-variant={small-caps},font-size={small}]{(uneq-lab)} to compare effect constants where the 
type arguments are 
not taken into account: intuitively we consider the effect constants as the `labels{'} 
of an effect record.
Most constants compare directly. The only exception 
in our system is the state effect where \mdSpan[class={math-inline}]{$\tst{h_1}\not\equiv \tst{h_2}$} does \mdEm{not} hold even 
if \mdSpan[class={math-inline}]{$h_1 \neq h_2$}.%
\end{mdP}%
\begin{mdP}[class={indent}]%
Using effect equality, we define \mdSpan[class={math-inline}]{$l \in \e$} iff \mdSpan[class={math-inline}]{$e \equiv \ext{l}{\e'}$} for some \mdSpan[class={math-inline}]{$\e'$}.%
\end{mdP}%
\mdHxxx[id=type-rules,label={[3.1]\{.heading-label\}},toc={},caption={Type rules}]{\mdSpan[class={heading-before}]{\mdSpan[class={heading-label}]{3.1}.{\enspace}}Type rules}\begin{mdDiv}[class={figure,align-center},id=fig-typerules,label={[3]\{.figure-label\}},elem={figure},toc={tof},toc-line={[3]\{.figure-label\}. General type rules with effects.},page-align={top},caption={General type rules with effects.}]%
\begin{mdP}%
{\strut}
 \mdSpan[class={rulename},font-variant={small-caps},font-size={small}]{(var)} \mdSpan[class={code,math-inline}]{$\inference{\Gamma(\mathid{x})\mathspace{1}=\mathspace{1}\s}{\infer{\Gamma}{\mathid{x}}{\s}{\e}}$} {\hspace{1em}}
 \mdSpan[class={rulename},font-variant={small-caps},font-size={small}]{(app)} \mdSpan[class={code,math-inline}]{$\inference{\infer{\Gamma}{\exp_1}{\tfun{\t_2}{\e}{\t}}{\e}\mathspace{1}\quad \infer{\Gamma}{\exp_2}{\t_2}{\e}}{\infer{\Gamma}{\exp_1\,\exp_2}{\t}{\e}}$} \mdBr
{\strut}\mdBr
 \mdSpan[class={rulename},font-variant={small-caps},font-size={small}]{(lam)} \mdSpan[class={code,math-inline}]{$\inference{\infer{\Gamma,\mathid{x}:\t_1}{\exp}{\t_2}{\e_2}}{\infer{\Gamma}{\lambda \mathid{x}.\,\exp}{\tfun{\t_1}{\e_2}{\t_2}}{\e}}$} {\hspace{1em}}
 \mdSpan[class={rulename},font-variant={small-caps},font-size={small}]{(let)} \mdSpan[class={code,math-inline}]{$\inference{\infer{\Gamma}{\exp_1}{\s}{{\langle}{\rangle}}\mathspace{1}\quad \infer{\Gamma,\mathid{x}:\s}{\exp_2}{\t}{\e}}{\infer{\Gamma}{\mathkw{let}\;\mathspace{1}\mathid{x}\mathspace{1}=\mathspace{1}\exp_1\;\mathkw{in}\;\exp_2}{\t}{\e}}$} \mdBr
{\strut}\mdBr
 \mdSpan[class={rulename},font-variant={small-caps},font-size={small}]{(gen)} \mdSpan[class={code,math-inline}]{$\inference{\infer{\Gamma}{\exp}{\t}{{\langle}{\rangle}}\mathspace{1}\quad \overline\alpha \not\in \ftv\Gamma}{\infer{\Gamma}{\exp}{\forall\overline\alpha.\,\t}{\e}}$} {\hspace{1em}}
 \mdSpan[class={rulename},font-variant={small-caps},font-size={small}]{(run)} \mdSpan[class={code,math-inline}]{$\inference{\infer{\Gamma}{\mathid{e}}{\t}{\ext{\tst{\xi}}{\e}}\mathspace{1}\quad \xi \not\in \ftv{\Gamma,\t,\e}}{\infer{\Gamma}{\run\,\mathid{e}}{\t}{\e}}$} \mdBr
 {\strut}\mdBr
 \mdSpan[class={rulename},font-variant={small-caps},font-size={small}]{(inst)} \mdSpan[class={code,math-inline}]{$\inference{\infer{\Gamma}{\exp}{\forall\overline\alpha.\,\t}{\e}}{\infer{\Gamma}{\exp}{[\overline\alpha :=\mathspace{1}\overline\t]\t}{\e}}$} {\hspace{1em}}
 \mdSpan[class={rulename},font-variant={small-caps},font-size={small}]{(catch)} \mdSpan[class={code,math-inline}]{$\inference{\infer{\Gamma}{\mathid{e}_1}{\t}{\ext{\ec{\mathid{exn}}}{\e}}\mathspace{1}\quad \infer{\Gamma}{\mathid{e}_2}{()\mathspace{1}{\rightarrow}\mathspace{1}\e\,\t}{\e}}{\infer{\Gamma}{\catch\,\mathid{e}_1\,\,\mathid{e}_2}{\t}{\e}}$} \mdBr%
\end{mdP}%
\begin{mdP}[class={indent}]%
{\strut}%
\end{mdP}%
\begin{mdTable}[class={madoko,block}]{3}{lll}

\mdTd[display={table-cell}]{\mdSpan[class={rulename},font-variant={small-caps},font-size={small}]{(alloc)}}&\mdTd[display={table-cell}]{\mdSpan[class={code,math-inline}]{$\Gamma\,\vdash\,\new$}}&\mdTd[display={table-cell}]{: \mdSpan[class={code,math-inline}]{$\tfun{\t}{\ext{\sth}{\e}}{\tref{\mathid{h}}{\t}}$} {\textbar} \mdSpan[class={code,math-inline}]{$\e'$}}\\
\mdTd[display={table-cell}]{\mdSpan[class={rulename},font-variant={small-caps},font-size={small}]{(read)}}&\mdTd[display={table-cell}]{\mdSpan[class={code,math-inline}]{$\Gamma\,\vdash\,(!)$}}&\mdTd[display={table-cell}]{: \mdSpan[class={code,math-inline}]{$\tfun{\tref{\mathid{h}}{\t}}{\ext{\tst{\mathid{h}},\ec{\mathid{div}}}{\e}}{\t}$} {\textbar} \mdSpan[class={code,math-inline}]{$\e'$}}\\
\mdTd[display={table-cell}]{\mdSpan[class={rulename},font-variant={small-caps},font-size={small}]{(write)}}&\mdTd[display={table-cell}]{\mdSpan[class={code,math-inline}]{$\Gamma\,\vdash\,(:=)$}}&\mdTd[display={table-cell}]{: \mdSpan[class={code,math-inline}]{$\tfun{(\tref{\mathid{h}}{\t},\t)}{\ext\sth\e}{()}$} {\textbar} \mdSpan[class={code,math-inline}]{$\e'$}}\\
\mdTd[display={table-cell}]{\mdSpan[class={rulename},font-variant={small-caps},font-size={small}]{(throw)}}&\mdTd[display={table-cell}]{\mdSpan[class={code,math-inline}]{$\Gamma\,\vdash\,\throw$}}&\mdTd[display={table-cell}]{: \mdSpan[class={code,math-inline}]{$\tfun{()}{\ext{\ec{\mathid{exn}}}{\e}}{\t}$} {\textbar} \mdSpan[class={code,math-inline}]{$\e'$}}\\
\mdTd[display={table-cell}]{\mdSpan[class={rulename},font-variant={small-caps},font-size={small}]{(unit)}}&\mdTd[display={table-cell}]{\mdSpan[class={code,math-inline}]{$\Gamma\,\vdash\,()$}}&\mdTd[display={table-cell}]{: \mdSpan[class={code,math-inline}]{$()$} {\textbar} \mdSpan[class={code,math-inline}]{$\e$}}\\
\mdTd[display={table-cell}]{\mdSpan[class={rulename},font-variant={small-caps},font-size={small}]{(fix)}}&\mdTd[display={table-cell}]{\mdSpan[class={code,math-inline}]{$\Gamma\,\vdash\,\mathkw{fix}$}}&\mdTd[display={table-cell}]{: \mdSpan[class={code,math-inline}]{$((\t_1\mathspace{1}{\rightarrow}\mathspace{1}\extdiv\,\t_2)\mathspace{1}{\rightarrow}\mathspace{1}(\t_1\mathspace{1}{\rightarrow}\mathspace{1}\extdiv\,\t_2))\mathspace{1}{\rightarrow}\mathspace{1}(\t_1\mathspace{1}{\rightarrow}\mathspace{1}\extdiv\,\t_2)$} {\textbar} \mdSpan[class={code,math-inline}]{$\e'$}}\\
\end{mdTable}
\mdHr[class={figureline,madoko}]{}\mdSpan[class={figure-caption}]{\mdSpan[class={caption-before}]{\mdStrong{Figure{\mdNbsp}\mdSpan[class={figure-label}]{3}.} }General type rules with effects.}%
\end{mdDiv}%
\begin{mdP}%
Figure{\mdNbsp}\mdA[class={localref},target-element={figure}]{fig-typerules}{}{\mdSpan[class={figure-label}]{3}} defines the formal type rules of our effect system.
The rules are defined over a small expression calculus:%
\end{mdP}%
\begin{mdTable}[class={madoko,block}]{4}{llll}

\mdTd[display={table-cell}]{\mdSpan[class={code,math-inline}]{$\mathid{e}$}}&\multicolumn{1}{c}{\mdTd[text-align={center},display={table-cell}]{\mdSpan[class={code,math-inline}]{$::=$}}}&\mdTd[display={table-cell}]{\mdSpan[class={code,math-inline}]{$\mathid{x}$} {\textbar} \mdSpan[class={code,math-inline}]{$\mathid{p}$} {\textbar} \mdSpan[class={code,math-inline}]{$\mathid{e}_1\mathspace{1}\mathid{e}_2$} {\textbar} \mdSpan[class={code,math-inline}]{$\lambda \mathid{x}.\mathspace{1}\mathid{e}$}}&\mdTd[display={table-cell}]{(variables, primitives, applications, functions)}\\
\mdTd[display={table-cell}]{}&\multicolumn{1}{c}{\mdTd[text-align={center},display={table-cell}]{{\textbar}}}&\mdTd[display={table-cell}]{\mdSpan[class={code,math-inline}]{$\bind{\mathid{x}}{\mathid{e}_1}{\mathid{e}_2}$} {\textbar} \mdSpan[class={code,math-inline}]{$\mathkw{let}\mathspace{1}\mathid{x}\mathspace{1}=\mathspace{1}\mathid{e}_1\mathspace{1}\mathkw{in}\mathspace{1}\mathid{e}_2$}}&\mdTd[display={table-cell}]{(sequence and let bindings)}\\
\mdTd[display={table-cell}]{}&\multicolumn{1}{c}{\mdTd[text-align={center},display={table-cell}]{{\textbar}}}&\mdTd[display={table-cell}]{\mdSpan[class={code,math-inline}]{$\mathkw{catch}\mathspace{1}\mathid{e}_1\mathspace{1}\mathid{e}_2$} {\textbar} \mdSpan[class={code,math-inline}]{$\mathkw{run}\mathspace{1}\exp$}}&\mdTd[display={table-cell}]{(catch exceptions and isolate state)}\\
\mdTd[display={table-cell}]{}&\multicolumn{1}{c}{\mdTd[text-align={center},display={table-cell}]{}}&\mdTd[display={table-cell}]{}&\mdTd[display={table-cell}]{}\\
\mdTd[display={table-cell}]{\mdSpan[class={code,math-inline}]{$\mathid{p}$}}&\multicolumn{1}{c}{\mdTd[text-align={center},display={table-cell}]{\mdSpan[class={code,math-inline}]{$::=$}}}&\mdTd[display={table-cell}]{\mdSpan[class={code,math-inline}]{$()$} {\textbar} \mdSpan[class={code,math-inline}]{$\mathkw{fix}$} {\textbar} \mdSpan[class={code,math-inline}]{$\mathkw{throw}$} {\textbar} \mdSpan[class={code,math-inline}]{$\mathkw{new}$} {\textbar} \mdSpan[class={code,math-inline}]{$(!)$} {\textbar} \mdSpan[class={code,math-inline}]{$(:=)$}}&\mdTd[display={table-cell}]{(primitives)}\\
\end{mdTable}
\begin{mdP}%
This expression syntax is meant as a surface syntax, but when we discuss the
semantics of the calculus, we will refine and extend the syntax further (see
Figure{\mdNbsp}\mdA[class={localref},target-element={figure}]{fig-syntax}{}{\mdSpan[class={figure-label}]{4}}). We use the bind expression
\mdSpan[class={math-inline}]{$\bind{x}{\exp_1}{\exp_2}$} for a  monomorphic binding of a variable \mdSpan[class={math-inline}]{$x$} to an
expression \mdSpan[class={math-inline}]{$\exp_1$} (which is really just syntactic sugar for the application
\mdSpan[class={code,math-inline}]{$(\lambda \mathid{x}.\mathspace{1}\mathid{e}_2)\mathspace{1}\mathid{e}_1$}). We write \mdSpan[class={code,math-inline}]{$\exp_1;\exp_2$} as a shorthand for the
expression \mdSpan[class={math-inline}]{$\bind{x}{\exp_1}{\exp_2}$} where \mdSpan[class={math-inline}]{$x \not\in \fv{\exp_2}$}.  We have
added \mdSpan[class={math-inline}]{$\run$} and \mdSpan[class={math-inline}]{$\catch$} as special expressions since this simplifies the
presentation where we can give direct type rules for them. Also, we
simplified both \mdSpan[class={math-inline}]{$\catch$} and \mdSpan[class={math-inline}]{$\throw$} by limiting the exception type
to the unit type (\mdSpan[class={math-inline}]{$()$}).%
\end{mdP}%
\begin{mdP}[class={indent}]%
The type rules are stated under a type environment \mdSpan[class={math-inline}]{$\Gamma$} which maps
variables to types. An environment can be extended using a comma. If \mdSpan[class={math-inline}]{$\Gamma'$}
is equal to \mdSpan[class={math-inline}]{$\Gamma,x : \s$} then \mdSpan[class={math-inline}]{$\Gamma'(x) = \s$} and \mdSpan[class={math-inline}]{$\Gamma'(y) =
\Gamma(y)$} for any \mdSpan[class={math-inline}]{$y \neq x$}. A type rule of the form
\mdSpan[class={math-inline}]{$\infer{\Gamma}{e}{\s}{\e}$} states that under an environment \mdSpan[class={math-inline}]{$\Gamma$} the
expression \mdSpan[class={math-inline}]{$e$} has type \mdSpan[class={math-inline}]{$\s$} with an effect{\mdNbsp}\mdSpan[class={math-inline}]{$\e$}.%
\end{mdP}%
\begin{mdP}[class={indent}]%
Most of the type rules in Figure{\mdNbsp}\mdA[class={localref},target-element={figure}]{fig-typerules}{}{\mdSpan[class={figure-label}]{3}} are quite standard.  The
rule \mdSpan[class={rulename},font-variant={small-caps},font-size={small}]{(var)} derives the type of a variable. The derived effect is
any arbitrary effect \mdSpan[class={math-inline}]{$\e$}. We may have expected to derive only the total
effect \mdSpan[class={code,math-inline}]{${\langle}{\rangle}$} since the evaluation of a variable has no effect at all (in our
strict setting). However, there is no rule that lets us upgrade the final
effect and instead we get to pick the final effect right away. Another way to
look at this is that since the variable evaluation has no effect, we are free
to assume any arbitrary effect.%
\end{mdP}%
\begin{mdP}[class={indent}]%
The \mdSpan[class={rulename},font-variant={small-caps},font-size={small}]{(lam)} rule is similar: the evaluation of a lambda expression is
a value and has no effect and we can assume any arbitrary effect \mdSpan[class={math-inline}]{$\e$}.
Interestingly, the effect derived for the body of the lambda expression,
\mdSpan[class={math-inline}]{$\e_2$}, shifts from the derivation on to the derived function type
\mdSpan[class={math-inline}]{$\tfun{\t_1}{\e_2}{\t_2}$}: indeed, calling this function and evaluating the
body causes the effect \mdSpan[class={math-inline}]{$\e_2$}. The \mdSpan[class={rulename},font-variant={small-caps},font-size={small}]{(app)} is also standard, and
derives an effect \mdSpan[class={math-inline}]{$\e$} requiring that its premises derive the same effect as
the function effect.%
\end{mdP}%
\begin{mdP}[class={indent}]%
Instantiation \mdSpan[class={rulename},font-variant={small-caps},font-size={small}]{(inst)} is standard and instantiates a type scheme.
The generalization rule \mdSpan[class={rulename},font-variant={small-caps},font-size={small}]{(gen)} has an interesting twist: it requires
the derived effect to be total \mdSpan[class={code,math-inline}]{${\langle}{\rangle}$} . It turns out this is required to ensure
a sound semantics as we show in Section{\mdNbsp}\mdA[class={localref},target-element={h1}]{sec-semantics}{}{\mdSpan[class={heading-label}]{4}}. Indeed, this is
essentially the equivalent of the value restriction in ML
\mdSpan[class={citations},target-element={bibitem}]{[\mdA[class={bibref,localref},target-element={bibitem}]{leroy:valuerestriction}{}{\mdSpan[class={bibitem-label}]{21}}]}.  Of course, in ML effects are not inferred by the
type system and the value restriction syntactically restricts the expression
over which one can generalize. In our setting we can directly express that we
only generalize over total expressions. The rule \mdSpan[class={rulename},font-variant={small-caps},font-size={small}]{(let)} binds
expressions with a polymorphic type scheme \mdSpan[class={math-inline}]{$\s$} and just like \mdSpan[class={rulename},font-variant={small-caps},font-size={small}]{(gen)}
requires that the bound expression has no effect. It turns out that is still
sound to allow more effects at generalization and let bindings. In particular,
we can allow \mdSpan[class={math-inline}]{$\ec{exn}$}, \mdSpan[class={math-inline}]{$\ec{div}$}. However, for the formal development we
will only consider the empty effect for now.%
\end{mdP}%
\begin{mdP}[class={indent}]%
All other rules are just type rules for the primitive constants.  Note that
all the effects for the primitive constants are open and can be freely chosen
(just as in the \mdSpan[class={rulename},font-variant={small-caps},font-size={small}]{(var)} rule). This is important as it allows us to
always assume more effects than induced by the operation. In this article we
limit exceptions to just the unit value \mdSpan[class={code,math-inline}]{$()$} but of course in a practical 
implementation this would be a proper exception type. However, note that
any exception type must be restricted to base values only and cannot contain
function values! If this would be allowed our divergence theorem{\mdNbsp}\mdA[class={localref},target-element={theorem}]{lem-div}{}{\mdSpan[class={theorem-label}]{4}} 
(in Section{\mdNbsp}\mdA[class={localref},target-element={h1}]{sec-semeffect}{}{\mdSpan[class={heading-label}]{6}}) would not hold {\textendash} indeed, unchecked exceptions 
that can include non-base values can be more expressive than
call/cc{\mdNbsp}\mdSpan[class={citations},target-element={bibitem}]{[\mdA[class={bibref,localref},target-element={bibitem}]{lillibridge:exncallcc}{}{\mdSpan[class={bibitem-label}]{22}}]}.%
\end{mdP}%
\begin{mdP}[class={indent}]%
Given these type rules, we can construct an efficient type inference algorithm
that infers principal types and is sound and complete with respect to the type
rules. This is described in detail in Appendix{\mdNbsp}\mdA[class={localref},target-element={h1}]{sec-inference}{}{\mdSpan[class={heading-label}]{A}} and a
separate technical report{\mdNbsp}\mdSpan[class={citations},target-element={bibitem}]{[\mdA[class={bibref,localref},target-element={bibitem}]{leijen:kokatr}{}{\mdSpan[class={bibitem-label}]{19}}]}.%
\end{mdP}%
\mdHxxx[id=simplifying-types,label={[3.2]\{.heading-label\}},toc={},caption={Simplifying types}]{\mdSpan[class={heading-before}]{\mdSpan[class={heading-label}]{3.2}.{\enspace}}Simplifying types}\begin{mdP}[class={para-continue}]%
The rule \mdSpan[class={rulename},font-variant={small-caps},font-size={small}]{(app)} is a little surprising since it requires both the effect
of the function and the argument to be the same. This only works because we set 
things up to always be able to infer effects on functions that are {\textquoteleft}open{\textquoteright} {\textendash} i.e.
have a polymorphic \mdSpan[class={code,math-inline}]{$\mu$} in their tail. For example, consider the identify function:%
\end{mdP}%
\begin{mdDiv}[class={para-block,para-block,input-mathpre},elem={pre}]%
\begin{mdDiv}[class={math-display}]%
\[\begin{mdMathprearray}
\mathid{id}\mathspace{1}=\mathspace{1}\lambda \mathid{x}.\mathspace{1}\mathid{x}
\end{mdMathprearray}\]%
\end{mdDiv}
\end{mdDiv}%
\begin{mdP}[class={para-continue}]%
If we assign the valid type \mdSpan[class={code,math-inline}]{$\forall\alpha.\mathspace{1}\alpha{}\mathspace{1}{\rightarrow}\mathspace{1}{\langle}{\rangle}\mathspace{1}\alpha$} to the \mdSpan[class={code,math-inline}]{$\mathid{id}$} function,
we get into trouble quickly. For example, the application \mdSpan[class={code,math-inline}]{$\mathid{id}\mathspace{1}(\mathid{throw}\mathspace{1}())$} would not
type check since the effect of \mdSpan[class={code,math-inline}]{$\mathid{id}$} is total while the effect of the argument
is \mdSpan[class={code,math-inline}]{${\langle}\mathid{exn}|\mu{\rangle}$}. Of course, the type inference algorithm always infers a most
general type for \mdSpan[class={code,math-inline}]{$\mathid{id}$}, namely:%
\end{mdP}%
\begin{mdDiv}[class={para-block,para-block,input-mathpre},elem={pre}]%
\begin{mdDiv}[class={math-display}]%
\[\begin{mdMathprearray}
\mathid{id}\mathspace{1}:\mathspace{1}\forall\alpha\mu.\mathspace{1}\alpha{}\mathspace{1}{\rightarrow}\mathspace{1}\mu{}\mathspace{1}\alpha
\end{mdMathprearray}\]%
\end{mdDiv}
\end{mdDiv}%
\begin{mdP}%
which has no such problems.%
\end{mdP}%
\begin{mdP}[class={indent,para-continue}]%
However, in practice we may wish to simplify the types more: for most users it is more
natural to state that \mdSpan[class={code,math-inline}]{$\mathid{id}$} is total, instead of stating that \mdSpan[class={code,math-inline}]{$\mathid{id}$} can be any assigned
any effect. One possible way that we explored in Koka was to display types in a simplified
form. In particular, we can \mdEm{close} a type of a let-bound variable of the form:%
\end{mdP}%
\begin{mdDiv}[class={para-block,para-block,input-mathpre},elem={pre}]%
\begin{mdDiv}[class={math-display}]%
\[\begin{mdMathprearray}
\forall\mu\alpha_1...\alpha_\mathid{n}.\mathspace{1}\tau_1\mathspace{1}{\rightarrow}\mathspace{1}{\langle}\mathid{l}_{1},...,\mathid{l}_\mathid{m}|\mu{\rangle}\mathspace{1}\tau_2)
\end{mdMathprearray}\]%
\end{mdDiv}
\end{mdDiv}%
\begin{mdP}[class={para-continue}]%
where \mdSpan[class={code,math-inline}]{$\mu \not\in \ftv{\tau_1,\tau_2,\mathid{l}_{1},...,\mathid{l}_\mathid{m}}$}, to:%
\end{mdP}%
\begin{mdDiv}[class={para-block,para-block,input-mathpre},elem={pre}]%
\begin{mdDiv}[class={math-display}]%
\[\begin{mdMathprearray}
\forall\alpha_1...\alpha_\mathid{n}.\mathspace{1}\tau_1\mathspace{1}{\rightarrow}\mathspace{1}{\langle}\mathid{l}_{1},...,\mathid{l}_\mathid{m}{\rangle}\mathspace{1}\tau_2
\end{mdMathprearray}\]%
\end{mdDiv}
\end{mdDiv}%
\begin{mdP}%
However, this is not an ideal solution since in practice users often annotate
functions with the more restrictive closed type. The solution we adopted in Koka is 
to actually close \mdEm{all} the types of \mdEm{let-bound} variables according the above rule,
and at instantiation of such let-bound variable, we \mdEm{open} the type again, introducing
a fresh \mdSpan[class={code,math-inline}]{$\mu$} for any function type with a closed effect.%
\end{mdP}%
\begin{mdP}[class={indent}]%
The way our system is set up in this paper, the most general type for a 
let-bound function will always have an open effect type, and therefore the
programs accepted under the Koka system will be exactly the same. In general
though, if we allow function constants with a closed effect type, then the
more liberal Koka system may accept more programs since it will instantiate
all such function types with a closed effect to an open effect. Of course,
this is still a sound extension since we can always soundly assume more
effects. Finally, note that we can only apply this form of simplification to
let-bound variables, and not to lambda-bound parameters or otherwise type
inference is no longer principal.%
\end{mdP}%
\mdHxx[id=sec-semantics,label={[4]\{.heading-label\}},toc={},caption={Semantics of effects}]{\mdSpan[class={heading-before}]{\mdSpan[class={heading-label}]{4}.{\enspace}}Semantics of effects}\begin{mdDiv}[class={figure,align-center},id=fig-syntax,label={[4]\{.figure-label\}},elem={figure},toc={tof},toc-line={[4]\{.figure-label\}. Full expression syntax},page-align={top},caption={Full expression syntax}]%
\begin{mdTable}[class={madoko,block}]{4}{llll}

\mdTd[display={table-cell}]{\mdSpan[class={code,math-inline}]{$\exp$}}&\multicolumn{1}{c}{\mdTd[text-align={center},display={table-cell}]{::=}}&\mdTd[display={table-cell}]{\mdSpan[class={code,math-inline}]{$\mathid{v}$} {\textbar} \mdSpan[class={code,math-inline}]{$\exp_1\;\exp_2$} {\textbar} \mdSpan[class={code,math-inline}]{$\mathkw{let}\mathspace{1}\mathid{x}\mathspace{1}=\mathspace{1}\exp_1\mathspace{1}\mathkw{in}\mathspace{1}\exp_2$}}&\mdTd[display={table-cell}]{(values, applications, let bindings)}\\
\mdTd[display={table-cell}]{}&\multicolumn{1}{c}{\mdTd[text-align={center},display={table-cell}]{{\textbar}}}&\mdTd[display={table-cell}]{\mdSpan[class={code,math-inline}]{$\hp\varphi\exp$} {\textbar} \mdSpan[class={code,math-inline}]{$\run\,\exp $}}&\mdTd[display={table-cell}]{(heap binding and isolation)}\\
\mdTd[display={table-cell}]{}&\multicolumn{1}{c}{\mdTd[text-align={center},display={table-cell}]{}}&\mdTd[display={table-cell}]{}&\mdTd[display={table-cell}]{}\\
\mdTd[display={table-cell}]{\mdSpan[class={code,math-inline}]{$\mathid{v}$}}&\multicolumn{1}{c}{\mdTd[text-align={center},display={table-cell}]{::=}}&\mdTd[display={table-cell}]{\mdSpan[class={code,math-inline}]{$\lambda \mathid{x}.\,\exp$} {\textbar} \mdSpan[class={code,math-inline}]{$\catch\,\exp $}}&\mdTd[display={table-cell}]{(functions and partial catch)}\\
\mdTd[display={table-cell}]{}&\multicolumn{1}{c}{\mdTd[text-align={center},display={table-cell}]{{\textbar}}}&\mdTd[display={table-cell}]{\mdSpan[class={code,math-inline}]{$\mathid{b}\mathspace{1}$}}&\mdTd[display={table-cell}]{(basic value (contains no \mdSpan[class={math-inline}]{$e$}))}\\
\mdTd[display={table-cell}]{}&\multicolumn{1}{c}{\mdTd[text-align={center},display={table-cell}]{}}&\mdTd[display={table-cell}]{}&\mdTd[display={table-cell}]{}\\
\mdTd[display={table-cell}]{\mdSpan[class={code,math-inline}]{$\mathid{b}$}}&\multicolumn{1}{c}{\mdTd[text-align={center},display={table-cell}]{::=}}&\mdTd[display={table-cell}]{\mdSpan[class={code,math-inline}]{$\mathid{x}$} {\textbar} \mdSpan[class={code,math-inline}]{$\mathid{c}$}}&\mdTd[display={table-cell}]{(variable and constants)}\\
\mdTd[display={table-cell}]{}&\multicolumn{1}{c}{\mdTd[text-align={center},display={table-cell}]{{\textbar}}}&\mdTd[display={table-cell}]{\mdSpan[class={code,math-inline}]{$\fix$} {\textbar} \mdSpan[class={code,math-inline}]{$\throw $} {\textbar} \mdSpan[class={code,math-inline}]{$\catch$}}&\mdTd[display={table-cell}]{(fixpoint and exceptions)}\\
\mdTd[display={table-cell}]{}&\multicolumn{1}{c}{\mdTd[text-align={center},display={table-cell}]{{\textbar}}}&\mdTd[display={table-cell}]{\mdSpan[class={code,math-inline}]{$\rx$}  {\textbar} \mdSpan[class={code,math-inline}]{$\new$}  {\textbar} \mdSpan[class={code,math-inline}]{$(!)$}  {\textbar} \mdSpan[class={code,math-inline}]{$(:=)$}  {\textbar} \mdSpan[class={code,math-inline}]{$(\rx:=)$}}&\mdTd[display={table-cell}]{(references)}\\
\mdTd[display={table-cell}]{}&\multicolumn{1}{c}{\mdTd[text-align={center},display={table-cell}]{}}&\mdTd[display={table-cell}]{}&\mdTd[display={table-cell}]{}\\
\mdTd[display={table-cell}]{\mdSpan[class={code,math-inline}]{$\mathid{w}$}}&\multicolumn{1}{c}{\mdTd[text-align={center},display={table-cell}]{::=}}&\mdTd[display={table-cell}]{\mdSpan[class={code,math-inline}]{$\mathid{b}$} {\textbar} \mdSpan[class={code,math-inline}]{$\throw\,\mathid{c}$}}&\mdTd[display={table-cell}]{(basic value or exception)}\\
\mdTd[display={table-cell}]{\mdSpan[class={code,math-inline}]{$\mathid{a}$}}&\multicolumn{1}{c}{\mdTd[text-align={center},display={table-cell}]{::=}}&\mdTd[display={table-cell}]{\mdSpan[class={code,math-inline}]{$\mathid{v}$} {\textbar} \mdSpan[class={code,math-inline}]{$\throw\,\mathid{c}$} {\textbar} \mdSpan[class={code,math-inline}]{$\hp\varphi{\mathid{v}}$} {\textbar} \mdSpan[class={code,math-inline}]{$\hp\varphi{\throw\,\mathid{c}}$}}&\mdTd[display={table-cell}]{(answers)}\\
\mdTd[display={table-cell}]{}&\multicolumn{1}{c}{\mdTd[text-align={center},display={table-cell}]{}}&\mdTd[display={table-cell}]{}&\mdTd[display={table-cell}]{}\\
\mdTd[display={table-cell}]{\mdSpan[class={code,math-inline}]{$\varphi$}}&\multicolumn{1}{c}{\mdTd[text-align={center},display={table-cell}]{::=}}&\mdTd[display={table-cell}]{\mdSpan[class={code,math-inline}]{${\langle}\rx_1\mathspace{1}\mapsto \mathid{v}_1{\rangle}\mathspace{1}...\mathspace{1}{\langle}\rx_\mathid{n}\mathspace{1}\mapsto \mathid{v}_\mathid{n}{\rangle}$}}&\mdTd[display={table-cell}]{(heap bindings)}\\
\end{mdTable}
\mdHr[class={figureline,madoko}]{}\mdSpan[class={figure-caption}]{\mdSpan[class={caption-before}]{\mdStrong{Figure{\mdNbsp}\mdSpan[class={figure-label}]{4}.} }Full expression syntax}%
\end{mdDiv}%
\begin{mdP}%
In this section we are going to define a precise semantics for our language,
and show that well-typed programs cannot go `wrong{'}. In contrast to our earlier
soundness and completeness result for the type inference algorithm, the soundness proof
of the type system in Hindley-Milner does not carry over easily in our setting: 
indeed, we are going to model many complex effects which is fraught with danger.%
\end{mdP}%
\begin{mdP}[class={indent}]%
First, we strengthen our expression syntax by separating out value expressions
\mdSpan[class={math-inline}]{$v$}, as shown in Figure{\mdNbsp}\mdA[class={localref},target-element={figure}]{fig-syntax}{}{\mdSpan[class={figure-label}]{4}}. We also define basic values \mdSpan[class={math-inline}]{$b$} as
values that cannot contain expressions themselves. Moreover, we added a few new expressions,
namely heap bindings (\mdSpan[class={math-inline}]{$\hp\varphi{e}$}), a partially applied catch (\mdSpan[class={math-inline}]{$\catch\,e$}), a
partially applied assignments \mdSpan[class={math-inline}]{$(v :=)$}, and general constants \mdSpan[class={math-inline}]{$(c)$}. 
Also, we denote heap variables using \mdSpan[class={math-inline}]{$r$}.
An expression \mdSpan[class={code,math-inline}]{$\hp{{\langle}\mathid{r}_1\mathspace{1}\mapsto \mathid{v}_1{\rangle},...,{\langle}\mathid{r}_\mathid{n}\mathspace{1}\mapsto \mathid{v}_\mathid{n}{\rangle}}{\mathid{e}}$} binds \mdSpan[class={math-inline}]{$r_1$} to \mdSpan[class={math-inline}]{$r_n$} in
\mdSpan[class={math-inline}]{$v_1,...,v_n$} and \mdSpan[class={math-inline}]{$e$}. By convention, we always require \mdSpan[class={math-inline}]{$r_1$} to \mdSpan[class={math-inline}]{$r_n$} to be distinct,
and consider heaps \mdSpan[class={math-inline}]{$\varphi$} equal modulo alpha-renaming.%
\end{mdP}%
\begin{mdDiv}[class={figure,align-center},id=fig-extratyperules,label={[5]\{.figure-label\}},elem={figure},toc={tof},toc-line={[5]\{.figure-label\}. Extra type rules for heap expressions and constants. We write \${\textbackslash}overline{\textbackslash}varphi\_h\$ for the conversion of a heap \${\textbackslash}varphi\$ to a type environment: if \${\textbackslash}varphi\$ equals `{\textless}r\_1 {\textbackslash}mapsto v\_1,..., r\_n {\textbackslash}mapsto v\_n{\textgreater}` then \${\textbackslash}overline{\textbackslash}varphi\_h = r\_1 : {\textbackslash}tref\{h\}\{{\textbackslash}t\_1\}, ..., r\_n : {\textbackslash}tref\{h\}\{{\textbackslash}t\_n\}\$ for some \${\textbackslash}t\_1\$ to \${\textbackslash}t\_n\$.},page-align={top},caption={Extra type rules for heap expressions and constants. We write \${\textbackslash}overline{\textbackslash}varphi\_h\$ for the conversion of a heap \${\textbackslash}varphi\$ to a type environment: if \${\textbackslash}varphi\$ equals `{\textless}r\_1 {\textbackslash}mapsto v\_1,..., r\_n {\textbackslash}mapsto v\_n{\textgreater}` then \${\textbackslash}overline{\textbackslash}varphi\_h = r\_1 : {\textbackslash}tref\{h\}\{{\textbackslash}t\_1\}, ..., r\_n : {\textbackslash}tref\{h\}\{{\textbackslash}t\_n\}\$ for some \${\textbackslash}t\_1\$ to \${\textbackslash}t\_n\$.}]%
\begin{mdP}%
\mdSpan[class={rulename},font-variant={small-caps},font-size={small}]{(heap)}  \mdSpan[class={code,math-inline}]{$\inference{\ontop{\forall{\langle}\mathid{r}_\mathid{i}\mathspace{1}\mapsto \mathid{v}_\mathid{i}{\rangle}\mathspace{1}\in \varphi.\;\;\infer{\Gamma,\overline\varphi_\mathid{h}}{\mathid{v}_\mathid{i}}{\t_\mathid{i}}{{\langle}{\rangle}}}{\infer{\Gamma,\overline\varphi_\mathid{h}}{\mathid{e}}{\t}{\ext\sth\e}}}{\infer{\Gamma}{\hp\varphi{\mathid{e}}}{\t}{\ext\sth\e}}$} 
{\enspace}
\mdSpan[class={rulename},font-variant={small-caps},font-size={small}]{(const)} \mdSpan[class={code,math-inline}]{$\inference{\typeof{\mathid{c}}\mathspace{1}=\mathspace{1}\sigma}{\infer{\Gamma}{\mathid{c}}{\sigma}{\e}}$}%
\end{mdP}%
\mdHr[class={figureline,madoko}]{}\mdSpan[class={figure-caption}]{\mdSpan[class={caption-before}]{\mdStrong{Figure{\mdNbsp}\mdSpan[class={figure-label}]{5}.} }Extra type rules for heap expressions and constants. We write \mdSpan[class={math-inline}]{$\overline\varphi_h$} for the conversion of a heap \mdSpan[class={math-inline}]{$\varphi$} to a type environment: if \mdSpan[class={math-inline}]{$\varphi$} equals \mdSpan[class={code,math-inline}]{${\langle}\mathid{r}_1\mathspace{1}\mapsto \mathid{v}_1,...,\mathspace{1}\mathid{r}_\mathid{n}\mathspace{1}\mapsto \mathid{v}_\mathid{n}{\rangle}$} then \mdSpan[class={math-inline}]{$\overline\varphi_h = r_1 : \tref{h}{\t_1}, ..., r_n : \tref{h}{\t_n}$} for some \mdSpan[class={math-inline}]{$\t_1$} to \mdSpan[class={math-inline}]{$\t_n$}.}%
\end{mdDiv}%
\begin{mdP}[class={indent}]%
The surface language never exposes the heap binding construct
\mdSpan[class={math-inline}]{$\hp\varphi{e}$} directly to the user but during evaluation the reductions on
heap operations create heaps and use them. In order to give a type to
such expression, we need an extra type rule for heap bindings, given in Figure
\mdA[class={localref},target-element={figure}]{fig-extratyperules}{}{\mdSpan[class={figure-label}]{5}}. Note how each heap value is typed under an enviroment
that contains types for all bindings (much like a recursive \mdSpan[class={code,math-inline}]{$\mathkw{let}$} binding).
Moreover, a heap binding induces the stateful effect \mdSpan[class={math-inline}]{$\sth$}. In the type rule for constants
 we assume a function \mdSpan[class={math-inline}]{$\typeof{c}$} that returns a closed
type scheme for each constant.%
\end{mdP}%
\mdHxxx[id=reductions,label={[4.1]\{.heading-label\}},toc={},caption={Reductions}]{\mdSpan[class={heading-before}]{\mdSpan[class={heading-label}]{4.1}.{\enspace}}Reductions}\begin{mdDiv}[class={figure,align-center},id=fig-reduction,label={[6]\{.figure-label\}},elem={figure},toc={tof},toc-line={[6]\{.figure-label\}. Reduction rules and evaluation contexts.},page-align={top},caption={Reduction rules and evaluation contexts.}]%
\begin{mdTable}[class={madoko,block}]{4}{llll}

\mdTd[text-align={left},display={table-cell}]{\mdSpan[class={rulename},font-variant={small-caps},font-size={small}]{(\mdSpan[class={math-inline}]{$\delta$})}}&\mdTd[text-align={left},display={table-cell}]{\mdSpan[class={code,math-inline}]{$\mathid{c}\,\mathid{v}$}}&\mdTd[display={table-cell}]{\mdSpan[class={code,math-inline}]{${\longrightarrow}$}}&\mdTd[text-align={left},display={table-cell}]{\mdSpan[class={code,math-inline}]{$\delta(\mathid{c},\mathid{v})\quad$} if \mdSpan[class={math-inline}]{$\delta(c,v)$} is defined}\\
\mdTd[text-align={left},display={table-cell}]{\mdSpan[class={rulename},font-variant={small-caps},font-size={small}]{(\mdSpan[class={math-inline}]{$\beta$})}}&\mdTd[text-align={left},display={table-cell}]{\mdSpan[class={code,math-inline}]{$(\lambda \mathid{x}.\mathspace{1}\mathid{e})\mathspace{1}\mathid{v}$}}&\mdTd[display={table-cell}]{\mdSpan[class={code,math-inline}]{${\longrightarrow}$}}&\mdTd[text-align={left},display={table-cell}]{\mdSpan[class={code,math-inline}]{$[\mathid{x}\mathspace{1}\mapsto \mathid{v}]\mathid{e}$}}\\
\mdTd[text-align={left},display={table-cell}]{\mdSpan[class={rulename},font-variant={small-caps},font-size={small}]{(let)}}&\mdTd[text-align={left},display={table-cell}]{\mdSpan[class={code,math-inline}]{$\mathkw{let}\mathspace{1}\mathid{x}\mathspace{1}=\mathspace{1}\mathid{v}\mathspace{1}\mathkw{in}\mathspace{1}\mathid{e}$}}&\mdTd[display={table-cell}]{\mdSpan[class={code,math-inline}]{${\longrightarrow}$}}&\mdTd[text-align={left},display={table-cell}]{\mdSpan[class={code,math-inline}]{$[\mathid{x}\mathspace{1}\mapsto \mathid{v}]\mathid{e}$}}\\
\mdTd[text-align={left},display={table-cell}]{\mdSpan[class={rulename},font-variant={small-caps},font-size={small}]{(fix)}}&\mdTd[text-align={left},display={table-cell}]{\mdSpan[class={code,math-inline}]{$\fix\,\mathid{v}$}}&\mdTd[display={table-cell}]{\mdSpan[class={code,math-inline}]{${\longrightarrow}$}}&\mdTd[text-align={left},display={table-cell}]{\mdSpan[class={code,math-inline}]{$\mathid{v}\,(\lambda \mathid{x}.\,(\fix\,\mathspace{1}\mathid{v})\,\mathid{x})$}}\\
\mdTd[text-align={left},display={table-cell}]{}&\mdTd[text-align={left},display={table-cell}]{}&\mdTd[display={table-cell}]{}&\mdTd[text-align={left},display={table-cell}]{}\\
\mdTd[text-align={left},display={table-cell}]{\mdSpan[class={rulename},font-variant={small-caps},font-size={small}]{(throw)}}&\mdTd[text-align={left},display={table-cell}]{\mdSpan[class={code,math-inline}]{$\mathid{X}[\throw\,\mathid{c}]$}}&\mdTd[display={table-cell}]{\mdSpan[class={code,math-inline}]{${\longrightarrow}$}}&\mdTd[text-align={left},display={table-cell}]{\mdSpan[class={code,math-inline}]{$\throw\,\mathid{c}\mathspace{1}\quad$} if \mdSpan[class={code,math-inline}]{$\mathid{X}\mathspace{1}\neq [\,]$}}\\
\mdTd[text-align={left},display={table-cell}]{\mdSpan[class={rulename},font-variant={small-caps},font-size={small}]{(catcht)}}&\mdTd[text-align={left},display={table-cell}]{\mdSpan[class={code,math-inline}]{$\catch\,(\throw\,\mathid{c})\,\exp$}}&\mdTd[display={table-cell}]{\mdSpan[class={code,math-inline}]{${\longrightarrow}$}}&\mdTd[text-align={left},display={table-cell}]{\mdSpan[class={code,math-inline}]{$\exp\,\mathid{c}$}}\\
\mdTd[text-align={left},display={table-cell}]{\mdSpan[class={rulename},font-variant={small-caps},font-size={small}]{(catchv)}}&\mdTd[text-align={left},display={table-cell}]{\mdSpan[class={code,math-inline}]{$\catch\,\mathid{v}\,\exp$}}&\mdTd[display={table-cell}]{\mdSpan[class={code,math-inline}]{${\longrightarrow}$}}&\mdTd[text-align={left},display={table-cell}]{\mdSpan[class={code,math-inline}]{$\mathid{v}$}}\\
\mdTd[text-align={left},display={table-cell}]{}&\mdTd[text-align={left},display={table-cell}]{}&\mdTd[display={table-cell}]{}&\mdTd[text-align={left},display={table-cell}]{}\\
\mdTd[text-align={left},display={table-cell}]{\mdSpan[class={rulename},font-variant={small-caps},font-size={small}]{(alloc)}}&\mdTd[text-align={left},display={table-cell}]{\mdSpan[class={code,math-inline}]{$\new\,\mathid{v}$}}&\mdTd[display={table-cell}]{\mdSpan[class={code,math-inline}]{${\longrightarrow}$}}&\mdTd[text-align={left},display={table-cell}]{\mdSpan[class={code,math-inline}]{$\hp{{\langle}\rx \mapsto \mathid{v}{\rangle}}{\rx}$}}\\
\mdTd[text-align={left},display={table-cell}]{\mdSpan[class={rulename},font-variant={small-caps},font-size={small}]{(read)}}&\mdTd[text-align={left},display={table-cell}]{\mdSpan[class={code,math-inline}]{$\hp{\varphi{\langle}\rx \mapsto \mathid{v}{\rangle}}{\R{!\rx}}$}}&\mdTd[display={table-cell}]{\mdSpan[class={code,math-inline}]{${\longrightarrow}$}}&\mdTd[text-align={left},display={table-cell}]{\mdSpan[class={code,math-inline}]{$\hp{\varphi{\langle}\rx \mapsto \mathid{v}{\rangle}}{\R{\mathid{v}}}$}}\\
\mdTd[text-align={left},display={table-cell}]{\mdSpan[class={rulename},font-variant={small-caps},font-size={small}]{(write)}}&\mdTd[text-align={left},display={table-cell}]{\mdSpan[class={code,math-inline}]{$\hp{\varphi{\langle}\rx \mapsto \mathid{v}_1{\rangle}}{\R{\rx :=\mathspace{1}\mathid{v}_2}}$}}&\mdTd[display={table-cell}]{\mdSpan[class={code,math-inline}]{${\longrightarrow}$}}&\mdTd[text-align={left},display={table-cell}]{\mdSpan[class={code,math-inline}]{$\hp{\varphi{\langle}\rx \mapsto \mathid{v}_2{\rangle}}{\R{()}}$}}\\
\mdTd[text-align={left},display={table-cell}]{\mdSpan[class={rulename},font-variant={small-caps},font-size={small}]{(merge)}}&\mdTd[text-align={left},display={table-cell}]{\mdSpan[class={code,math-inline}]{$\hp{\varphi_1}{\hp{\varphi_2}\mathid{e}}$}}&\mdTd[display={table-cell}]{\mdSpan[class={code,math-inline}]{${\longrightarrow}$}}&\mdTd[text-align={left},display={table-cell}]{\mdSpan[class={code,math-inline}]{$\hp{\varphi_1\varphi_2}{\mathid{e}}$}}\\
\mdTd[text-align={left},display={table-cell}]{\mdSpan[class={rulename},font-variant={small-caps},font-size={small}]{(lift)}}&\mdTd[text-align={left},display={table-cell}]{\mdSpan[class={code,math-inline}]{$\R{\hp\varphi{\mathid{e}}}$}}&\mdTd[display={table-cell}]{\mdSpan[class={code,math-inline}]{${\longrightarrow}$}}&\mdTd[text-align={left},display={table-cell}]{\mdSpan[class={code,math-inline}]{$\hp{\varphi}{\R{\mathid{e}}}\mathspace{1}\quad\textrm{if}\;\mathspace{1}\mathid{R}\mathspace{1}\neq [\,]$}}\\
\mdTd[text-align={left},display={table-cell}]{}&\mdTd[text-align={left},display={table-cell}]{}&\mdTd[display={table-cell}]{}&\mdTd[text-align={left},display={table-cell}]{}\\
\mdTd[text-align={left},display={table-cell}]{\mdSpan[class={rulename},font-variant={small-caps},font-size={small}]{(runl)}}&\mdTd[text-align={left},display={table-cell}]{\mdSpan[class={code,math-inline}]{$\run\,[\hp\varphi]\,\lambda \mathid{x}.\,\mathid{e}$}}&\mdTd[display={table-cell}]{\mdSpan[class={code,math-inline}]{${\longrightarrow}$}}&\mdTd[text-align={left},display={table-cell}]{\mdSpan[class={code,math-inline}]{$\lambda \mathid{x}.\,\run\,([\hp\varphi]\,\mathid{e})$}}\\
\mdTd[text-align={left},display={table-cell}]{\mdSpan[class={rulename},font-variant={small-caps},font-size={small}]{(runc)}}&\mdTd[text-align={left},display={table-cell}]{\mdSpan[class={code,math-inline}]{$\run\,[\hp\varphi]\,\catch\,\mathid{e}$}}&\mdTd[display={table-cell}]{\mdSpan[class={code,math-inline}]{${\longrightarrow}$}}&\mdTd[text-align={left},display={table-cell}]{\mdSpan[class={code,math-inline}]{$\catch\,(\run\,([\hp\varphi]\,\mathid{e}))$}}\\
\mdTd[text-align={left},display={table-cell}]{\mdSpan[class={rulename},font-variant={small-caps},font-size={small}]{(runh)}}&\mdTd[text-align={left},display={table-cell}]{\mdSpan[class={code,math-inline}]{$\run\,[\hp\varphi]\,\mathid{w}$}}&\mdTd[display={table-cell}]{\mdSpan[class={code,math-inline}]{${\longrightarrow}$}}&\mdTd[text-align={left},display={table-cell}]{\mdSpan[class={code,math-inline}]{$\mathid{w}\mathspace{1}\quad\textrm{if $\frv{w}\mathspace{1}\disjoint\,\mathspace{1}\dom\varphi$}$}}\\
\end{mdTable}
\begin{mdP}%
Evaluation contexts:%
\end{mdP}%
\begin{mdTable}[class={madoko,block}]{3}{lll}

\mdTd[text-align={left},display={table-cell}]{\mdSpan[class={code,math-inline}]{$\mathid{X}$}}&\multicolumn{1}{c}{\mdTd[text-align={center},display={table-cell}]{::=}}&\mdTd[text-align={left},display={table-cell}]{\mdSpan[class={code,math-inline}]{$[\,]$} {\textbar} \mdSpan[class={code,math-inline}]{$\mathid{X}\mathspace{1}\exp$} {\textbar} \mdSpan[class={code,math-inline}]{$\mathid{v}\;\mathid{X}$} {\textbar} \mdSpan[class={code,math-inline}]{$\mathkw{let}\mathspace{1}\mathid{x}\mathspace{1}=\mathspace{1}\mathid{X}\mathspace{1}\mathkw{in}\mathspace{1}\exp$}}\\
\mdTd[text-align={left},display={table-cell}]{\mdSpan[class={code,math-inline}]{$\mathid{R}$}}&\multicolumn{1}{c}{\mdTd[text-align={center},display={table-cell}]{::=}}&\mdTd[text-align={left},display={table-cell}]{\mdSpan[class={code,math-inline}]{$[\,]$} {\textbar} \mdSpan[class={code,math-inline}]{$\mathid{R}\mathspace{1}\exp$} {\textbar} \mdSpan[class={code,math-inline}]{$\mathid{v}\;\mathid{R}$} {\textbar} \mdSpan[class={code,math-inline}]{$\mathkw{let}\mathspace{1}\mathid{x}\mathspace{1}=\mathspace{1}\mathid{R}\mathspace{1}\mathkw{in}\mathspace{1}\exp$} {\textbar} \mdSpan[class={code,math-inline}]{$\catch\,\mathid{R}\,\exp$}}\\
\mdTd[text-align={left},display={table-cell}]{\mdSpan[class={code,math-inline}]{$\mathid{E}$}}&\multicolumn{1}{c}{\mdTd[text-align={center},display={table-cell}]{::=}}&\mdTd[text-align={left},display={table-cell}]{\mdSpan[class={code,math-inline}]{$[\,]$} {\textbar} \mdSpan[class={code,math-inline}]{$\mathid{E}\mathspace{1}\exp$} {\textbar} \mdSpan[class={code,math-inline}]{$\mathid{v}\;\mathid{E}$} {\textbar} \mdSpan[class={code,math-inline}]{$\mathkw{let}\mathspace{1}\mathid{x}\mathspace{1}=\mathspace{1}\mathid{E}\mathspace{1}\mathkw{in}\mathspace{1}\exp$} {\textbar} \mdSpan[class={code,math-inline}]{$\catch\,\mathid{E}\,\exp$} {\textbar} \mdSpan[class={code,math-inline}]{$\hp{\varphi}{\mathid{E}}$} {\textbar} \mdSpan[class={code,math-inline}]{$\run\,\mathid{E}$}}\\
\end{mdTable}
\mdHr[class={figureline,madoko}]{}\mdSpan[class={figure-caption}]{\mdSpan[class={caption-before}]{\mdStrong{Figure{\mdNbsp}\mdSpan[class={figure-label}]{6}.} }Reduction rules and evaluation contexts.}%
\end{mdDiv}%
\begin{mdP}%
We can now consider primitive reductions for the various expressions as shown
in Figure{\mdNbsp}\mdA[class={localref},target-element={figure}]{fig-reduction}{}{\mdSpan[class={figure-label}]{6}}. The first four reductions are standard for the
lambda calculus. To abstract away from a particular set of constants, we assume
a function \mdSpan[class={math-inline}]{$\delta$} which takes a constant and a closed value to a closed value.
If \mdSpan[class={code,math-inline}]{$\typeof{\mathid{c}}\mathspace{1}=\mathspace{1}\forall\overline\alpha.\,\t_1\mathspace{1}{\rightarrow}\mathspace{1}\e \t_2$}, with 
\mdSpan[class={math-inline}]{$\sub = [\overline\alpha \mapsto \overline\t]$} and \mdSpan[class={code,math-inline}]{$\infer\cdot{\mathid{v}}{\sub\t_1}{{\langle}{\rangle}}$}, then 
\mdSpan[class={math-inline}]{$\delta(c,v)$} is defined, and \mdSpan[class={code,math-inline}]{$\infer\cdot{\delta(\mathid{c},\mathid{v})}{\sub\t_2}{\sub\e}$}.
The reductions \mdSpan[class={math-inline}]{$\beta$}, \mdSpan[class={rulename},font-variant={small-caps},font-size={small}]{(let)} and \mdSpan[class={rulename},font-variant={small-caps},font-size={small}]{(fix)} are all standard.%
\end{mdP}%
\begin{mdP}[class={indent}]%
The next three rules deal with exceptions. In particular, the rule
\mdSpan[class={rulename},font-variant={small-caps},font-size={small}]{(throw)} progates exceptions under a context \mdSpan[class={math-inline}]{$X$}. Since \mdSpan[class={math-inline}]{$X$} does not
include \mdSpan[class={math-inline}]{$\catch\,e_1\,e_2$}, \mdSpan[class={math-inline}]{$\hp\varphi{e}$} or \mdSpan[class={math-inline}]{$\run\,e$}, this propagates
the exception to the nearest  exception handler or state block. The
\mdSpan[class={rulename},font-variant={small-caps},font-size={small}]{(catcht)} reduction catches exceptions and passes them on to the
handler. If the handler raises an exception itself, that  exception will then
propagate to its nearest enclosing exception hander.%
\end{mdP}%
\begin{mdP}[class={indent}]%
The next five rules
model heap reductions. Allocation creates a heap, while \mdSpan[class={math-inline}]{$(!)$} and \mdSpan[class={math-inline}]{$(:=)$} read
and write from the a heap. Through the \mdSpan[class={math-inline}]{$R$} context, these always operate on
the nearest enclosing heap since \mdSpan[class={math-inline}]{$R$} does not contain \mdSpan[class={math-inline}]{$\hp\varphi\,e$} or
\mdSpan[class={math-inline}]{$\run\,e$} expressions. The rules \mdSpan[class={rulename},font-variant={small-caps},font-size={small}]{(lift)} and \mdSpan[class={rulename},font-variant={small-caps},font-size={small}]{(merge)} let
us lift heaps out of expressions to ensure that all references can be bound in
the nearest enclosing heap.%
\end{mdP}%
\begin{mdP}[class={indent}]%
The final three rules deal with state isolation through \mdSpan[class={math-inline}]{$\run$}. We write
\mdSpan[class={math-inline}]{$[\hp\varphi]$} to denote an optional heap binding (so we really define six
rules for state isolation). The first two rules \mdSpan[class={rulename},font-variant={small-caps},font-size={small}]{(runl)} and
\mdSpan[class={rulename},font-variant={small-caps},font-size={small}]{(runc)} push a \mdSpan[class={math-inline}]{$\run$} operation down into a lambda-expression or
partial catch expression.
The final rule \mdSpan[class={rulename},font-variant={small-caps},font-size={small}]{(runh)} captures the essence of state isolation and 
reduces to a new value (or exception) discarding the heap \mdSpan[class={math-inline}]{$\varphi$}. 
The side condition \mdSpan[class={math-inline}]{$\frv{w} \disjoint\, \dom\varphi$} is necessary to ensure
well-formedness where a reference should not `escape{'} its binding.%
\end{mdP}%
\begin{mdP}[class={indent}]%
Using the reduction rules, we can now define an evaluation function.
Using the evaluation context \mdSpan[class={math-inline}]{$E$} defined in Figure{\mdNbsp}\mdA[class={localref},target-element={figure}]{fig-reduction}{}{\mdSpan[class={figure-label}]{6}}, we define
\mdSpan[class={math-inline}]{$\E{e} \longmapsto \E{e'}  \quad\mathrm{iff}\quad e \longrightarrow\, e'$}.
The evaluation context ensures strict semantics where only the leftmost-
outermost reduction is applicable in an expression. We define the relation
\mdSpan[class={code,math-inline}]{${\dlongmapsto}$} as the reflexive and transtive closure of \mdSpan[class={math-inline}]{$\longmapsto$}. We can show
that \mdSpan[class={code,math-inline}]{${\dlongmapsto}$} is a function even though we need a simple diamond theorem since
the order in which \mdSpan[class={rulename},font-variant={small-caps},font-size={small}]{(lift)} and \mdSpan[class={rulename},font-variant={small-caps},font-size={small}]{(merge)} reductions happen
is not fixed{\mdNbsp}\mdSpan[class={citations},target-element={bibitem}]{[\mdA[class={bibref,localref},target-element={bibitem}]{wrightfelleisen}{}{\mdSpan[class={bibitem-label}]{44}}]}.%
\end{mdP}%
\begin{mdP}[class={indent}]%
The final results, or answers \mdSpan[class={math-inline}]{$a$}, that expressions evaluate to, are either
values \mdSpan[class={math-inline}]{$v$}, exceptions \mdSpan[class={math-inline}]{$\throw\,c$}, heap bound values \mdSpan[class={math-inline}]{$\hp\varphi{v}$} or heap
bound exceptions \mdSpan[class={math-inline}]{$\hp\varphi{\throw\,c}$} (as defined in Figure{\mdNbsp}\mdA[class={localref},target-element={figure}]{fig-syntax}{}{\mdSpan[class={figure-label}]{4}}).%
\end{mdP}%
\begin{mdP}[class={indent}]%
Our modeling of the heap is sligthly unconventional. Usually, a heap is
defined either as a parameter to the semantics, or always scoped on the
outside. This has the advantage that one doesn{'}t need operations like
\mdSpan[class={rulename},font-variant={small-caps},font-size={small}]{(merge)} and \mdSpan[class={rulename},font-variant={small-caps},font-size={small}]{(lift)}. However, for us it is important that
the \mdSpan[class={code,math-inline}]{$\mathkw{run}$} operation can completely discard the heap which is hard to do
in the conventional approach. In particular, in Section{\mdNbsp}\mdA[class={localref},target-element={h1}]{sec-semeffect}{}{\mdSpan[class={heading-label}]{6}} we want to
state the \mdEm{purity} Theorem{\mdNbsp}\mdA[class={localref},target-element={theorem}]{lem-st}{}{\mdSpan[class={theorem-label}]{3}} that says that  if
\mdSpan[class={math-inline}]{$\infer{\Gamma}{\exp}{\tau}\e$} where \mdSpan[class={math-inline}]{$\sth \not\in \e$} then we never  have
\mdSpan[class={code,math-inline}]{$\exp {\dlongmapsto}\mathspace{1}\hp\varphi{\mathid{v}}$}. Such theorem is difficult to state if we modeled the
heap more conventionally.%
\end{mdP}%
\mdHxx[id=sec-semsound,label={[5]\{.heading-label\}},toc={},caption={Semantic soundness}]{\mdSpan[class={heading-before}]{\mdSpan[class={heading-label}]{5}.{\enspace}}Semantic soundness}\begin{mdP}%
We now show that well-typed programs cannot go `wrong{'}. Our proof closely
follows the subject reduction proofs of Wright and Felleisen
\mdSpan[class={citations},target-element={bibitem}]{[\mdA[class={bibref,localref},target-element={bibitem}]{wrightfelleisen}{}{\mdSpan[class={bibitem-label}]{44}}]}. Our main theorem is:%
\end{mdP}%
\begin{mdDiv}[class={theorem,block},id=th-semsound,label={[1]\{.theorem-label\}},elem={theorem}]%
\begin{mdP}%
\mdSpan[class={theorem-before}]{\mdStrong{Theorem{\mdNbsp}\mdSpan[class={theorem-label}]{1}.} }(\mdEm{Semantic soundness})\mdBr
If \mdSpan[class={math-inline}]{$\infer{\cdot}{e}{\t}{\e}$} then either \mdSpan[class={math-inline}]{$e \Uparrow$} or \mdSpan[class={code,math-inline}]{$\mathid{e}\mathspace{1}{\dlongmapsto}\mathspace{1}\mathid{a}$}
where \mdSpan[class={math-inline}]{$\infer{\cdot}{a}{\t}{\e}$}.%
\end{mdP}
\end{mdDiv}%
\begin{mdP}[class={para-continue}]%
where we use the notation \mdSpan[class={math-inline}]{$e \Uparrow$} for a never ending
reduction.  The proof of this theorem consists of showing two main lemmas:%
\end{mdP}%
\begin{mdUl}[class={compact}]%
\begin{mdLi}%
Show that reduction in the operational semantics preserves well-typing 
(called subject reduction).%
\end{mdLi}%
\begin{mdLi}%
Show that \mdEm{faulty} expressions are not typeable.%
\end{mdLi}
\end{mdUl}%
\begin{mdP}%
If programs are closed and well-typed, we know from subject reduction that we
can only reduce to well-typed terms, which can be either faulty, an answer, or
an expression containing a further redex. Since faulty expressions are not
typeable it must be that evaluation either produces a well-typed answer or
diverges. Often, proofs of soundness are carried out using \mdEm{progress} instead
of \mdEm{faulty} expressions but it turns out that for proving the soundness
of state isolation, our current approach works better.%
\end{mdP}%
\mdHxxx[id=sec-subject,label={[5.1]\{.heading-label\}},toc={},caption={Subject reduction}]{\mdSpan[class={heading-before}]{\mdSpan[class={heading-label}]{5.1}.{\enspace}}Subject reduction}\begin{mdP}%
The subject reduction theorem states that a well-typed term remains well-typed
under reduction:%
\end{mdP}%
\begin{mdDiv}[class={lemma,block},id=th-subject,label={[1]\{.lemma-label\}},elem={lemma}]%
\begin{mdP}%
\mdSpan[class={lemma-before}]{\mdStrong{Lemma{\mdNbsp}\mdSpan[class={lemma-label}]{1}.} }(\mdEm{Subject reduction})\mdBr
If \mdSpan[class={math-inline}]{$\infer{\Gamma}{e_1}\t\e$} and \mdSpan[class={code,math-inline}]{$\mathid{e}_1\mathspace{1}{\longrightarrow}\mathspace{1}\mathid{e}_2$} then \mdSpan[class={math-inline}]{$\infer\Gamma{e_2}\t\e$}.%
\end{mdP}
\end{mdDiv}%
\begin{mdP}%
To show that subject reduction holds, we need to establish various lemmas.  Two
particularly important lemmas are the substitution and extension lemmas:%
\end{mdP}%
\begin{mdDiv}[class={lemma,block},id=th-subst,label={[2]\{.lemma-label\}},elem={lemma}]%
\begin{mdP}%
\mdSpan[class={lemma-before}]{\mdStrong{Lemma{\mdNbsp}\mdSpan[class={lemma-label}]{2}.} }(\mdEm{Substitution})\mdBr
If \mdSpan[class={math-inline}]{$\infer{\Gamma,x:\forall\overline\alpha.\,\t}{e}{\t'}\e$} where \mdSpan[class={math-inline}]{$x \not\in \dom{\Gamma}$},
\mdSpan[class={code,math-inline}]{$\infer{\Gamma}{\mathid{v}}{\t}{{\langle}{\rangle}}$}, and \mdSpan[class={math-inline}]{$\overline\alpha \disjoint\, \ftv{\Gamma}$}, then 
\mdSpan[class={math-inline}]{$\infer{\Gamma}{[x \mapsto v]e}{\t'}{\e}$}.%
\end{mdP}
\end{mdDiv}%
\begin{mdDiv}[class={lemma,block},id=th-extension,label={[3]\{.lemma-label\}},elem={lemma}]%
\begin{mdP}%
\mdSpan[class={lemma-before}]{\mdStrong{Lemma{\mdNbsp}\mdSpan[class={lemma-label}]{3}.} }(\mdEm{Extension})\mdBr
If \mdSpan[class={math-inline}]{$\infer{\Gamma}{e}{\t}{\e}$} and for all \mdSpan[class={math-inline}]{$x \in \fv{e}$} we have \mdSpan[class={math-inline}]{$\Gamma(x) = \Gamma'(x)$},
then \mdSpan[class={math-inline}]{$\infer{\Gamma'}{e}{\t}{\e}$}.%
\end{mdP}
\end{mdDiv}%
\begin{mdP}%
The proofs of these lemmas from{\mdNbsp}\mdSpan[class={citations},target-element={bibitem}]{[\mdA[class={bibref,localref},target-element={bibitem}]{wrightfelleisen}{}{\mdSpan[class={bibitem-label}]{44}}]} carry over directly to
our system. However, to show subject reduction, we require an extra lemma to
reason about state effects.%
\end{mdP}%
\begin{mdDiv}[class={lemma,block},id=th-heapr,label={[4]\{.lemma-label\}},elem={lemma}]%
\begin{mdP}%
\mdSpan[class={lemma-before}]{\mdStrong{Lemma{\mdNbsp}\mdSpan[class={lemma-label}]{4}.} }(\mdEm{Stateful effects})\mdBr
If \mdSpan[class={math-inline}]{$\infer{\Gamma}{e}{\t}{\ext\sth\e}$} and \mdSpan[class={math-inline}]{$\infer\Gamma{\R{e}}{\t'}{\e'}$} then
\mdSpan[class={math-inline}]{$\sth \in \e'$}.%
\end{mdP}
\end{mdDiv}%
\begin{mdP}%
The above lemma essentially states that a stateful effect cannot be discarded
in an \mdSpan[class={math-inline}]{$R$} context. Later we will generalize this lemma to arbitrary contexts
and effects but for subject reduction this lemma is strong enough.%
\end{mdP}%
\begin{mdDiv}[class={proof,block},elem={proof}]%
\begin{mdP}%
\mdSpan[class={proof-before}]{\mdStrong{Proof}. }
(Lemma{\mdNbsp}\mdA[class={localref},target-element={lemma}]{th-heapr}{}{\mdSpan[class={lemma-label}]{4}})
We proceed by induction on the structure of \mdSpan[class={math-inline}]{$R$}:
\mdBr
\mdStrong{Case} \mdSpan[class={math-inline}]{$R=[]$}: By definition \mdSpan[class={math-inline}]{$\sth \in \ext\sth\e$}.
\mdBr
\mdStrong{Case} \mdSpan[class={math-inline}]{$R=R'\,e_2$}: We have \mdSpan[class={math-inline}]{$\infer{\Gamma}{(\RX{e})\,e_2}{\t'}{\e'}$} and by \mdSpan[class={rulename},font-variant={small-caps},font-size={small}]{(app)}
we have \mdSpan[class={code,math-inline}]{$\infer{\Gamma}{\RX{\mathid{e}}}{\t_2\mathspace{1}{\rightarrow}\mathspace{1}\e'\,\t'}{\e'}$}. By induction, \mdSpan[class={math-inline}]{$\sth \in \e'$}.
\mdBr
\mdStrong{Case} \mdSpan[class={math-inline}]{$R=v\,R'$}: Similar to previous case.
\mdBr
\mdStrong{Case} \mdSpan[class={code,math-inline}]{$\mathid{R}=\mathkw{let}\mathspace{1}\mathid{x}\mathspace{1}=\mathspace{1}\mathid{R}'\mathspace{1}\mathkw{in}\mathspace{1}\mathid{e}_2$}: By \mdSpan[class={rulename},font-variant={small-caps},font-size={small}]{(let)} we have 
\mdSpan[class={code,math-inline}]{$\infer{\Gamma}{\RX{\mathid{e}}}{\t_1}{{\langle}{\rangle}}$} but that contradicts our assumption.
\mdBr
\mdStrong{Case} \mdSpan[class={math-inline}]{$R=\catch\,R'\,e_2$}: By \mdSpan[class={rulename},font-variant={small-caps},font-size={small}]{(catch)} we have 
\mdSpan[class={math-inline}]{$\infer{\Gamma}{\catch\,\RX{e}\,e_2}{\t'}{\e'}$} where \mdSpan[class={math-inline}]{$\infer{\Gamma}{\RX{e}}{\t'}{\ext{\ec{exn}}{\e'}}$}. By induction \mdSpan[class={math-inline}]{$\sth \in \ext{\ec{exn}}{\e'}$} which implies that 
\mdSpan[class={math-inline}]{$\sth \in \e'$}.%
\end{mdP}
\end{mdDiv}%
\begin{mdP}%
Before proving subject reduction we need one more type rule on our 
internal language. The \mdSpan[class={rulename},font-variant={small-caps},font-size={small}]{(extend)} rule states that we can always
assume a stateful effect for an expression:%
\end{mdP}%
\begin{mdDiv}[class={center,block,align-center},elem={center}]%
\begin{mdP}%
\mdSpan[class={rulename},font-variant={small-caps},font-size={small}]{(st-extend)} \mdSpan[class={code,math-inline}]{$\inference{\infer\Gamma{\mathid{e}}{\t}{\e}}{\infer\Gamma{\mathid{e}}{\t}{\ext{\sth}{\e}}}$}%
\end{mdP}
\end{mdDiv}%
\begin{mdP}%
Now we are ready to prove the subject reduction theorem:%
\end{mdP}%
\begin{mdDiv}[class={proof,block},elem={proof}]%
\begin{mdP}[class={para-continue}]%
\mdSpan[class={proof-before}]{\mdStrong{Proof}. }
(Lemma{\mdNbsp}\mdA[class={localref},target-element={lemma}]{th-subject}{}{\mdSpan[class={lemma-label}]{1}})
We prove this by induction on the reduction rules of \mdSpan[class={code,math-inline}]{${\longrightarrow}$}. We will not
repeat all cases here and refer to{\mdNbsp}\mdSpan[class={citations},target-element={bibitem}]{[\mdA[class={bibref,localref},target-element={bibitem}]{wrightfelleisen}{}{\mdSpan[class={bibitem-label}]{44}}]}, but instead concentrate
on the interesting cases, especially with regard to state isolation.
\mdBr
\mdStrong{Case} \mdSpan[class={code,math-inline}]{$\letb{\mathid{x}\mathspace{1}=\mathspace{1}\mathid{v}}{\mathid{e}}\mathspace{1}{\longrightarrow}\mathspace{1}[\mathid{x}\mathspace{1}\mapsto \mathid{v}]\mathid{e}$}: From \mdSpan[class={rulename},font-variant={small-caps},font-size={small}]{(let)} we have
\mdSpan[class={code,math-inline}]{$\infer\Gamma{\mathid{v}}{\t'}{{\langle}{\rangle}}$} and \mdSpan[class={math-inline}]{$\infer{\Gamma,x : \gen{\Gamma}{\t'}}{e}{\t}{\e}$}.
By definition, \mdSpan[class={math-inline}]{$\gen{\Gamma}{\t'} = \forall\overline\alpha.\,\t'$} where 
\mdSpan[class={math-inline}]{$\overline\alpha \not\in \ftv{\Gamma}$} and by Lemma{\mdNbsp}\mdA[class={localref},target-element={lemma}]{th-subst}{}{\mdSpan[class={lemma-label}]{2}} we have
\mdSpan[class={math-inline}]{$\infer{\Gamma}{[x \mapsto v]e}\t\e$}.
\mdBr
\mdStrong{Case} \mdSpan[class={code,math-inline}]{$\R{\hp\varphi{\mathid{e}}}\mathspace{1}{\longrightarrow}\mathspace{1}\hp\varphi\R{\mathid{e}}$}: This is case is proven
by induction over the structure of \mdSpan[class={math-inline}]{$R$}:%
\end{mdP}%
\begin{mdDiv}[class={subcase,para-block},elem={subcase},margin-left={1ex},margin-top={0ex},margin-bottom={0ex}]%
\mdStrong{case}{\mdNbsp}\mdSpan[class={code,math-inline}]{$\mathid{R}=[]$}: Does not apply due to the side condition on \mdSpan[class={code,math-inline}]{${\longrightarrow}$}.%
\end{mdDiv}%
\begin{mdDiv}[class={subcase,para-block},elem={subcase},margin-left={1ex},margin-top={0ex},margin-bottom={0ex}]%
\mdStrong{case}{\mdNbsp}\mdSpan[class={code,math-inline}]{$\mathid{R}=\mathid{R}'\,\mathid{e}'$}: Then \mdSpan[class={code,math-inline}]{$\infer{\Gamma}{\context{\mathid{R}'}{\hp\varphi{\mathid{e}}}\,\mathid{e}'}{\t}{\e}$}
and by \mdSpan[class={rulename},font-variant={small-caps},font-size={small}]{(app)} we have \mdSpan[class={code,math-inline}]{$\infer{\Gamma}{\context{\mathid{R}'}{\hp\varphi{\mathid{e}}}}{\t_2\mathspace{1}{\rightarrow}\mathspace{1}\e\,\t}{\e}$} 
\mdStrong{(1)} and \mdSpan[class={math-inline}]{$\infer\Gamma{e'}{\t_2}\e$} \mdStrong{(2)}. By the induction hypothesis and (1), we have 
\mdSpan[class={code,math-inline}]{$\infer\Gamma{\hp\varphi{\context{\mathid{R}'}{\mathid{e}}}}{\t_2\mathspace{1}{\rightarrow}\mathspace{1}\e\,\t}{\e}$}.
Then by \mdSpan[class={rulename},font-variant={small-caps},font-size={small}]{(heap)} we know \mdSpan[class={code,math-inline}]{$\infer{\Gamma,\overline\varphi_\mathid{h}}{\mathid{v}_\mathid{j}}{\t_\mathid{j}}{{\langle}{\rangle}}$} \mdStrong{(3)} and
\mdSpan[class={code,math-inline}]{$\infer{\Gamma,\overline\varphi_\mathid{h}}{\context{\mathid{R}'}{\mathid{e}}}{\t_2\mathspace{1}{\rightarrow}\mathspace{1}\e\,\t}{\e}$} \mdStrong{(4)} where 
\mdSpan[class={code,math-inline}]{$\varphi =\mathspace{1}{\langle}\mathid{r}_1\mathspace{1}\mapsto \mathid{v}_1,...,\mathid{r}_\mathid{n}\mathspace{1}\mapsto \mathid{v}_\mathid{n}{\rangle}$}. 
Since \mdSpan[class={math-inline}]{$r_1,...,r_n \not\in \fv{e'}$} we can use (2) and{\mdNbsp}\mdA[class={localref},target-element={lemma}]{th-extension}{}{\mdSpan[class={lemma-label}]{3}} to conclude
\mdSpan[class={math-inline}]{$\infer{\Gamma,\overline\varphi}{e'}{\t_2}{\e}$} \mdStrong{(5)}. Using \mdSpan[class={rulename},font-variant={small-caps},font-size={small}]{(app)} with (4)
and (5),
we have \mdSpan[class={math-inline}]{$\infer{\Gamma,\overline\varphi}{\context{R'}{e}\,e'}{\t}\e$} where we can use
\mdSpan[class={rulename},font-variant={small-caps},font-size={small}]{(heap)} with (3) to conclude \mdSpan[class={math-inline}]{$\infer{\Gamma}{\hp\varphi{\context{R'}{e}\,e'}}{\t}{\e}$}.%
\end{mdDiv}%
\begin{mdDiv}[class={subcase,para-block},elem={subcase},margin-left={1ex},margin-top={0ex},margin-bottom={0ex}]%
\mdStrong{case}{\mdNbsp}\mdSpan[class={code,math-inline}]{$\mathid{R}=\mathid{v}\,\mathid{R}'$}: Similar to the previous case.%
\end{mdDiv}%
\begin{mdDiv}[class={subcase,para-block},elem={subcase},margin-left={1ex},margin-top={0ex},margin-bottom={0ex}]%
\mdStrong{case}{\mdNbsp}\mdSpan[class={code,math-inline}]{$\mathid{R}=\letb{\mathid{x}\mathspace{1}=\mathspace{1}\mathid{R}'}{\mathid{e}'}$}: If this is well-typed, then 
by rule \mdSpan[class={rulename},font-variant={small-caps},font-size={small}]{(let)} we must have
\mdSpan[class={code,math-inline}]{$\infer{\Gamma}{\context{\mathid{R}'}{\hp\varphi{\mathid{e}}}}{\t'}{{\langle}{\rangle}}$}. However, due to Lemma{\mdNbsp}\mdA[class={localref},target-element={lemma}]{th-heapr}{}{\mdSpan[class={lemma-label}]{4}} 
and \mdSpan[class={rulename},font-variant={small-caps},font-size={small}]{(heap)}, we have  \mdSpan[class={code,math-inline}]{$\sth \in {\langle}{\rangle}$} which is a contradiction.
Note that this case is essential, as it prevents generalization of stateful references.
For ML, this is also the tricky proof case that only works if one defines special
{\textquoteleft}imperative type variables{\textquoteright}{\mdNbsp}\mdSpan[class={citations},target-element={bibitem}]{[\mdA[class={bibref,localref},target-element={bibitem}]{wrightfelleisen}{}{\mdSpan[class={bibitem-label}]{44}}]} or the value restriction. In our case
the effect system ensures safety.%
\end{mdDiv}%
\begin{mdP}[class={para-continue}]%
\mdStrong{Case} \mdSpan[class={code,math-inline}]{$\run\,([\hp\varphi]\,\lambda \mathid{x}.\,\mathid{e})\mathspace{1}{\longrightarrow}\mathspace{1}\lambda \mathid{x}.\,\run\,([\hp\varphi]\,\mathid{e})$}: 
By rule \mdSpan[class={rulename},font-variant={small-caps},font-size={small}]{(run)} and \mdSpan[class={rulename},font-variant={small-caps},font-size={small}]{(heap)} we have that 
\mdSpan[class={code,math-inline}]{$\infer{\Gamma}{\lambda \mathid{x}.\,\mathid{e}}{\t}{\ext\sth\e}$} where \mdSpan[class={code,math-inline}]{$\mathid{h}\mathspace{1}\not\in \ftv{\Gamma,\t,\e}$} \mdStrong{(1)}. 
Applying \mdSpan[class={rulename},font-variant={small-caps},font-size={small}]{(lam)} gives \mdSpan[class={code,math-inline}]{$\infer{\Gamma,\mathid{x}:\t_1}{\mathid{e}}{\t_2}{\e_2}$} with \mdSpan[class={code,math-inline}]{$\t =\mathspace{1}\t_1\mathspace{1}{\rightarrow}\mathspace{1}\e_2\,\t_2$}. 
Using \mdSpan[class={rulename},font-variant={small-caps},font-size={small}]{(st-extend)} we can also derive \mdSpan[class={math-inline}]{$\infer{\Gamma,x:\t_1}{e}{\t_2}{\ext\sth{\e_2}}$}.
Due to (1) and \mdSpan[class={math-inline}]{$h \not\in \t_1$}, we can apply \mdSpan[class={rulename},font-variant={small-caps},font-size={small}]{(run)} and \mdSpan[class={rulename},font-variant={small-caps},font-size={small}]{(heap)} again to 
infer \mdSpan[class={math-inline}]{$\infer{\Gamma,x:\t_1}{\run\,([\hp\varphi]\,e)}{\t_2}{\e_2}$} and finally \mdSpan[class={rulename},font-variant={small-caps},font-size={small}]{(lam)} again to 
conclude \mdSpan[class={math-inline}]{$\infer{\Gamma}{\lambda x.\,(\run\,([\hp\varphi]\,e))}{\t}{\e}$}.
\mdBr
\mdStrong{Case} \mdSpan[class={code,math-inline}]{$\run\,([\hp\varphi]\,\catch\,\mathid{e})\mathspace{1}{\longrightarrow}\mathspace{1}\catch\,(\run\,([\hp\varphi]\,\mathid{e}))$}: 
Similar to the previous case.
\mdBr
\mdStrong{Case} \mdSpan[class={code,math-inline}]{$\run\,([\hp\varphi]\,\mathid{w})\mathspace{1}{\longrightarrow}\mathspace{1}\mathid{w}$} with \mdSpan[class={code,math-inline}]{$\frv{\mathid{w}}\disjoint \dom\varphi$} \mdStrong{(1)}: 
By rule \mdSpan[class={rulename},font-variant={small-caps},font-size={small}]{(run)} and \mdSpan[class={rulename},font-variant={small-caps},font-size={small}]{(heap)} we have that 
\mdSpan[class={math-inline}]{$\infer{\Gamma,\overline\varphi_h}{w}{\t}{\ext\sth\e}$} where \mdSpan[class={math-inline}]{$h \not\in \ftv{\Gamma,\t,\e}$} \mdStrong{(2)}. 
By (1) it must also be that \mdSpan[class={math-inline}]{$\infer{\Gamma}{w}{\t}{\ext\sth\e}$} \mdStrong{(3)} (this follows directly
if there was no heap binding \mdSpan[class={math-inline}]{$\hp\varphi{}$}).
We proceed over the structure of \mdSpan[class={math-inline}]{$w$}:%
\end{mdP}%
\begin{mdDiv}[class={subcase,para-block},elem={subcase},margin-left={1ex},margin-top={0ex},margin-bottom={0ex}]%
\mdStrong{case}{\mdNbsp}\mdSpan[class={code,math-inline}]{$\mathid{w}\mathspace{1}=\mathspace{1}\throw\,\mathid{c}$}: Then by (3) we have \mdSpan[class={math-inline}]{$\infer{\Gamma}{\throw\,c}{\t}{\ext\sth\e}$},
but also \mdSpan[class={math-inline}]{$\infer{\Gamma}{\throw\,c}{\t}{\e}$} since we can choose the result effect freely
in \mdSpan[class={rulename},font-variant={small-caps},font-size={small}]{(throw)}.%
\end{mdDiv}%
\begin{mdDiv}[class={subcase,para-block},elem={subcase},margin-left={1ex},margin-top={0ex},margin-bottom={0ex}]%
\mdStrong{case}{\mdNbsp}\mdSpan[class={code,math-inline}]{$\mathid{w}=\mathid{r}$}: By \mdSpan[class={rulename},font-variant={small-caps},font-size={small}]{(var)} and (3), we have \mdSpan[class={math-inline}]{$\infer{\Gamma}{r}{\tref{h'}{\t'}}{\ext\sth\e}$}.
where \mdSpan[class={math-inline}]{$h \neq\ h'$} satisfying (2). Since the result effect is free in \mdSpan[class={rulename},font-variant={small-caps},font-size={small}]{(var)},
we can also derive \mdSpan[class={math-inline}]{$\infer{\Gamma}{r}{\tref{h'}{\t'}}{\e}$}%
\end{mdDiv}%
\begin{mdDiv}[class={subcase,para-block},elem={subcase},margin-left={1ex},margin-top={0ex},margin-bottom={0ex}]%
\mdStrong{case}{\mdNbsp}\mdSpan[class={code,math-inline}]{$\mathid{w}=(\mathid{r}:=)$}: As the previous case.%
\end{mdDiv}%
\begin{mdDiv}[class={subcase,para-block},elem={subcase},margin-left={1ex},margin-top={0ex},margin-bottom={0ex}]%
\mdStrong{case}{\mdNbsp}\mdSpan[class={code,math-inline}]{$\mathid{w}=\mathid{x}$}: By \mdSpan[class={rulename},font-variant={small-caps},font-size={small}]{(var)} and (3), we have \mdSpan[class={math-inline}]{$\infer{\Gamma}{x}{\t}{\ext\sth\e}$} but in
\mdSpan[class={rulename},font-variant={small-caps},font-size={small}]{(var)} the result effect is free, so we can also derive \mdSpan[class={math-inline}]{$\infer{\Gamma}{x}{\t}{\e}$}.%
\end{mdDiv}%
\begin{mdDiv}[class={subcase,para-block},elem={subcase},margin-left={1ex},margin-top={0ex},margin-bottom={0ex}]%
\mdStrong{case}{\mdNbsp}other: Similarly.%
\end{mdDiv}
\end{mdDiv}%
\mdHxxx[id=sec-faulty,label={[5.2]\{.heading-label\}},toc={},caption={Faulty expressions}]{\mdSpan[class={heading-before}]{\mdSpan[class={heading-label}]{5.2}.{\enspace}}Faulty expressions}\begin{mdP}[class={para-continue}]%
The main purpose of type checking is of course to guarantee
that certain bad expressions cannot occur. Apart from the usual
errors, like adding a number to a string, we particularly would
like to avoid state errors. There are two aspects to this.
One of them is notorious where polymorphic types in combination
with state can be unsound (which is not the case in our system
because of Lemma{\mdNbsp}\mdA[class={localref},target-element={lemma}]{th-subject}{}{\mdSpan[class={lemma-label}]{1}}). But in addition, we would like
to show that in our system it is not possible to read or write
to locations outside the local heap (encapsulated by \mdSpan[class={math-inline}]{$\run$}), nor
is it possible to let local references escape. 
To make this precise, the \mdEm{faulty} expressions are defined as:%
\end{mdP}%
\begin{mdUl}[class={compact}]%
\begin{mdLi}%
Undefined: \mdSpan[class={math-inline}]{$c\,v$} where \mdSpan[class={math-inline}]{$\delta(c,v)$} is not defined.%
\end{mdLi}%
\begin{mdLi}%
Escaping read: \mdSpan[class={math-inline}]{$\run\,(\hp\varphi{\R{!r}})$} where \mdSpan[class={math-inline}]{$r \not\in \dom\varphi$}.%
\end{mdLi}%
\begin{mdLi}%
Escaping write: \mdSpan[class={math-inline}]{$\run\,(\hp\varphi{\R{r := v}})$} where \mdSpan[class={math-inline}]{$r \not\in \dom\varphi$}.%
\end{mdLi}%
\begin{mdLi}%
Escaping reference: \mdSpan[class={math-inline}]{$\run\,(\hp\varphi{w})$} where \mdSpan[class={math-inline}]{$\frv{w} \cap \dom\varphi \neq \varnothing$}.%
\end{mdLi}%
\begin{mdLi}%
Not a function: \mdSpan[class={math-inline}]{$v\,e$} where \mdSpan[class={math-inline}]{$v$} is not a constant or lambda expression.%
\end{mdLi}%
\begin{mdLi}%
Not a reference: \mdSpan[class={math-inline}]{$!v$} or \mdSpan[class={math-inline}]{$v := e$} where \mdSpan[class={math-inline}]{$v$} is not a reference.%
\end{mdLi}%
\begin{mdLi}%
Not an exception: \mdSpan[class={math-inline}]{$\throw\,c$} where \mdSpan[class={math-inline}]{$c$} is not the unit value.%
\end{mdLi}
\end{mdUl}%
\begin{mdDiv}[class={lemma,block},id=th-faulty,label={[5]\{.lemma-label\}},elem={lemma}]%
\begin{mdP}%
\mdSpan[class={lemma-before}]{\mdStrong{Lemma{\mdNbsp}\mdSpan[class={lemma-label}]{5}.} }(\mdEm{Faulty expressions are untypeable})\mdBr
If an expression \mdSpan[class={math-inline}]{$e$} is faulty, it cannot be typed, i.e. there exists no
\mdSpan[class={math-inline}]{$\Gamma$}, \mdSpan[class={math-inline}]{$\t$}, and \mdSpan[class={math-inline}]{$\e$} such that \mdSpan[class={math-inline}]{$\infer\Gamma{e}\t\e$}.%
\end{mdP}
\end{mdDiv}%
\begin{mdP}%
In the following proof, especially the case for escaping state is interesting.
To our knowledge, this is the first proof of the safety of \mdSpan[class={code,math-inline}]{$\run$} in a strict
setting (and in combination with other effects like exceptions).%
\end{mdP}%
\begin{mdDiv}[class={proof,block},elem={proof}]%
\begin{mdP}%
\mdSpan[class={proof-before}]{\mdStrong{Proof}. }
(Lemma{\mdNbsp}\mdA[class={localref},target-element={lemma}]{th-faulty}{}{\mdSpan[class={lemma-label}]{5}})
Each faulty expression is handled separately. We show here the interesting
cases for escaping reads, writes, and references:
\mdBr
\mdStrong{Case} \mdSpan[class={code,math-inline}]{$\run\,(\hp\varphi\R{!\mathid{r}})$} with \mdSpan[class={code,math-inline}]{$\mathid{r}\mathspace{1}\not\in \dom\varphi$} \mdStrong{(1)}: To be typed in a context
\mdSpan[class={math-inline}]{$\Gamma$} we apply \mdSpan[class={rulename},font-variant={small-caps},font-size={small}]{(run)} and \mdSpan[class={rulename},font-variant={small-caps},font-size={small}]{(heap)}
and need to show \mdSpan[class={math-inline}]{$\infer{\Gamma,\overline\varphi_h}{\R{!r}}{\t}{\ext\sth\e}$} \mdStrong{(2)},
where \mdSpan[class={math-inline}]{$h \not\in \ftv{\Gamma,\t,\e}$} \mdStrong{(3)}. For \mdSpan[class={math-inline}]{$\R{!r}$} to be well-typed, we 
also need \mdSpan[class={math-inline}]{$\infer{\Gamma,\overline\varphi_h}{!r}{\t'}{\ext{\tst{h'}}{\e'}}$} \mdStrong{(4)} where
\mdSpan[class={math-inline}]{$\infer{\Gamma,\overline\varphi_h}{r}{\tref{h'}{\t'}}{\ext{\tst{h'}}{\e'}}$} \mdStrong{(5)}.
From Lemma{\mdNbsp}\mdA[class={localref},target-element={lemma}]{th-heapr}{}{\mdSpan[class={lemma-label}]{4}}, (4), and (2), it must be that \mdSpan[class={math-inline}]{$h' = h$} \mdStrong{(6)}. 
But since \mdSpan[class={math-inline}]{$r \not\in \dom\varphi$}
(1), it follows by (5) and (6), that \mdSpan[class={math-inline}]{$\infer{\Gamma}{r}{\tref{h}{\t'}}{\ext{\tst{h}}{\e'}}$}.
But that means \mdSpan[class={math-inline}]{$h \in \ftv{\Gamma}$} contradicting (3).
\mdBr
\mdStrong{Case} \mdSpan[class={code,math-inline}]{$\run\,(\hp\varphi\R{(\mathid{r}:=)})$} with \mdSpan[class={code,math-inline}]{$\mathid{r}\mathspace{1}\not\in \dom\varphi$}:
Similar to the previous case.
\mdBr
\mdStrong{Case} \mdSpan[class={code,math-inline}]{$\run\,(\hp\varphi{\mathid{w}})$} where \mdSpan[class={code,math-inline}]{$\frv{\mathid{w}}\mathspace{1}\cap \dom\varphi \neq \varnothing$} \mdStrong{(1)} 
To be typed in a context
\mdSpan[class={math-inline}]{$\Gamma$} we need to show by \mdSpan[class={rulename},font-variant={small-caps},font-size={small}]{(heap)} and \mdSpan[class={rulename},font-variant={small-caps},font-size={small}]{(run)} that
\mdSpan[class={math-inline}]{$\infer{\Gamma,\overline\varphi_h}{w}{\t}{\ext\sth\e}$} where \mdSpan[class={math-inline}]{$h \not\in \ftv{\Gamma,\t,\e}$}{\mdNbsp}\mdStrong{(2)}. 
If \mdSpan[class={math-inline}]{$w = \throw\,c$} then by \mdSpan[class={rulename},font-variant={small-caps},font-size={small}]{(throw)} the type of \mdSpan[class={math-inline}]{$c$} is \mdSpan[class={math-inline}]{$()$} and thus \mdSpan[class={math-inline}]{$c$} is the unit constant.
But \mdSpan[class={math-inline}]{$\frv{()} = \varnothing$} contradicting our assumption.
Otherwise, \mdSpan[class={math-inline}]{$w = b$} and cannot contain an arbitrary \mdSpan[class={math-inline}]{$e$}. Since \mdSpan[class={math-inline}]{$\frv{w} \neq \varnothing$} (1), 
it must be that \mdSpan[class={math-inline}]{$w$} is either one of \mdSpan[class={math-inline}]{$r$} or \mdSpan[class={math-inline}]{$(r:=)$} with 
\mdSpan[class={math-inline}]{$r \in \dom\varphi$}.
To be well-typed, \mdSpan[class={math-inline}]{$\infer{\Gamma,\overline\varphi_h}{r}{\tref{h}{\t'}}{\e'}$} must 
hold. However, the possible types for
\mdSpan[class={math-inline}]{$r$} and \mdSpan[class={math-inline}]{$(r :=)$} are \mdSpan[class={math-inline}]{$\tref{h}{\t'}$} and \mdSpan[class={code,math-inline}]{$\t'\mathspace{1}{\rightarrow}\mathspace{1}\sth\,()$} and in both cases \mdSpan[class={math-inline}]{$h \in \ftv{\t}$}
which contradicts{\mdNbsp}(2).%
\end{mdP}
\end{mdDiv}%
\mdHxx[id=sec-semeffect,label={[6]\{.heading-label\}},toc={},caption={Effectful semantics}]{\mdSpan[class={heading-before}]{\mdSpan[class={heading-label}]{6}.{\enspace}}Effectful semantics}\begin{mdP}%
Up till now, we have used the effect types to good effect and showed that our
system is semantically sound, even though state and polymorphic types are notoriously 
tricky to combine. Moreover, we showed that local state isolation through \mdSpan[class={math-inline}]{$\run$} is sound
and statically prevents references from escaping.%
\end{mdP}%
\begin{mdP}[class={indent}]%
But the true power of the effect system is really to enable more reasoning about the
behavior of a program at a higher level. In particular, the absence of certain effects
determines the absence of certain answers. For example, if the exception effect
is not inferred, then evaluating the program will never produce an answer of the form
\mdSpan[class={math-inline}]{$\throw\,c$} or \mdSpan[class={math-inline}]{$\hp\varphi{\throw\,c}$}! It would not be entirely correct to say that
such program never throws an exception: indeed, a local catch block can handle such
exceptions. We can state the exception property formally as:%
\end{mdP}%
\begin{mdDiv}[class={theorem,block},id=lem-exn,label={[2]\{.theorem-label\}},elem={theorem}]%
\begin{mdP}%
\mdSpan[class={theorem-before}]{\mdStrong{Theorem{\mdNbsp}\mdSpan[class={theorem-label}]{2}.} }(\mdEm{Exceptions})\mdBr
If \mdSpan[class={math-inline}]{$\infer{\Gamma}{\exp}{\tau}{\e}$} where \mdSpan[class={math-inline}]{$\ec{exn} \not\in \e$} then either \mdSpan[class={math-inline}]{$\exp\Uparrow$},
 \mdSpan[class={code,math-inline}]{$\exp {\dlongmapsto}\mathspace{1}\mathid{v}$} or \mdSpan[class={code,math-inline}]{$\exp {\dlongmapsto}\mathspace{1}\hp\varphi \mathid{v}$}.%
\end{mdP}
\end{mdDiv}%
\begin{mdDiv}[class={proof,block},elem={proof}]%
\begin{mdP}%
\mdSpan[class={proof-before}]{\mdStrong{Proof}. }
(Theorem{\mdNbsp}\mdA[class={localref},target-element={theorem}]{lem-exn}{}{\mdSpan[class={theorem-label}]{2}}) By contradiction over the result terms:
\mdBr
\mdStrong{Case} \mdSpan[class={code,math-inline}]{$\mathid{e}\mathspace{1}{\dlongmapsto}\mathspace{1}\throw\,\mathid{c}$}: By subject reduction (Lemma{\mdNbsp}\mdA[class={localref},target-element={lemma}]{th-subject}{}{\mdSpan[class={lemma-label}]{1}}), 
it must be \mdSpan[class={math-inline}]{$\infer\Gamma{\throw\,c}{\tau}{\e}$}.
Using the type rule for \mdSpan[class={math-inline}]{$\throw$} with \mdSpan[class={rulename},font-variant={small-caps},font-size={small}]{(app)}, it must be the case that 
\mdSpan[class={math-inline}]{$\e \equiv \ext{\ec{exn}}{\e'}$} contradicting our assumption.
\mdBr
\mdStrong{Case} \mdSpan[class={code,math-inline}]{$\mathid{e}\mathspace{1}{\dlongmapsto}\mathspace{1}\hp\varphi{\throw\,\mathid{c}}$}: Similar to the previous case.%
\end{mdP}
\end{mdDiv}%
\begin{mdP}%
Similarly to the exception case, we can state such theorem over heap effects too.
In particular, if the \mdSpan[class={math-inline}]{$\sth$} effect is absent, then evaluation will not produce an 
answer that contains a heap, i.e. \mdSpan[class={math-inline}]{$\hp\varphi{w}$}. Again, it would not be right
to say that the program itself never performs any stateful operations
the state can be encapsulated inside a \mdSpan[class={math-inline}]{$\run$} construct and its stateful behavior is
not observable from outside. Formally, we can state this as:%
\end{mdP}%
\begin{mdDiv}[class={theorem,block},id=lem-st,label={[3]\{.theorem-label\}},elem={theorem}]%
\begin{mdP}%
\mdSpan[class={theorem-before}]{\mdStrong{Theorem{\mdNbsp}\mdSpan[class={theorem-label}]{3}.} }(\mdEm{State})\mdBr
Iff \mdSpan[class={math-inline}]{$\infer{\Gamma}{\exp}{\tau}\e$} where \mdSpan[class={math-inline}]{$\sth \not\in \e$} then either \mdSpan[class={math-inline}]{$\exp\Uparrow$},
\mdSpan[class={code,math-inline}]{$\exp {\dlongmapsto}\mathspace{1}\mathid{v}$} or \mdSpan[class={code,math-inline}]{$\exp {\dlongmapsto}\mathspace{1}\throw\,\mathid{c}$}.%
\end{mdP}
\end{mdDiv}%
\begin{mdDiv}[class={proof,block},elem={proof}]%
\begin{mdP}%
\mdSpan[class={proof-before}]{\mdStrong{Proof}. }
(Theorem{\mdNbsp}\mdA[class={localref},target-element={theorem}]{lem-st}{}{\mdSpan[class={theorem-label}]{3}}) We prove by contradiction over the result terms:
\mdBr
\mdStrong{Case} \mdSpan[class={code,math-inline}]{$\mathid{e}\mathspace{1}{\dlongmapsto}\mathspace{1}\hp\varphi{\mathid{v}}$}: By subject reduction (Lemma{\mdNbsp}\mdA[class={localref},target-element={lemma}]{th-subject}{}{\mdSpan[class={lemma-label}]{1}}), it must be 
\mdSpan[class={math-inline}]{$\infer\Gamma{\hp\varphi{v}}{\tau}{\e}$}.
Using \mdSpan[class={rulename},font-variant={small-caps},font-size={small}]{(heap)}, it must be the case that 
\mdSpan[class={math-inline}]{$\e \equiv \ext{\sth}{\e'}$} contradicting our assumption.
\mdBr
\mdStrong{Case} \mdSpan[class={code,math-inline}]{$\mathid{e}\mathspace{1}{\dlongmapsto}\mathspace{1}\hp\varphi{\throw\,\mathid{c}}$}: Similar to the previous case.%
\end{mdP}
\end{mdDiv}%
\begin{mdP}%
Finally, our most powerful theorem is about the divergence effect; in particular, if the
divergent effect is absent, then evaluation is guaranteed to terminate!%
\end{mdP}%
\begin{mdDiv}[class={theorem,block},id=lem-div,label={[4]\{.theorem-label\}},elem={theorem}]%
\begin{mdP}%
\mdSpan[class={theorem-before}]{\mdStrong{Theorem{\mdNbsp}\mdSpan[class={theorem-label}]{4}.} }(\mdEm{Divergence})\mdBr
If \mdSpan[class={math-inline}]{$\infer{\Gamma}{\exp}{\tau}\e$} where \mdSpan[class={math-inline}]{$\ec{div}\notin\e$} then \mdSpan[class={code,math-inline}]{$\exp {\dlongmapsto}\mathspace{1}\mathid{a}$}.%
\end{mdP}
\end{mdDiv}%
\begin{mdP}%
The proof of the divergence theorem (Theorem{\mdNbsp}\mdA[class={localref},target-element={theorem}]{lem-div}{}{\mdSpan[class={theorem-label}]{4}}) is more complicated 
as we cannot
use subject reduction to show this by contradiction. Instead, we need to
do this proof using induction over logical relations{\mdNbsp}\mdSpan[class={citations},target-element={bibitem}]{[\mdA[class={bibref,localref},target-element={bibitem}]{girard:prot}{}{\mdSpan[class={bibitem-label}]{9}}]}.%
\end{mdP}%
\begin{mdP}[class={indent}]%
In our case, we say that if \mdSpan[class={math-inline}]{$\infer\cdot{e}{\t}{\e}$},
then \mdSpan[class={math-inline}]{$e$} is in the set \mdSpan[class={math-inline}]{$\red{\t}{\e}$}, {\textquotedblleft}the reducible terms of
type \mdSpan[class={math-inline}]{$\t$} with effect \mdSpan[class={math-inline}]{$\e$}{\textquotedblright}, if \mdSpan[class={math-inline}]{$\ec{div} \not\in \e$} and (1)
when \mdSpan[class={math-inline}]{$\t$} is a non-arrow type, if \mdSpan[class={math-inline}]{$e$} halts, and (2) when 
\mdSpan[class={code,math-inline}]{$\t =\mathspace{1}\t_1\mathspace{1}{\rightarrow}\mathspace{1}\e_2\mathspace{1}\t_2$}, if \mdSpan[class={math-inline}]{$e$} halts and if for all \mdSpan[class={math-inline}]{$e_1 \in \red{\t_1}{\e}$},
we have that \mdSpan[class={math-inline}]{$e\,e_1 \in \red{\t_2}{\e_2}$}.%
\end{mdP}%
\begin{mdP}[class={indent}]%
The proof of Theorem{\mdNbsp}\mdA[class={localref},target-element={theorem}]{lem-div}{}{\mdSpan[class={theorem-label}]{4}} is a standard result{\mdNbsp}\mdSpan[class={citations},target-element={bibitem}]{[\mdA[class={bibref,localref},target-element={bibitem}]{girard:prot}{}{\mdSpan[class={bibitem-label}]{9}}]}
and is a corollary from the following two main lemmas:%
\end{mdP}%
\begin{mdDiv}[class={lemma,block},id=lem-rpreserve,label={[6]\{.lemma-label\}},elem={lemma}]%
\begin{mdP}%
\mdSpan[class={lemma-before}]{\mdStrong{Lemma{\mdNbsp}\mdSpan[class={lemma-label}]{6}.} }(\mdEm{\mdSpan[class={math-inline}]{$\mathcal{R}$} is preserved by reduction})
Iff \mdSpan[class={math-inline}]{$\infer\cdot{e}{\t}{\e}$}, \mdSpan[class={math-inline}]{$e \in \red{\t}{\e}$}, and \mdSpan[class={math-inline}]{$e \longmapsto e'$}, then
also \mdSpan[class={math-inline}]{$e' \in \red{\t}{\e}$}.%
\end{mdP}
\end{mdDiv}%
\begin{mdDiv}[class={lemma,block},id=lem-rinfer,label={[7]\{.lemma-label\}},elem={lemma}]%
\begin{mdP}%
\mdSpan[class={lemma-before}]{\mdStrong{Lemma{\mdNbsp}\mdSpan[class={lemma-label}]{7}.} }(\mdEm{A well-typed term is in \mdSpan[class={math-inline}]{$\mathcal{R}$}})
If \mdSpan[class={math-inline}]{$\infer\cdot{e}{\t}{\e}$} and \mdSpan[class={math-inline}]{$\ec{div} \not\in \e$}, then \mdSpan[class={math-inline}]{$e \in \red{\e}{\t}$}.%
\end{mdP}
\end{mdDiv}%
\begin{mdDiv}[class={proof,block},elem={proof}]%
\begin{mdP}%
\mdSpan[class={proof-before}]{\mdStrong{Proof}. }
(Lemma{\mdNbsp}\mdA[class={localref},target-element={lemma}]{lem-rpreserve}{}{\mdSpan[class={lemma-label}]{6}})
This is shown by induction over the type \mdSpan[class={math-inline}]{$\t$}. For atomic types, this
holds by definition. For arrow types, \mdSpan[class={code,math-inline}]{$\t_1\mathspace{1}{\rightarrow}\mathspace{1}\e_2\mathspace{1}\t_2$}  we must show
for a given \mdSpan[class={math-inline}]{$e_1 \in \red{\t_1}{\e}$} that if \mdSpan[class={math-inline}]{$e\,e_1 \in \red{\t_2}{\e_2}$},
then also \mdSpan[class={math-inline}]{$e'\,e_1 \in \red{\t_2}{\e_2}$} (and the other direction). By
\mdSpan[class={rulename},font-variant={small-caps},font-size={small}]{(app)}, it must be \mdSpan[class={math-inline}]{$\e_2 = \e$} \mdStrong{(1)}. We can now proceed over the
structure of reductions on \mdSpan[class={math-inline}]{$e\,e_1$}:
\mdBr
\mdStrong{Case} \mdSpan[class={code,math-inline}]{$(!)\,\mathid{e}_1$}: In this case, since \mdSpan[class={rulename},font-variant={small-caps},font-size={small}]{(read)} has a \mdSpan[class={math-inline}]{$\ec{div}$} effect, we have by 
(1) that \mdSpan[class={math-inline}]{$\ec{div} \in \e$} contradicting our assumption. Note that if we
would have cheated and not include \mdSpan[class={math-inline}]{$\ec{div}$} in the type, we would have gotten 
a reduction to some \mdSpan[class={math-inline}]{$v$} which we could not show to be strongly normalizing,
and thus if it is an element of \mdSpan[class={math-inline}]{$\red{\t_2}{\e_2}$}.
\mdBr
\mdStrong{Case} \mdSpan[class={code,math-inline}]{$\fix\,\e_1$}: As the previous case.
\mdBr
\mdStrong{Case} \mdSpan[class={code,math-inline}]{$(\lambda \mathid{x}.\mathspace{1}\mathid{e}_2)\,\mathid{e}_1$}: In this case, we can reduce to \mdSpan[class={math-inline}]{$e'\,e_1$}, and by the
induction hypothesis \mdSpan[class={math-inline}]{$e'\,e_1 \in \red{\t}{\e_2}$} since \mdSpan[class={math-inline}]{$\t_2$} is smaller.%
\end{mdP}
\end{mdDiv}%
\begin{mdDiv}[class={proof,block},elem={proof}]%
\begin{mdP}[class={para-continue}]%
\mdSpan[class={proof-before}]{\mdStrong{Proof}. }
(Lemma{\mdNbsp}\mdA[class={localref},target-element={lemma}]{lem-rinfer}{}{\mdSpan[class={lemma-label}]{7}})
This is proven over the structure of the type derivation. 
However, as usual, we need to strengthen our induction hypothesis to include the environment. 
We extend \mdSpan[class={math-inline}]{$\mathcal{R}$} over environments to be a set of substitutions:%
\end{mdP}%
\begin{mdDiv}[class={math,para-block,input-math},elem={math}]%
\begin{mdDiv}[class={math,math-display}]%
\[\redg{\Gamma} = \{\sub\;|\; \dom\Gamma = \dom\sub \wedge \forall (x:\t \in \Gamma).\, \sub x \in \red{\t}{\total} \}\]%
\end{mdDiv}
\end{mdDiv}%
\begin{mdP}[class={para-continue}]%
where we assume a monomorphic environment for simplicity but we can extend this easily to
a (first-order) polymorphic setting. Our strengthened lemma we use for our proof is:%
\end{mdP}%
\begin{mdDiv}[class={math,para-block,input-math},elem={math}]%
\begin{mdDiv}[class={math,math-display}]%
\[\mathrm{if\ } \infer\Gamma{e}{\t}{\e} \wedge \sub \in \redg\Gamma \wedge \ec{div}  \not\in \e \mathrm{\ then\ } \sub e \in \red{\t}{\e} \]%
\end{mdDiv}
\end{mdDiv}%
\begin{mdP}%
The induction is standard, and we show only some sample cases:
\mdBr
\mdStrong{Case} \mdSpan[class={rulename},font-variant={small-caps},font-size={small}]{(fix)}: Since the result effect is free, we can choose any \mdSpan[class={math-inline}]{$\e$} 
such that \mdSpan[class={math-inline}]{$\ec{div} \not\in \e$}. Indeed, just an occurrence of \mdSpan[class={math-inline}]{$\fix$} is ok {\textendash} only an 
application may diverge.
\mdBr
\mdStrong{Case} \mdSpan[class={rulename},font-variant={small-caps},font-size={small}]{(app)}: By the induction hypothesis and \mdSpan[class={rulename},font-variant={small-caps},font-size={small}]{(app)} we have
\mdSpan[class={code,math-inline}]{$\sub \mathid{e}_1\mathspace{1}\in \red{\t_2\mathspace{1}{\rightarrow}\mathspace{1}\e\,\t}{\e}$} and \mdSpan[class={math-inline}]{$\sub e_2 \in \red{\t_2}{\e}$}. By definition
of \mdSpan[class={code,math-inline}]{$\red{\t_2\mathspace{1}{\rightarrow}\mathspace{1}\e\,\t}{\e}$}, \mdSpan[class={math-inline}]{$\sub e_1\,\sub e_2 \in \red{\t_2}{\e}$} and therefore
\mdSpan[class={math-inline}]{$\sub (e_1\,e_2) \in \red{\t_2}{\e}$}. Note that the induction hypothesis ensures
that \mdSpan[class={math-inline}]{$\ec{div} \not\in \e$} and therefore we cannot apply a potentially divergent function
(like \mdSpan[class={math-inline}]{$\fix$} or \mdSpan[class={math-inline}]{$(!)$}).
\mdBr
\mdStrong{Case} other: Standard, except that for effect elimination rules, we need to show that 
\mdSpan[class={math-inline}]{$\ec{div}$} is not eliminated.%
\end{mdP}
\end{mdDiv}%
\begin{mdP}%
Contrast the results of this section to many other effect  systems
\mdSpan[class={citations},target-element={bibitem}]{[\mdA[class={bibref,localref},target-element={bibitem}]{marino:genericeffects}{}{\mdSpan[class={bibitem-label}]{25}}, \mdA[class={bibref,localref},target-element={bibitem}]{scala:effects}{}{\mdSpan[class={bibitem-label}]{33}}]} that just use effect types as
syntactic `labels{'}. In our case the theorems of this section are truly
significant as they show there is a deep correspondence between the effect
types and the dynamic semantics which eventually enables us to reason about
our programs much more effectively.%
\end{mdP}%
\mdHxx[id=related-work,label={[7]\{.heading-label\}},toc={},caption={Related work}]{\mdSpan[class={heading-before}]{\mdSpan[class={heading-label}]{7}.{\enspace}}Related work}\begin{mdP}%
A main contribution of this paper is showing that our notion of mutable state
is sound, in particular the combination of mutable state and polymorphic let-
bindings is tricky as shown by Tofte{\mdNbsp}\mdSpan[class={citations},target-element={bibitem}]{[\mdA[class={bibref,localref},target-element={bibitem}]{tofte:refs}{}{\mdSpan[class={bibitem-label}]{39}}]} for the ML
language. Later, variants of the ML value restriction are studied by Leroy
\mdSpan[class={citations},target-element={bibitem}]{[\mdA[class={bibref,localref},target-element={bibitem}]{leroy:valuerestriction}{}{\mdSpan[class={bibitem-label}]{21}}]}.%
\end{mdP}%
\begin{mdP}[class={indent}]%
Safe state encapsulation using a lazy state monad was first proven formally by
Launchbury and Sabry{\mdNbsp}\mdSpan[class={citations},target-element={bibitem}]{[\mdA[class={bibref,localref},target-element={bibitem}]{launchbury:monadstate}{}{\mdSpan[class={bibitem-label}]{16}}]}. Their formalization is
quite different though from ours and applies to a lazy store in a monadic
setting. In particular, in their formalization there is no separate heap
binding, but heaps are always bound at the outer \mdSpan[class={code,math-inline}]{$\mathid{run}$}. We tried this, but it
proved difficult in our setting; for example, it is hard to state the stateful
lemma since answers would never contain an explicit heap. Very similar to our
state encapsulation is region inference{\mdNbsp}\mdSpan[class={citations},target-element={bibitem}]{[\mdA[class={bibref,localref},target-element={bibitem}]{tofte:regioninf}{}{\mdSpan[class={bibitem-label}]{40}}]}. Our \mdSpan[class={code,math-inline}]{$\mathid{run}$}
operation essentially delimits a heap region. Regions are values
though, and we can for example not access references in several regions at
once.%
\end{mdP}%
\begin{mdP}[class={indent}]%
Independently of our work, Lindley and Cheney{\mdNbsp}\mdSpan[class={citations},target-element={bibitem}]{[\mdA[class={bibref,localref},target-element={bibitem}]{lindley:effects}{}{\mdSpan[class={bibitem-label}]{23}}]} also
used row polymorphism for effect types. Their approach is based on
presence/absence flags{\mdNbsp}\mdSpan[class={citations},target-element={bibitem}]{[\mdA[class={bibref,localref},target-element={bibitem}]{remy:records}{}{\mdSpan[class={bibitem-label}]{32}}]} to give effect types to database
operations in the context of the Links web programming language. The main
effects of the database operations are \mdEm{wild}, \mdEm{tame}, and \mdEm{hear}, for
arbitrary effects including divergence, pure queries, and asynchronous
messages respectively. They report on practical experience exposing effect
types to the programmer and discuss various syntax forms to easily denote
effect types.%
\end{mdP}%
\begin{mdP}[class={indent}]%
The problems with arbitrary effects have been widely recognized, and there is
a large body of work studying how to delimit the scope of effects. There have
been many effect typing disciplines proposed. Early work is by Gifford and
Lucassen{\mdNbsp}\mdSpan[class={citations},target-element={bibitem}]{[\mdA[class={bibref,localref},target-element={bibitem}]{gifford:imperative}{}{\mdSpan[class={bibitem-label}]{8}}, \mdA[class={bibref,localref},target-element={bibitem}]{lucassen:polyeffect}{}{\mdSpan[class={bibitem-label}]{24}}]} which was later
extended by Talpin{\mdNbsp}\mdSpan[class={citations},target-element={bibitem}]{[\mdA[class={bibref,localref},target-element={bibitem}]{talpin:thesis}{}{\mdSpan[class={bibitem-label}]{37}}]} and others
\mdSpan[class={citations},target-element={bibitem}]{[\mdA[class={bibref,localref},target-element={bibitem}]{nielson:polyeffect}{}{\mdSpan[class={bibitem-label}]{28}}, \mdA[class={bibref,localref},target-element={bibitem}]{talpin:effects}{}{\mdSpan[class={bibitem-label}]{36}}]}.  These systems are closely related
since they describe polymorphic effect systems and use type constraints to
give principal types. The system described by Nielson \mdEm{et al.}
\mdSpan[class={citations},target-element={bibitem}]{[\mdA[class={bibref,localref},target-element={bibitem}]{nielson:polyeffect}{}{\mdSpan[class={bibitem-label}]{28}}]}  also requires the effects to form a complete
lattice with meets and joins.%
\end{mdP}%
\begin{mdP}[class={indent}]%
Java contains a simple effect system where each method is labeled with the
exceptions it might raise{\mdNbsp}\mdSpan[class={citations},target-element={bibitem}]{[\mdA[class={bibref,localref},target-element={bibitem}]{gosling:java}{}{\mdSpan[class={bibitem-label}]{10}}]}. A system for finding uncaught
exceptions was developed for ML by Pessaux \mdEm{et al.}{\mdNbsp}\mdSpan[class={citations},target-element={bibitem}]{[\mdA[class={bibref,localref},target-element={bibitem}]{pessaux:exceptions}{}{\mdSpan[class={bibitem-label}]{29}}]}. 
A more powerful system for tracking effects was
developed by Benton{\mdNbsp}\mdSpan[class={citations},target-element={bibitem}]{[\mdA[class={bibref,localref},target-element={bibitem}]{benton:exceptioneffects}{}{\mdSpan[class={bibitem-label}]{2}}]} who also studies the
semantics of such effect systems{\mdNbsp}\mdSpan[class={citations},target-element={bibitem}]{[\mdA[class={bibref,localref},target-element={bibitem}]{benton:relsemantics}{}{\mdSpan[class={bibitem-label}]{3}}]}. Recent work on
effects in Scala{\mdNbsp}\mdSpan[class={citations},target-element={bibitem}]{[\mdA[class={bibref,localref},target-element={bibitem}]{scala:effects}{}{\mdSpan[class={bibitem-label}]{33}}]} shows how even a restricted form of
polymorphic effect types can track effects for many programs in
practice.%
\end{mdP}%
\begin{mdP}[class={indent}]%
Tolmach{\mdNbsp}\mdSpan[class={citations},target-element={bibitem}]{[\mdA[class={bibref,localref},target-element={bibitem}]{tolmach:monadhierarchy}{}{\mdSpan[class={bibitem-label}]{41}}]} describes an effect analysis for ML
in terms of effect monads, namely \mdSpan[class={code,math-inline}]{$\mathid{Total}$}, \mdSpan[class={code,math-inline}]{$\mathid{Partial}$}, \mdSpan[class={code,math-inline}]{$\mathid{Divergent}$} and \mdSpan[class={code,math-inline}]{$\mathid{ST}$}.
This is system is not polymorphic though and meant more for internal compiler
analysis. In the context proof systems there has been work to show absence of
observable side effects for object-oriented programming languages, for example
by Naumann{\mdNbsp}\mdSpan[class={citations},target-element={bibitem}]{[\mdA[class={bibref,localref},target-element={bibitem}]{naumann:purity}{}{\mdSpan[class={bibitem-label}]{27}}]}.%
\end{mdP}%
\begin{mdP}[class={indent}]%
Marino \mdEm{et al.} recently produced a generic type-and-effect system
\mdSpan[class={citations},target-element={bibitem}]{[\mdA[class={bibref,localref},target-element={bibitem}]{marino:genericeffects}{}{\mdSpan[class={bibitem-label}]{25}}]}. This system uses privilege checking to describe
analytical effect systems. For example, an effect system could use 
try-catch statements to grant
the \mdSpan[class={code,math-inline}]{$\mathid{canThrow}$} privilege inside try blocks. \mdSpan[class={code,math-inline}]{$\mathid{throw}$} statements are then only
permitted when this privilege is present. Their system is very general and can
express many properties but has no semantics on its own. For example, it would
be sound for the effect system to have {\textquotedblleft}+{\textquotedblright} grant the \mdSpan[class={code,math-inline}]{$\mathid{canThrow}$} privilege to
its arguments, and one has to do an additional proof to show that the effects
in these systems actually correspond to an intended meaning.%
\end{mdP}%
\begin{mdP}[class={indent}]%
Wadler and Thiemann showed the close relation between effect systems and
monads{\mdNbsp}\mdSpan[class={citations},target-element={bibitem}]{[\mdA[class={bibref,localref},target-element={bibitem}]{wadler:marriage}{}{\mdSpan[class={bibitem-label}]{42}}]} and showed how any effect system can be translated
to a monadic version. For our particular system though a monadic translation
is quite involved due to polymorphic effects; essentially we need dependently
typed operations and we leave a proper monadic semantics for future work.%
\end{mdP}%
\begin{mdP}[class={indent}]%
Recently, \mdEm{effect handlers} are proposed to program with effects{\mdNbsp}\mdSpan[class={citations},target-element={bibitem}]{[\mdA[class={bibref,localref},target-element={bibitem}]{brady:handlers}{}{\mdSpan[class={bibitem-label}]{4}}, \mdA[class={bibref,localref},target-element={bibitem}]{kammar:handlers}{}{\mdSpan[class={bibitem-label}]{14}}, \mdA[class={bibref,localref},target-element={bibitem}]{kammarplotkin:handlers}{}{\mdSpan[class={bibitem-label}]{15}}]}. In this work, computational effects are 
modeled as operations of an algebraic theory. Even though algebraic effects are
subsumed by monads, they can be combined more easily and the specification of
handlers offers new ways of programming. This work is quite different than what
we described in this paper since we only considered effects that are intrinsic
to the language while effect handlers deal specifically with {\textquoteleft}user-defined{\textquoteright} effects.
However, we are currently working on integrating user-defined effects in Koka
using effect handlers, and investigating how this can work well with the
current effect type system.%
\end{mdP}%
\mdHxx[id=references,caption={References}]{References}\begin{mdDiv}[class={tex,input-tex},elem={tex}]%
\begin{mdBibliography}[class={bibliography,bib-numeric},elem={bibliography},bibstyle={eptcs},bibdata={daan},caption={10}]%
\begin{mdBibitem}[class={bibitem},id=abel:termination,label={[1]\{.bibitem-label\}},elem={bibitem},cite-label={[1]\{.bibitem-label\}},caption={[Andreas Abel]\{ .bibinfo field={'}author{'}\} ([1998]\{ .bibinfo field={'}year{'}\}): \_[Foetus -- A Termination Checker for Simple Functional Programs]\{ .bibinfo field={'}title{'}\}\_. \\[Unpublished note]\{ .bibinfo field={'}note{'}\}.},searchterm={+Andreas+Abel+Foetus+--+Termination+Checker+Simple+Functional+Programs+Unpublished+note++}]%
\mdSpan[class={bibitem-before}]{[\mdSpan[class={bibitem-label}]{1}]{\mdNbsp}{\mdNbsp}}\mdSpan[class={bibinfo},field={author}]{Andreas Abel} (\mdSpan[class={bibinfo},field={year}]{1998}):
  \mdEm{\mdSpan[class={bibinfo},field={title}]{Foetus {\textendash} A Termination Checker for Simple Functional
  Programs}}.
\mdSpan[class={newblock}]{} \mdSpan[class={bibinfo},field={note}]{Unpublished note}.%
\end{mdBibitem}%
\begin{mdBibitem}[class={bibitem},id=benton:exceptioneffects,label={[2]\{.bibitem-label\}},elem={bibitem},cite-label={[2]\{.bibitem-label\}},caption={[Nick Benton]\{ .bibinfo field={'}author{'}\} {\&} [Peter Buchlovsky]\{ .bibinfo field={'}author{'}\} ([2007]\{ .bibinfo field={'}year{'}\}): \_[Semantics of an effect analysis for exceptions]\{ .bibinfo field={'}title{'}\}\_. \\In: [ [TLDI {'}07: Proceedings of the 2007 ACM SIGPLAN international workshop on Types in languages design and implementation]\{ .bibinfo field={'}booktitle{'}\}]\{font-style=oblique\}, pp. [15--26]\{ .bibinfo field={'}pages{'}\}, doi:[10.1145/1190315.1190320](http://dx.doi.org/10.1145/1190315.1190320).},searchterm={+Nick+Benton+Peter+Buchlovsky+Semantics+effect+analysis+exceptions+TLDI+Proceedings+SIGPLAN+international+workshop+Types+languages+design+implementation+http++}]%
\mdSpan[class={bibitem-before}]{[\mdSpan[class={bibitem-label}]{2}]{\mdNbsp}{\mdNbsp}}\mdSpan[class={bibinfo},field={author}]{Nick Benton} {\&}
  \mdSpan[class={bibinfo},field={author}]{Peter Buchlovsky}
  (\mdSpan[class={bibinfo},field={year}]{2007}): \mdEm{\mdSpan[class={bibinfo},field={title}]{Semantics of an effect analysis
  for exceptions}}.
\mdSpan[class={newblock}]{} In: \mdSpan[font-style={oblique}]{ \mdSpan[class={bibinfo},field={booktitle}]{TLDI {'}07: Proceedings of the 2007 ACM
  SIGPLAN international workshop on Types in languages design and
  implementation}}, pp. \mdSpan[class={bibinfo},field={pages}]{15{\textendash}26}, doi:\mdA{http://dx.doi.org/10.1145/1190315.1190320}{}{10.1145/1190315.1190320}.%
\end{mdBibitem}%
\begin{mdBibitem}[class={bibitem},id=benton:relsemantics,label={[3]\{.bibitem-label\}},elem={bibitem},cite-label={[3]\{.bibitem-label\}},caption={[Nick Benton]\{ .bibinfo field={'}author{'}\}, [Andrew Kennedy]\{ .bibinfo field={'}author{'}\}, [Lennart Beringer]\{ .bibinfo field={'}author{'}\} {\&} [Martin Hofmann]\{ .bibinfo field={'}author{'}\} ([2007]\{ .bibinfo field={'}year{'}\}): \_[Relational semantics for effect-based program transformations with dynamic allocation]\{ .bibinfo field={'}title{'}\}\_. \\In: [ [PPDP {'}07: Proc. of the 9th ACM SIGPLAN int. conf. on Principles and Practice of Declarative Prog.]\{ .bibinfo field={'}booktitle{'}\}]\{font-style=oblique\}, pp. [87--96]\{ .bibinfo field={'}pages{'}\}, doi:[10.1145/1273920.1273932](http://dx.doi.org/10.1145/1273920.1273932).},searchterm={+Nick+Benton+Andrew+Kennedy+Lennart+Beringer+Martin+Hofmann+Relational+semantics+effect+based+program+transformations+with+dynamic+allocation+PPDP+Proc+SIGPLAN+conf+Principles+Practice+Declarative+Prog+http++}]%
\mdSpan[class={bibitem-before}]{[\mdSpan[class={bibitem-label}]{3}]{\mdNbsp}{\mdNbsp}}\mdSpan[class={bibinfo},field={author}]{Nick Benton}, \mdSpan[class={bibinfo},field={author}]{Andrew
  Kennedy}, \mdSpan[class={bibinfo},field={author}]{Lennart Beringer} {\&} \mdSpan[class={bibinfo},field={author}]{Martin Hofmann} (\mdSpan[class={bibinfo},field={year}]{2007}): \mdEm{\mdSpan[class={bibinfo},field={title}]{Relational
  semantics for effect-based program transformations with dynamic allocation}}.
\mdSpan[class={newblock}]{} In: \mdSpan[font-style={oblique}]{ \mdSpan[class={bibinfo},field={booktitle}]{PPDP {'}07: Proc. of the 9th ACM SIGPLAN
  int. conf. on Principles and Practice of Declarative Prog.}}, pp.
  \mdSpan[class={bibinfo},field={pages}]{87{\textendash}96}, doi:\mdA{http://dx.doi.org/10.1145/1273920.1273932}{}{10.1145/1273920.1273932}.%
\end{mdBibitem}%
\begin{mdBibitem}[class={bibitem},id=brady:handlers,label={[4]\{.bibitem-label\}},elem={bibitem},cite-label={[4]\{.bibitem-label\}},caption={[Edwin Brady]\{ .bibinfo field={'}author{'}\} ([2013]\{ .bibinfo field={'}year{'}\}): \_[Programming and Reasoning with Algebraic Effects and Dependent Types]\{ .bibinfo field={'}title{'}\}\_. \\In: [ [Proceedings of the 18th ACM SIGPLAN International Conference on Functional Programming]\{ .bibinfo field={'}booktitle{'}\}]\{font-style=oblique\}, [ICFP {'}13]\{ .bibinfo field={'}series{'}\}, [ACM]\{ .bibinfo field={'}publisher{'}\}, [New York, NY, USA]\{ .bibinfo field={'}address{'}\}, pp. [133--144]\{ .bibinfo field={'}pages{'}\}, doi:[10.1145/2500365.2500581](http://dx.doi.org/10.1145/2500365.2500581).},searchterm={+Edwin+Brady+Programming+Reasoning+with+Algebraic+Effects+Dependent+Types+Proceedings+SIGPLAN+International+Conference+Functional+Programming+ICFP+York+http++}]%
\mdSpan[class={bibitem-before}]{[\mdSpan[class={bibitem-label}]{4}]{\mdNbsp}{\mdNbsp}}\mdSpan[class={bibinfo},field={author}]{Edwin Brady} (\mdSpan[class={bibinfo},field={year}]{2013}):
  \mdEm{\mdSpan[class={bibinfo},field={title}]{Programming and Reasoning with Algebraic Effects and
  Dependent Types}}.
\mdSpan[class={newblock}]{} In: \mdSpan[font-style={oblique}]{ \mdSpan[class={bibinfo},field={booktitle}]{Proceedings of the 18th ACM SIGPLAN
  International Conference on Functional Programming}}, \mdSpan[class={bibinfo},field={series}]{ICFP
  {'}13}, \mdSpan[class={bibinfo},field={publisher}]{ACM}, \mdSpan[class={bibinfo},field={address}]{New York, NY, USA}, pp.
  \mdSpan[class={bibinfo},field={pages}]{133{\textendash}144}, doi:\mdA{http://dx.doi.org/10.1145/2500365.2500581}{}{10.1145/2500365.2500581}.%
\end{mdBibitem}%
\begin{mdBibitem}[class={bibitem},id=damasmilner:hm,label={[5]\{.bibitem-label\}},elem={bibitem},cite-label={[5]\{.bibitem-label\}},caption={[Luis Damas]\{ .bibinfo field={'}author{'}\} {\&} [Robin Milner]\{ .bibinfo field={'}author{'}\} ([1982]\{ .bibinfo field={'}year{'}\}): \_[Principal type-schemes for functional programs]\{ .bibinfo field={'}title{'}\}\_. \\In: [ [9th ACM symp. on Principles of Programming Languages (POPL{'}82)]\{ .bibinfo field={'}booktitle{'}\}]\{font-style=oblique\}, pp. [207--212]\{ .bibinfo field={'}pages{'}\}, doi:[10.1145/582153.582176](http://dx.doi.org/10.1145/582153.582176).},searchterm={+Luis+Damas+Robin+Milner+Principal+type+schemes+functional+programs+symp+Principles+Programming+Languages+POPL+http++}]%
\mdSpan[class={bibitem-before}]{[\mdSpan[class={bibitem-label}]{5}]{\mdNbsp}{\mdNbsp}}\mdSpan[class={bibinfo},field={author}]{Luis Damas} {\&} \mdSpan[class={bibinfo},field={author}]{Robin
  Milner} (\mdSpan[class={bibinfo},field={year}]{1982}):
  \mdEm{\mdSpan[class={bibinfo},field={title}]{Principal type-schemes for functional programs}}.
\mdSpan[class={newblock}]{} In: \mdSpan[font-style={oblique}]{ \mdSpan[class={bibinfo},field={booktitle}]{9th ACM symp. on Principles of
  Programming Languages (POPL{'}82)}}, pp. \mdSpan[class={bibinfo},field={pages}]{207{\textendash}212},
  doi:\mdA{http://dx.doi.org/10.1145/582153.582176}{}{10.1145/582153.582176}.%
\end{mdBibitem}%
\begin{mdBibitem}[class={bibitem},id=filliatre:garsiawachs,label={[6]\{.bibitem-label\}},elem={bibitem},cite-label={[6]\{.bibitem-label\}},caption={[Jean-Christophe Filli{\^{a}}tre]\{ .bibinfo field={'}author{'}\} ([2008]\{ .bibinfo field={'}year{'}\}): \_[A Functional Implementation of the Garsia--wachs Algorithm: (Functional Pearl)]\{ .bibinfo field={'}title{'}\}\_. \\In: [ [Proceedings of the 2008 ACM SIGPLAN Workshop on ML]\{ .bibinfo field={'}booktitle{'}\}]\{font-style=oblique\}, [ML {'}08]\{ .bibinfo field={'}series{'}\}, [ACM]\{ .bibinfo field={'}publisher{'}\}, [New York, NY, USA]\{ .bibinfo field={'}address{'}\}, pp. [91--96]\{ .bibinfo field={'}pages{'}\}, doi:[10.1145/1411304.1411317](http://dx.doi.org/10.1145/1411304.1411317).},searchterm={+Jean+Christophe+Filli+Functional+Implementation+Garsia+wachs+Algorithm+Functional+Pearl+Proceedings+SIGPLAN+Workshop+York+http++}]%
\mdSpan[class={bibitem-before}]{[\mdSpan[class={bibitem-label}]{6}]{\mdNbsp}{\mdNbsp}}\mdSpan[class={bibinfo},field={author}]{Jean-Christophe Filli{\^{a}}tre}
  (\mdSpan[class={bibinfo},field={year}]{2008}): \mdEm{\mdSpan[class={bibinfo},field={title}]{A Functional Implementation of
  the Garsia{\textendash}wachs Algorithm: (Functional Pearl)}}.
\mdSpan[class={newblock}]{} In: \mdSpan[font-style={oblique}]{ \mdSpan[class={bibinfo},field={booktitle}]{Proceedings of the 2008 ACM SIGPLAN
  Workshop on ML}}, \mdSpan[class={bibinfo},field={series}]{ML {'}08}, \mdSpan[class={bibinfo},field={publisher}]{ACM},
  \mdSpan[class={bibinfo},field={address}]{New York, NY, USA}, pp. \mdSpan[class={bibinfo},field={pages}]{91{\textendash}96},
  doi:\mdA{http://dx.doi.org/10.1145/1411304.1411317}{}{10.1145/1411304.1411317}.%
\end{mdBibitem}%
\begin{mdBibitem}[class={bibitem},id=gasterjones:trex,label={[7]\{.bibitem-label\}},elem={bibitem},cite-label={[7]\{.bibitem-label\}},caption={[Ben{\mdNbsp}R. Gaster]\{ .bibinfo field={'}author{'}\} {\&} [Mark{\mdNbsp}P. Jones]\{ .bibinfo field={'}author{'}\} ([1996]\{ .bibinfo field={'}year{'}\}): \_[A Polymorphic Type System for Extensible Records and Variants]\{ .bibinfo field={'}title{'}\}\_. \\[Technical Report]\{ .bibinfo field={'}type{'}\} [NOTTCS-TR-96-3]\{ .bibinfo field={'}number{'}\}, [University of Nottingham]\{ .bibinfo field={'}institution{'}\}.},searchterm={+Gaster+Mark+Jones+Polymorphic+Type+System+Extensible+Records+Variants+Technical+Report+NOTTCS+University+Nottingham++}]%
\mdSpan[class={bibitem-before}]{[\mdSpan[class={bibitem-label}]{7}]{\mdNbsp}{\mdNbsp}}\mdSpan[class={bibinfo},field={author}]{Ben{\mdNbsp}R. Gaster} {\&}
  \mdSpan[class={bibinfo},field={author}]{Mark{\mdNbsp}P. Jones}
  (\mdSpan[class={bibinfo},field={year}]{1996}): \mdEm{\mdSpan[class={bibinfo},field={title}]{A Polymorphic Type System for
  Extensible Records and Variants}}.
\mdSpan[class={newblock}]{} \mdSpan[class={bibinfo},field={type}]{Technical Report} \mdSpan[class={bibinfo},field={number}]{NOTTCS-TR-96-3},
  \mdSpan[class={bibinfo},field={institution}]{University of Nottingham}.%
\end{mdBibitem}%
\begin{mdBibitem}[class={bibitem},id=gifford:imperative,label={[8]\{.bibitem-label\}},elem={bibitem},cite-label={[8]\{.bibitem-label\}},caption={[David{\mdNbsp}K. Gifford]\{ .bibinfo field={'}author{'}\} {\&} [John{\mdNbsp}M. Lucassen]\{ .bibinfo field={'}author{'}\} ([1986]\{ .bibinfo field={'}year{'}\}): \_[Integrating functional and imperative programming]\{ .bibinfo field={'}title{'}\}\_. \\In: [ [LFP {'}86: Proceedings of the 1986 ACM conference on LISP and functional programming]\{ .bibinfo field={'}booktitle{'}\}]\{font-style=oblique\}, pp. [28--38]\{ .bibinfo field={'}pages{'}\}, doi:[10.1145/319838.319848](http://dx.doi.org/10.1145/319838.319848).},searchterm={+David+Gifford+John+Lucassen+Integrating+functional+imperative+programming+Proceedings+conference+LISP+functional+programming+http++}]%
\mdSpan[class={bibitem-before}]{[\mdSpan[class={bibitem-label}]{8}]{\mdNbsp}{\mdNbsp}}\mdSpan[class={bibinfo},field={author}]{David{\mdNbsp}K. Gifford} {\&}
  \mdSpan[class={bibinfo},field={author}]{John{\mdNbsp}M. Lucassen}
  (\mdSpan[class={bibinfo},field={year}]{1986}): \mdEm{\mdSpan[class={bibinfo},field={title}]{Integrating functional and
  imperative programming}}.
\mdSpan[class={newblock}]{} In: \mdSpan[font-style={oblique}]{ \mdSpan[class={bibinfo},field={booktitle}]{LFP {'}86: Proceedings of the 1986 ACM
  conference on LISP and functional programming}}, pp. \mdSpan[class={bibinfo},field={pages}]{28{\textendash}38},
  doi:\mdA{http://dx.doi.org/10.1145/319838.319848}{}{10.1145/319838.319848}.%
\end{mdBibitem}%
\begin{mdBibitem}[class={bibitem},id=girard:prot,label={[9]\{.bibitem-label\}},elem={bibitem},cite-label={[9]\{.bibitem-label\}},caption={[Jean-Yves Girard]\{ .bibinfo field={'}author{'}\}, [Paul Taylor]\{ .bibinfo field={'}author{'}\} {\&} [Yves Lafont]\{ .bibinfo field={'}author{'}\} ([1989]\{ .bibinfo field={'}year{'}\}): \_[Proofs and types]\{ .bibinfo field={'}title{'}\}\_. \\[Cambridge University Press]\{ .bibinfo field={'}publisher{'}\}.},searchterm={+Jean+Yves+Girard+Paul+Taylor+Yves+Lafont+Proofs+types+Cambridge+University+Press++}]%
\mdSpan[class={bibitem-before}]{[\mdSpan[class={bibitem-label}]{9}]{\mdNbsp}{\mdNbsp}}\mdSpan[class={bibinfo},field={author}]{Jean-Yves Girard},
  \mdSpan[class={bibinfo},field={author}]{Paul Taylor} {\&}
  \mdSpan[class={bibinfo},field={author}]{Yves Lafont}
  (\mdSpan[class={bibinfo},field={year}]{1989}): \mdEm{\mdSpan[class={bibinfo},field={title}]{Proofs and types}}.
\mdSpan[class={newblock}]{} \mdSpan[class={bibinfo},field={publisher}]{Cambridge University Press}.%
\end{mdBibitem}%
\begin{mdBibitem}[class={bibitem},id=gosling:java,label={[10]\{.bibitem-label\}},elem={bibitem},cite-label={[10]\{.bibitem-label\}},caption={[James Gosling]\{ .bibinfo field={'}author{'}\}, [Bill Joy]\{ .bibinfo field={'}author{'}\} {\&} [Guy Steele]\{ .bibinfo field={'}author{'}\} ([1996]\{ .bibinfo field={'}year{'}\}): \_[The Java Language Specification]\{ .bibinfo field={'}title{'}\}\_. \\[Addison-Wesley]\{ .bibinfo field={'}publisher{'}\}.},searchterm={+James+Gosling+Bill+Steele+Java+Language+Specification+Addison+Wesley++}]%
\mdSpan[class={bibitem-before}]{[\mdSpan[class={bibitem-label}]{10}]{\mdNbsp}{\mdNbsp}}\mdSpan[class={bibinfo},field={author}]{James Gosling}, \mdSpan[class={bibinfo},field={author}]{Bill
  Joy} {\&} \mdSpan[class={bibinfo},field={author}]{Guy Steele} (\mdSpan[class={bibinfo},field={year}]{1996}): \mdEm{\mdSpan[class={bibinfo},field={title}]{The Java
  Language Specification}}.
\mdSpan[class={newblock}]{} \mdSpan[class={bibinfo},field={publisher}]{Addison-Wesley}.%
\end{mdBibitem}%
\begin{mdBibitem}[class={bibitem},id=hindley:types,label={[11]\{.bibitem-label\}},elem={bibitem},cite-label={[11]\{.bibitem-label\}},caption={[J.R. Hindley]\{ .bibinfo field={'}author{'}\} ([1969]\{ .bibinfo field={'}year{'}\}): \_[The principal type scheme of an object in combinatory logic]\{ .bibinfo field={'}title{'}\}\_. \\[ [Trans. of the American Mathematical Society]\{ .bibinfo field={'}journal{'}\}]\{font-style=oblique\} [146]\{ .bibinfo field={'}volume{'}\}, pp. [29--60]\{ .bibinfo field={'}pages{'}\}, doi:[10.2307/1995158](http://dx.doi.org/10.2307/1995158).},searchterm={+Hindley+principal+type+scheme+object+combinatory+logic+Trans+American+Mathematical+Society+http++}]%
\mdSpan[class={bibitem-before}]{[\mdSpan[class={bibitem-label}]{11}]{\mdNbsp}{\mdNbsp}}\mdSpan[class={bibinfo},field={author}]{J.R. Hindley} (\mdSpan[class={bibinfo},field={year}]{1969}):
  \mdEm{\mdSpan[class={bibinfo},field={title}]{The principal type scheme of an object in combinatory
  logic}}.
\mdSpan[class={newblock}]{} \mdSpan[font-style={oblique}]{ \mdSpan[class={bibinfo},field={journal}]{Trans. of the American Mathematical Society}}
  \mdSpan[class={bibinfo},field={volume}]{146}, pp. \mdSpan[class={bibinfo},field={pages}]{29{\textendash}60}, doi:\mdA{http://dx.doi.org/10.2307/1995158}{}{10.2307/1995158}.%
\end{mdBibitem}%
\begin{mdBibitem}[class={bibitem},id=jones:qualifiedtypes,label={[12]\{.bibitem-label\}},elem={bibitem},cite-label={[12]\{.bibitem-label\}},caption={[Mark{\mdNbsp}P. Jones]\{ .bibinfo field={'}author{'}\} ([1992]\{ .bibinfo field={'}year{'}\}): \_[A theory of qualified types]\{ .bibinfo field={'}title{'}\}\_. \\In: [ [4th. European Symposium on Programming (ESOP{'}92)]\{ .bibinfo field={'}booktitle{'}\}]\{font-style=oblique\}, [ [Lecture Notes in Computer Science]\{ .bibinfo field={'}series{'}\}]\{font-style=oblique\} [582]\{ .bibinfo field={'}volume{'}\}, [Springer-Verlag]\{ .bibinfo field={'}publisher{'}\}, pp. [287--306]\{ .bibinfo field={'}pages{'}\}, doi:[10.1007/3-540-55253-7\_17](http://dx.doi.org/10.1007/3-540-55253-7\_17).},searchterm={+Mark+Jones+theory+qualified+types+European+Symposium+Programming+ESOP+Lecture+Notes+Computer+Science+Springer+Verlag+http++}]%
\mdSpan[class={bibitem-before}]{[\mdSpan[class={bibitem-label}]{12}]{\mdNbsp}{\mdNbsp}}\mdSpan[class={bibinfo},field={author}]{Mark{\mdNbsp}P. Jones}
  (\mdSpan[class={bibinfo},field={year}]{1992}): \mdEm{\mdSpan[class={bibinfo},field={title}]{A theory of qualified types}}.
\mdSpan[class={newblock}]{} In: \mdSpan[font-style={oblique}]{ \mdSpan[class={bibinfo},field={booktitle}]{4th. European Symposium on Programming
  (ESOP{'}92)}}, \mdSpan[font-style={oblique}]{ \mdSpan[class={bibinfo},field={series}]{Lecture Notes in Computer Science}}
  \mdSpan[class={bibinfo},field={volume}]{582}, \mdSpan[class={bibinfo},field={publisher}]{Springer-Verlag}, pp.
  \mdSpan[class={bibinfo},field={pages}]{287{\textendash}306}, doi:\mdA{http://dx.doi.org/10.1007/3-540-55253-7\_17}{}{10.1007/3-540-55253-7\_17}.%
\end{mdBibitem}%
\begin{mdBibitem}[class={bibitem},id=jones:constructorclasses,label={[13]\{.bibitem-label\}},elem={bibitem},cite-label={[13]\{.bibitem-label\}},caption={[Mark{\mdNbsp}P. Jones]\{ .bibinfo field={'}author{'}\} ([1995]\{ .bibinfo field={'}year{'}\}): \_[A system of constructor classes: overloading and implicit higher-order polymorphism]\{ .bibinfo field={'}title{'}\}\_. \\[ [Journal of Functional Programming]\{ .bibinfo field={'}journal{'}\}]\{font-style=oblique\} [5]\{ .bibinfo field={'}volume{'}\}([1]\{ .bibinfo field={'}number{'}\}), pp. [1--35]\{ .bibinfo field={'}pages{'}\}, doi:[10.1145/165180.165190](http://dx.doi.org/10.1145/165180.165190).},searchterm={+Mark+Jones+system+constructor+classes+overloading+implicit+higher+order+polymorphism+Journal+Functional+Programming+http++}]%
\mdSpan[class={bibitem-before}]{[\mdSpan[class={bibitem-label}]{13}]{\mdNbsp}{\mdNbsp}}\mdSpan[class={bibinfo},field={author}]{Mark{\mdNbsp}P. Jones}
  (\mdSpan[class={bibinfo},field={year}]{1995}): \mdEm{\mdSpan[class={bibinfo},field={title}]{A system of constructor
  classes: overloading and implicit higher-order polymorphism}}.
\mdSpan[class={newblock}]{} \mdSpan[font-style={oblique}]{ \mdSpan[class={bibinfo},field={journal}]{Journal of Functional Programming}}
  \mdSpan[class={bibinfo},field={volume}]{5}(\mdSpan[class={bibinfo},field={number}]{1}), pp. \mdSpan[class={bibinfo},field={pages}]{1{\textendash}35},
  doi:\mdA{http://dx.doi.org/10.1145/165180.165190}{}{10.1145/165180.165190}.%
\end{mdBibitem}%
\begin{mdBibitem}[class={bibitem},id=kammar:handlers,label={[14]\{.bibitem-label\}},elem={bibitem},cite-label={[14]\{.bibitem-label\}},caption={[Ohad Kammar]\{ .bibinfo field={'}author{'}\}, [Sam Lindley]\{ .bibinfo field={'}author{'}\} {\&} [Nicolas Oury]\{ .bibinfo field={'}author{'}\} ([2013]\{ .bibinfo field={'}year{'}\}): \_[Handlers in Action]\{ .bibinfo field={'}title{'}\}\_. \\In: [ [Proceedings of the 18th ACM SIGPLAN International Conference on Functional Programming]\{ .bibinfo field={'}booktitle{'}\}]\{font-style=oblique\}, [ICFP {'}13]\{ .bibinfo field={'}series{'}\}, [ACM]\{ .bibinfo field={'}publisher{'}\}, [New York, NY, USA]\{ .bibinfo field={'}address{'}\}, pp. [145--158]\{ .bibinfo field={'}pages{'}\}, doi:[10.1145/2500365.2500590](http://dx.doi.org/10.1145/2500365.2500590).},searchterm={+Ohad+Kammar+Lindley+Nicolas+Oury+Handlers+Action+Proceedings+SIGPLAN+International+Conference+Functional+Programming+ICFP+York+http++}]%
\mdSpan[class={bibitem-before}]{[\mdSpan[class={bibitem-label}]{14}]{\mdNbsp}{\mdNbsp}}\mdSpan[class={bibinfo},field={author}]{Ohad Kammar}, \mdSpan[class={bibinfo},field={author}]{Sam
  Lindley} {\&} \mdSpan[class={bibinfo},field={author}]{Nicolas Oury} (\mdSpan[class={bibinfo},field={year}]{2013}): \mdEm{\mdSpan[class={bibinfo},field={title}]{Handlers in
  Action}}.
\mdSpan[class={newblock}]{} In: \mdSpan[font-style={oblique}]{ \mdSpan[class={bibinfo},field={booktitle}]{Proceedings of the 18th ACM SIGPLAN
  International Conference on Functional Programming}}, \mdSpan[class={bibinfo},field={series}]{ICFP
  {'}13}, \mdSpan[class={bibinfo},field={publisher}]{ACM}, \mdSpan[class={bibinfo},field={address}]{New York, NY, USA}, pp.
  \mdSpan[class={bibinfo},field={pages}]{145{\textendash}158}, doi:\mdA{http://dx.doi.org/10.1145/2500365.2500590}{}{10.1145/2500365.2500590}.%
\end{mdBibitem}%
\begin{mdBibitem}[class={bibitem},id=kammarplotkin:handlers,label={[15]\{.bibitem-label\}},elem={bibitem},cite-label={[15]\{.bibitem-label\}},caption={[Ohad Kammar]\{ .bibinfo field={'}author{'}\} {\&} [Gordon{\mdNbsp}D. Plotkin]\{ .bibinfo field={'}author{'}\} ([2012]\{ .bibinfo field={'}year{'}\}): \_[Algebraic Foundations for Effect-dependent Optimisations]\{ .bibinfo field={'}title{'}\}\_. \\In: [ [Proceedings of the 39th Annual ACM SIGPLAN-SIGACT Symposium on Principles of Programming Languages]\{ .bibinfo field={'}booktitle{'}\}]\{font-style=oblique\}, [POPL {'}12]\{ .bibinfo field={'}series{'}\}, [ACM]\{ .bibinfo field={'}publisher{'}\}, [New York, NY, USA]\{ .bibinfo field={'}address{'}\}, pp. [349--360]\{ .bibinfo field={'}pages{'}\}, doi:[10.1145/2103656.2103698](http://dx.doi.org/10.1145/2103656.2103698).},searchterm={+Ohad+Kammar+Gordon+Plotkin+Algebraic+Foundations+Effect+dependent+Optimisations+Proceedings+Annual+SIGPLAN+SIGACT+Symposium+Principles+Programming+Languages+POPL+York+http++}]%
\mdSpan[class={bibitem-before}]{[\mdSpan[class={bibitem-label}]{15}]{\mdNbsp}{\mdNbsp}}\mdSpan[class={bibinfo},field={author}]{Ohad Kammar} {\&}
  \mdSpan[class={bibinfo},field={author}]{Gordon{\mdNbsp}D. Plotkin}
  (\mdSpan[class={bibinfo},field={year}]{2012}): \mdEm{\mdSpan[class={bibinfo},field={title}]{Algebraic Foundations for
  Effect-dependent Optimisations}}.
\mdSpan[class={newblock}]{} In: \mdSpan[font-style={oblique}]{ \mdSpan[class={bibinfo},field={booktitle}]{Proceedings of the 39th Annual ACM
  SIGPLAN-SIGACT Symposium on Principles of Programming Languages}},
  \mdSpan[class={bibinfo},field={series}]{POPL {'}12}, \mdSpan[class={bibinfo},field={publisher}]{ACM}, \mdSpan[class={bibinfo},field={address}]{New
  York, NY, USA}, pp. \mdSpan[class={bibinfo},field={pages}]{349{\textendash}360}, doi:\mdA{http://dx.doi.org/10.1145/2103656.2103698}{}{10.1145/2103656.2103698}.%
\end{mdBibitem}%
\begin{mdBibitem}[class={bibitem},id=launchbury:monadstate,label={[16]\{.bibitem-label\}},elem={bibitem},cite-label={[16]\{.bibitem-label\}},caption={[John Launchbury]\{ .bibinfo field={'}author{'}\} {\&} [Amr Sabry]\{ .bibinfo field={'}author{'}\} ([1997]\{ .bibinfo field={'}year{'}\}): \_[Monadic State: Axiomatization and Type Safety]\{ .bibinfo field={'}title{'}\}\_. \\In: [ [ICFP{'}97]\{ .bibinfo field={'}booktitle{'}\}]\{font-style=oblique\}, pp. [227--238]\{ .bibinfo field={'}pages{'}\}, doi:[10.1145/258948.258970](http://dx.doi.org/10.1145/258948.258970).},searchterm={+John+Launchbury+Sabry+Monadic+State+Axiomatization+Type+Safety+ICFP+http++}]%
\mdSpan[class={bibitem-before}]{[\mdSpan[class={bibitem-label}]{16}]{\mdNbsp}{\mdNbsp}}\mdSpan[class={bibinfo},field={author}]{John Launchbury} {\&}
  \mdSpan[class={bibinfo},field={author}]{Amr Sabry} (\mdSpan[class={bibinfo},field={year}]{1997}):
  \mdEm{\mdSpan[class={bibinfo},field={title}]{Monadic State: Axiomatization and Type Safety}}.
\mdSpan[class={newblock}]{} In: \mdSpan[font-style={oblique}]{ \mdSpan[class={bibinfo},field={booktitle}]{ICFP{'}97}}, pp.
  \mdSpan[class={bibinfo},field={pages}]{227{\textendash}238}, doi:\mdA{http://dx.doi.org/10.1145/258948.258970}{}{10.1145/258948.258970}.%
\end{mdBibitem}%
\begin{mdBibitem}[class={bibitem},id=leijen:scopedlabels,label={[17]\{.bibitem-label\}},elem={bibitem},cite-label={[17]\{.bibitem-label\}},caption={[Daan Leijen]\{ .bibinfo field={'}author{'}\} ([2005]\{ .bibinfo field={'}year{'}\}): \_[Extensible records with scoped labels]\{ .bibinfo field={'}title{'}\}\_. \\In: [ [In: Proceedings of the 2005 Symposium on Trends in Functional Programming]\{ .bibinfo field={'}booktitle{'}\}]\{font-style=oblique\}, pp. [297--312]\{ .bibinfo field={'}pages{'}\}.},searchterm={+Daan+Leijen+Extensible+records+with+scoped+labels+Proceedings+Symposium+Trends+Functional+Programming++}]%
\mdSpan[class={bibitem-before}]{[\mdSpan[class={bibitem-label}]{17}]{\mdNbsp}{\mdNbsp}}\mdSpan[class={bibinfo},field={author}]{Daan Leijen} (\mdSpan[class={bibinfo},field={year}]{2005}):
  \mdEm{\mdSpan[class={bibinfo},field={title}]{Extensible records with scoped labels}}.
\mdSpan[class={newblock}]{} In: \mdSpan[font-style={oblique}]{ \mdSpan[class={bibinfo},field={booktitle}]{In: Proceedings of the 2005 Symposium on
  Trends in Functional Programming}}, pp. \mdSpan[class={bibinfo},field={pages}]{297{\textendash}312}.%
\end{mdBibitem}%
\begin{mdBibitem}[class={bibitem},id=koka,label={[18]\{.bibitem-label\}},elem={bibitem},cite-label={[18]\{.bibitem-label\}},caption={[Daan Leijen]\{ .bibinfo field={'}author{'}\} ([2012]\{ .bibinfo field={'}year{'}\}): \_[Try Koka online]\{ .bibinfo field={'}title{'}\}\_. \\[[http://rise4fun.com/koka/tutorial](http://rise4fun.com/koka/tutorial) and [http://koka.codeplex.com](http://koka.codeplex.com)]\{ .bibinfo field={'}note{'}\}.},searchterm={+Daan+Leijen+Koka+online+http+rise4fun+koka+tutorial+http+rise4fun+koka+tutorial+http+koka+codeplex+http+koka+codeplex++}]%
\mdSpan[class={bibitem-before}]{[\mdSpan[class={bibitem-label}]{18}]{\mdNbsp}{\mdNbsp}}\mdSpan[class={bibinfo},field={author}]{Daan Leijen} (\mdSpan[class={bibinfo},field={year}]{2012}):
  \mdEm{\mdSpan[class={bibinfo},field={title}]{Try Koka online}}.
\mdSpan[class={newblock}]{} \mdSpan[class={bibinfo},field={note}]{\mdA{http://rise4fun.com/koka/tutorial}{}{http://rise4fun.com/koka/tutorial} and
{\mdNbsp}\mdA{http://koka.codeplex.com}{}{http://koka.codeplex.com}}.%
\end{mdBibitem}%
\begin{mdBibitem}[class={bibitem},id=leijen:kokatr,label={[19]\{.bibitem-label\}},elem={bibitem},cite-label={[19]\{.bibitem-label\}},caption={[Daan Leijen]\{ .bibinfo field={'}author{'}\} ([2013]\{ .bibinfo field={'}year{'}\}): \_[Koka: Programming with Row-Polymorphic Effect Types]\{ .bibinfo field={'}title{'}\}\_. \\[Technical Report]\{ .bibinfo field={'}type{'}\} [MSR-TR-2013-79]\{ .bibinfo field={'}number{'}\}, [Microsoft Research]\{ .bibinfo field={'}institution{'}\}.},searchterm={+Daan+Leijen+Koka+Programming+with+Polymorphic+Effect+Types+Technical+Report+Microsoft+Research++}]%
\mdSpan[class={bibitem-before}]{[\mdSpan[class={bibitem-label}]{19}]{\mdNbsp}{\mdNbsp}}\mdSpan[class={bibinfo},field={author}]{Daan Leijen} (\mdSpan[class={bibinfo},field={year}]{2013}):
  \mdEm{\mdSpan[class={bibinfo},field={title}]{Koka: Programming with Row-Polymorphic Effect
  Types}}.
\mdSpan[class={newblock}]{} \mdSpan[class={bibinfo},field={type}]{Technical Report} \mdSpan[class={bibinfo},field={number}]{MSR-TR-2013-79},
  \mdSpan[class={bibinfo},field={institution}]{Microsoft Research}.%
\end{mdBibitem}%
\begin{mdBibitem}[class={bibitem},id=madoko,label={[20]\{.bibitem-label\}},elem={bibitem},cite-label={[20]\{.bibitem-label\}},caption={[Daan Leijen]\{ .bibinfo field={'}author{'}\} ([2014]\{ .bibinfo field={'}year{'}\}): \_[Madoko: A Scholarly Markdown processor]\{ .bibinfo field={'}title{'}\}\_. \\[[http://madoko.codeplex.com](http://madoko.codeplex.com)]\{ .bibinfo field={'}note{'}\}.},searchterm={+Daan+Leijen+Madoko+Scholarly+Markdown+processor+http+madoko+codeplex+http+madoko+codeplex++}]%
\mdSpan[class={bibitem-before}]{[\mdSpan[class={bibitem-label}]{20}]{\mdNbsp}{\mdNbsp}}\mdSpan[class={bibinfo},field={author}]{Daan Leijen} (\mdSpan[class={bibinfo},field={year}]{2014}):
  \mdEm{\mdSpan[class={bibinfo},field={title}]{Madoko: A Scholarly Markdown processor}}.
\mdSpan[class={newblock}]{} \mdSpan[class={bibinfo},field={note}]{\mdA{http://madoko.codeplex.com}{}{http://madoko.codeplex.com}}.%
\end{mdBibitem}%
\begin{mdBibitem}[class={bibitem},id=leroy:valuerestriction,label={[21]\{.bibitem-label\}},elem={bibitem},cite-label={[21]\{.bibitem-label\}},caption={[Xavier Leroy]\{ .bibinfo field={'}author{'}\} ([1993]\{ .bibinfo field={'}year{'}\}): \_[Polymorphism by name for references and continuations]\{ .bibinfo field={'}title{'}\}\_. \\In: [ [POPL {'}93: Proc. of the 20th ACM SIGPLAN-SIGACT symposium on Principles of programming languages]\{ .bibinfo field={'}booktitle{'}\}]\{font-style=oblique\}, pp. [220--231]\{ .bibinfo field={'}pages{'}\}, doi:[10.1145/158511.158632](http://dx.doi.org/10.1145/158511.158632).},searchterm={+Xavier+Leroy+Polymorphism+name+references+continuations+POPL+Proc+SIGPLAN+SIGACT+symposium+Principles+programming+languages+http++}]%
\mdSpan[class={bibitem-before}]{[\mdSpan[class={bibitem-label}]{21}]{\mdNbsp}{\mdNbsp}}\mdSpan[class={bibinfo},field={author}]{Xavier Leroy} (\mdSpan[class={bibinfo},field={year}]{1993}):
  \mdEm{\mdSpan[class={bibinfo},field={title}]{Polymorphism by name for references and
  continuations}}.
\mdSpan[class={newblock}]{} In: \mdSpan[font-style={oblique}]{ \mdSpan[class={bibinfo},field={booktitle}]{POPL {'}93: Proc. of the 20th ACM
  SIGPLAN-SIGACT symposium on Principles of programming languages}}, pp.
  \mdSpan[class={bibinfo},field={pages}]{220{\textendash}231}, doi:\mdA{http://dx.doi.org/10.1145/158511.158632}{}{10.1145/158511.158632}.%
\end{mdBibitem}%
\begin{mdBibitem}[class={bibitem},id=lillibridge:exncallcc,label={[22]\{.bibitem-label\}},elem={bibitem},cite-label={[22]\{.bibitem-label\}},caption={[Mark Lillibridge]\{ .bibinfo field={'}author{'}\} ([1999]\{ .bibinfo field={'}year{'}\}): \_[Unchecked Exceptions Can Be Strictly More Powerful Than Call/CC]\{ .bibinfo field={'}title{'}\}\_. \\[ [Higher-Order and Symbolic Computation]\{ .bibinfo field={'}journal{'}\}]\{font-style=oblique\} [12]\{ .bibinfo field={'}volume{'}\}([1]\{ .bibinfo field={'}number{'}\}), pp. [75--104]\{ .bibinfo field={'}pages{'}\}, doi:[10.1023/A:1010020917337](http://dx.doi.org/10.1023/A:1010020917337).},searchterm={+Mark+Lillibridge+Unchecked+Exceptions+Strictly+More+Powerful+Than+Call+Higher+Order+Symbolic+Computation+http++}]%
\mdSpan[class={bibitem-before}]{[\mdSpan[class={bibitem-label}]{22}]{\mdNbsp}{\mdNbsp}}\mdSpan[class={bibinfo},field={author}]{Mark Lillibridge}
  (\mdSpan[class={bibinfo},field={year}]{1999}): \mdEm{\mdSpan[class={bibinfo},field={title}]{Unchecked Exceptions Can Be
  Strictly More Powerful Than Call/CC}}.
\mdSpan[class={newblock}]{} \mdSpan[font-style={oblique}]{ \mdSpan[class={bibinfo},field={journal}]{Higher-Order and Symbolic Computation}}
  \mdSpan[class={bibinfo},field={volume}]{12}(\mdSpan[class={bibinfo},field={number}]{1}), pp. \mdSpan[class={bibinfo},field={pages}]{75{\textendash}104},
  doi:\mdA{http://dx.doi.org/10.1023/A:1010020917337}{}{10.1023/A:1010020917337}.%
\end{mdBibitem}%
\begin{mdBibitem}[class={bibitem},id=lindley:effects,label={[23]\{.bibitem-label\}},elem={bibitem},cite-label={[23]\{.bibitem-label\}},caption={[Sam Lindley]\{ .bibinfo field={'}author{'}\} {\&} [James Cheney]\{ .bibinfo field={'}author{'}\} ([2012]\{ .bibinfo field={'}year{'}\}): \_[Row-based effect types for database integration]\{ .bibinfo field={'}title{'}\}\_. \\In: [ [TLDI{'}12]\{ .bibinfo field={'}booktitle{'}\}]\{font-style=oblique\}, pp. [91--102]\{ .bibinfo field={'}pages{'}\}, doi:[10.1145/2103786.2103798](http://dx.doi.org/10.1145/2103786.2103798).},searchterm={+Lindley+James+Cheney+based+effect+types+database+integration+TLDI+http++}]%
\mdSpan[class={bibitem-before}]{[\mdSpan[class={bibitem-label}]{23}]{\mdNbsp}{\mdNbsp}}\mdSpan[class={bibinfo},field={author}]{Sam Lindley} {\&}
  \mdSpan[class={bibinfo},field={author}]{James Cheney}
  (\mdSpan[class={bibinfo},field={year}]{2012}): \mdEm{\mdSpan[class={bibinfo},field={title}]{Row-based effect types for
  database integration}}.
\mdSpan[class={newblock}]{} In: \mdSpan[font-style={oblique}]{ \mdSpan[class={bibinfo},field={booktitle}]{TLDI{'}12}}, pp. \mdSpan[class={bibinfo},field={pages}]{91{\textendash}102},
  doi:\mdA{http://dx.doi.org/10.1145/2103786.2103798}{}{10.1145/2103786.2103798}.%
\end{mdBibitem}%
\begin{mdBibitem}[class={bibitem},id=lucassen:polyeffect,label={[24]\{.bibitem-label\}},elem={bibitem},cite-label={[24]\{.bibitem-label\}},caption={[J.{\mdNbsp}M. Lucassen]\{ .bibinfo field={'}author{'}\} {\&} [D.{\mdNbsp}K. Gifford]\{ .bibinfo field={'}author{'}\} ([1988]\{ .bibinfo field={'}year{'}\}): \_[Polymorphic effect systems]\{ .bibinfo field={'}title{'}\}\_. \\In: [ [POPL {'}88]\{ .bibinfo field={'}booktitle{'}\}]\{font-style=oblique\}, pp. [47--57]\{ .bibinfo field={'}pages{'}\}, doi:[10.1145/73560.73564s](http://dx.doi.org/10.1145/73560.73564s).},searchterm={+Lucassen+Gifford+Polymorphic+effect+systems+POPL+http++}]%
\mdSpan[class={bibitem-before}]{[\mdSpan[class={bibitem-label}]{24}]{\mdNbsp}{\mdNbsp}}\mdSpan[class={bibinfo},field={author}]{J.{\mdNbsp}M. Lucassen} {\&}
  \mdSpan[class={bibinfo},field={author}]{D.{\mdNbsp}K. Gifford}
  (\mdSpan[class={bibinfo},field={year}]{1988}): \mdEm{\mdSpan[class={bibinfo},field={title}]{Polymorphic effect systems}}.
\mdSpan[class={newblock}]{} In: \mdSpan[font-style={oblique}]{ \mdSpan[class={bibinfo},field={booktitle}]{POPL {'}88}}, pp. \mdSpan[class={bibinfo},field={pages}]{47{\textendash}57},
  doi:\mdA{http://dx.doi.org/10.1145/73560.73564s}{}{10.1145/73560.73564s}.%
\end{mdBibitem}%
\begin{mdBibitem}[class={bibitem},id=marino:genericeffects,label={[25]\{.bibitem-label\}},elem={bibitem},cite-label={[25]\{.bibitem-label\}},caption={[Daniel Marino]\{ .bibinfo field={'}author{'}\} {\&} [Todd Millstein]\{ .bibinfo field={'}author{'}\} ([2009]\{ .bibinfo field={'}year{'}\}): \_[A generic type-and-effect system]\{ .bibinfo field={'}title{'}\}\_. \\In: [ [TLDI {'}09: Proceedings of the 4th international workshop on Types in language design and implementation]\{ .bibinfo field={'}booktitle{'}\}]\{font-style=oblique\}, pp. [39--50]\{ .bibinfo field={'}pages{'}\}, doi:[10.1145/1481861.1481868](http://dx.doi.org/10.1145/1481861.1481868).},searchterm={+Daniel+Marino+Todd+Millstein+generic+type+effect+system+TLDI+Proceedings+international+workshop+Types+language+design+implementation+http++}]%
\mdSpan[class={bibitem-before}]{[\mdSpan[class={bibitem-label}]{25}]{\mdNbsp}{\mdNbsp}}\mdSpan[class={bibinfo},field={author}]{Daniel Marino} {\&}
  \mdSpan[class={bibinfo},field={author}]{Todd Millstein}
  (\mdSpan[class={bibinfo},field={year}]{2009}): \mdEm{\mdSpan[class={bibinfo},field={title}]{A generic type-and-effect
  system}}.
\mdSpan[class={newblock}]{} In: \mdSpan[font-style={oblique}]{ \mdSpan[class={bibinfo},field={booktitle}]{TLDI {'}09: Proceedings of the 4th
  international workshop on Types in language design and implementation}}, pp.
  \mdSpan[class={bibinfo},field={pages}]{39{\textendash}50}, doi:\mdA{http://dx.doi.org/10.1145/1481861.1481868}{}{10.1145/1481861.1481868}.%
\end{mdBibitem}%
\begin{mdBibitem}[class={bibitem},id=milner:types,label={[26]\{.bibitem-label\}},elem={bibitem},cite-label={[26]\{.bibitem-label\}},caption={[Robin Milner]\{ .bibinfo field={'}author{'}\} ([1978]\{ .bibinfo field={'}year{'}\}): \_[A theory of type polymorphism in programming]\{ .bibinfo field={'}title{'}\}\_. \\[ [Journal of Computer and System Sciences]\{ .bibinfo field={'}journal{'}\}]\{font-style=oblique\} [17]\{ .bibinfo field={'}volume{'}\}, pp. [248--375]\{ .bibinfo field={'}pages{'}\}, doi:[10.1016/0022-007890014-4](http://dx.doi.org/10.1016/0022-007890014-4).},searchterm={+Robin+Milner+theory+type+polymorphism+programming+Journal+Computer+System+Sciences+http++}]%
\mdSpan[class={bibitem-before}]{[\mdSpan[class={bibitem-label}]{26}]{\mdNbsp}{\mdNbsp}}\mdSpan[class={bibinfo},field={author}]{Robin Milner} (\mdSpan[class={bibinfo},field={year}]{1978}):
  \mdEm{\mdSpan[class={bibinfo},field={title}]{A theory of type polymorphism in programming}}.
\mdSpan[class={newblock}]{} \mdSpan[font-style={oblique}]{ \mdSpan[class={bibinfo},field={journal}]{Journal of Computer and System Sciences}}
  \mdSpan[class={bibinfo},field={volume}]{17}, pp. \mdSpan[class={bibinfo},field={pages}]{248{\textendash}375},
  doi:\mdA{http://dx.doi.org/10.1016/0022-007890014-4}{}{10.1016/0022-007890014-4}.%
\end{mdBibitem}%
\begin{mdBibitem}[class={bibitem},id=naumann:purity,label={[27]\{.bibitem-label\}},elem={bibitem},cite-label={[27]\{.bibitem-label\}},caption={[David{\mdNbsp}A. Naumann]\{ .bibinfo field={'}author{'}\} ([2007]\{ .bibinfo field={'}year{'}\}): \_[Observational purity and encapsulation]\{ .bibinfo field={'}title{'}\}\_. \\[ [Theor. Comput. Sci.]\{ .bibinfo field={'}journal{'}\}]\{font-style=oblique\} [376]\{ .bibinfo field={'}volume{'}\}([3]\{ .bibinfo field={'}number{'}\}), pp. [205--224]\{ .bibinfo field={'}pages{'}\}, doi:[10.1007/978-3-540-31984-9\_15](http://dx.doi.org/10.1007/978-3-540-31984-9\_15).},searchterm={+David+Naumann+Observational+purity+encapsulation+Theor+Comput+http++}]%
\mdSpan[class={bibitem-before}]{[\mdSpan[class={bibitem-label}]{27}]{\mdNbsp}{\mdNbsp}}\mdSpan[class={bibinfo},field={author}]{David{\mdNbsp}A. Naumann}
  (\mdSpan[class={bibinfo},field={year}]{2007}): \mdEm{\mdSpan[class={bibinfo},field={title}]{Observational purity and
  encapsulation}}.
\mdSpan[class={newblock}]{} \mdSpan[font-style={oblique}]{ \mdSpan[class={bibinfo},field={journal}]{Theor. Comput. Sci.}}
  \mdSpan[class={bibinfo},field={volume}]{376}(\mdSpan[class={bibinfo},field={number}]{3}), pp. \mdSpan[class={bibinfo},field={pages}]{205{\textendash}224},
  doi:\mdA{http://dx.doi.org/10.1007/978-3-540-31984-9\_15}{}{10.1007/978-3-540-31984-9\_15}.%
\end{mdBibitem}%
\begin{mdBibitem}[class={bibitem},id=nielson:polyeffect,label={[28]\{.bibitem-label\}},elem={bibitem},cite-label={[28]\{.bibitem-label\}},caption={[Hanne{\mdNbsp}Riis Nielson]\{ .bibinfo field={'}author{'}\}, [Flemming Nielson]\{ .bibinfo field={'}author{'}\} {\&} [Torben Amtoft]\{ .bibinfo field={'}author{'}\} ([1997]\{ .bibinfo field={'}year{'}\}): \_[Polymorphic Subtyping for Effect Analysis: The Static Semantics]\{ .bibinfo field={'}title{'}\}\_. \\In: [ [Selected papers from the 5th LOMAPS Workshop on Analysis and Verification of Multiple-Agent Languages]\{ .bibinfo field={'}booktitle{'}\}]\{font-style=oblique\}, pp. [141--171]\{ .bibinfo field={'}pages{'}\}, doi:[10.1007/3-540-62503-8\_8](http://dx.doi.org/10.1007/3-540-62503-8\_8).},searchterm={+Hanne+Riis+Nielson+Flemming+Nielson+Torben+Amtoft+Polymorphic+Subtyping+Effect+Analysis+Static+Semantics+Selected+papers+from+LOMAPS+Workshop+Analysis+Verification+Multiple+Agent+Languages+http++}]%
\mdSpan[class={bibitem-before}]{[\mdSpan[class={bibitem-label}]{28}]{\mdNbsp}{\mdNbsp}}\mdSpan[class={bibinfo},field={author}]{Hanne{\mdNbsp}Riis Nielson},
  \mdSpan[class={bibinfo},field={author}]{Flemming Nielson} {\&}
  \mdSpan[class={bibinfo},field={author}]{Torben Amtoft}
  (\mdSpan[class={bibinfo},field={year}]{1997}): \mdEm{\mdSpan[class={bibinfo},field={title}]{Polymorphic Subtyping for
  Effect Analysis: The Static Semantics}}.
\mdSpan[class={newblock}]{} In: \mdSpan[font-style={oblique}]{ \mdSpan[class={bibinfo},field={booktitle}]{Selected papers from the 5th LOMAPS
  Workshop on Analysis and Verification of Multiple-Agent Languages}}, pp.
  \mdSpan[class={bibinfo},field={pages}]{141{\textendash}171}, doi:\mdA{http://dx.doi.org/10.1007/3-540-62503-8\_8}{}{10.1007/3-540-62503-8\_8}.%
\end{mdBibitem}%
\begin{mdBibitem}[class={bibitem},id=pessaux:exceptions,label={[29]\{.bibitem-label\}},elem={bibitem},cite-label={[29]\{.bibitem-label\}},caption={[Fran{\c{c}}ois Pessaux]\{ .bibinfo field={'}author{'}\} {\&} [Xavier Leroy]\{ .bibinfo field={'}author{'}\} ([1999]\{ .bibinfo field={'}year{'}\}): \_[Type-based analysis of uncaught exceptions]\{ .bibinfo field={'}title{'}\}\_. \\In: [ [POPL {'}99]\{ .bibinfo field={'}booktitle{'}\}]\{font-style=oblique\}, pp. [276--290]\{ .bibinfo field={'}pages{'}\}, doi:[10.1145/292540.292565](http://dx.doi.org/10.1145/292540.292565).},searchterm={+Fran+Pessaux+Xavier+Leroy+Type+based+analysis+uncaught+exceptions+POPL+http++}]%
\mdSpan[class={bibitem-before}]{[\mdSpan[class={bibitem-label}]{29}]{\mdNbsp}{\mdNbsp}}\mdSpan[class={bibinfo},field={author}]{Fran{\c{c}}ois Pessaux} {\&}
  \mdSpan[class={bibinfo},field={author}]{Xavier Leroy}
  (\mdSpan[class={bibinfo},field={year}]{1999}): \mdEm{\mdSpan[class={bibinfo},field={title}]{Type-based analysis of uncaught
  exceptions}}.
\mdSpan[class={newblock}]{} In: \mdSpan[font-style={oblique}]{ \mdSpan[class={bibinfo},field={booktitle}]{POPL {'}99}}, pp.
  \mdSpan[class={bibinfo},field={pages}]{276{\textendash}290}, doi:\mdA{http://dx.doi.org/10.1145/292540.292565}{}{10.1145/292540.292565}.%
\end{mdBibitem}%
\begin{mdBibitem}[class={bibitem},id=stateinhaskell,label={[30]\{.bibitem-label\}},elem={bibitem},cite-label={[30]\{.bibitem-label\}},caption={[Simon{\mdNbsp}L Peyton{\mdNbsp}Jones]\{ .bibinfo field={'}author{'}\} {\&} [John Launchbury]\{ .bibinfo field={'}author{'}\} ([1995]\{ .bibinfo field={'}year{'}\}): \_[State in Haskell]\{ .bibinfo field={'}title{'}\}\_. \\[ [Lisp and Symbolic Comp.]\{ .bibinfo field={'}journal{'}\}]\{font-style=oblique\} [8]\{ .bibinfo field={'}volume{'}\}([4]\{ .bibinfo field={'}number{'}\}), pp. [293--341]\{ .bibinfo field={'}pages{'}\}, doi:[10.1007/BF01018827](http://dx.doi.org/10.1007/BF01018827).},searchterm={+Simon+Peyton+Jones+John+Launchbury+State+Haskell+Lisp+Symbolic+Comp+BF01018827+http+BF01018827++}]%
\mdSpan[class={bibitem-before}]{[\mdSpan[class={bibitem-label}]{30}]{\mdNbsp}{\mdNbsp}}\mdSpan[class={bibinfo},field={author}]{Simon{\mdNbsp}L Peyton{\mdNbsp}Jones} {\&}
  \mdSpan[class={bibinfo},field={author}]{John Launchbury}
  (\mdSpan[class={bibinfo},field={year}]{1995}): \mdEm{\mdSpan[class={bibinfo},field={title}]{State in Haskell}}.
\mdSpan[class={newblock}]{} \mdSpan[font-style={oblique}]{ \mdSpan[class={bibinfo},field={journal}]{Lisp and Symbolic Comp.}}
  \mdSpan[class={bibinfo},field={volume}]{8}(\mdSpan[class={bibinfo},field={number}]{4}), pp. \mdSpan[class={bibinfo},field={pages}]{293{\textendash}341},
  doi:\mdA{http://dx.doi.org/10.1007/BF01018827}{}{10.1007/BF01018827}.%
\end{mdBibitem}%
\begin{mdBibitem}[class={bibitem},id=remy:mlart,label={[31]\{.bibitem-label\}},elem={bibitem},cite-label={[31]\{.bibitem-label\}},caption={[Didier Remy]\{ .bibinfo field={'}author{'}\} ([1994]\{ .bibinfo field={'}year{'}\}): \_[Programming Objects with ML-ART, an Extension to ML with Abstract and Record Types]\{ .bibinfo field={'}title{'}\}\_. \\In: [ [TACS {'}94: Proc. Int. Conf. on Theoretical Aspects of Computer Software]\{ .bibinfo field={'}booktitle{'}\}]\{font-style=oblique\}, pp. [321--346]\{ .bibinfo field={'}pages{'}\}, doi:[10.1007/3-540-57887-0\_102](http://dx.doi.org/10.1007/3-540-57887-0\_102).},searchterm={+Didier+Remy+Programming+Objects+with+Extension+with+Abstract+Record+Types+TACS+Proc+Conf+Theoretical+Aspects+Computer+Software+http++}]%
\mdSpan[class={bibitem-before}]{[\mdSpan[class={bibitem-label}]{31}]{\mdNbsp}{\mdNbsp}}\mdSpan[class={bibinfo},field={author}]{Didier Remy} (\mdSpan[class={bibinfo},field={year}]{1994}):
  \mdEm{\mdSpan[class={bibinfo},field={title}]{Programming Objects with ML-ART, an Extension to ML
  with Abstract and Record Types}}.
\mdSpan[class={newblock}]{} In: \mdSpan[font-style={oblique}]{ \mdSpan[class={bibinfo},field={booktitle}]{TACS {'}94: Proc. Int. Conf. on
  Theoretical Aspects of Computer Software}}, pp. \mdSpan[class={bibinfo},field={pages}]{321{\textendash}346},
  doi:\mdA{http://dx.doi.org/10.1007/3-540-57887-0\_102}{}{10.1007/3-540-57887-0\_102}.%
\end{mdBibitem}%
\begin{mdBibitem}[class={bibitem},id=remy:records,label={[32]\{.bibitem-label\}},elem={bibitem},cite-label={[32]\{.bibitem-label\}},caption={[Didier R{\'{e}}my]\{ .bibinfo field={'}author{'}\} ([1994]\{ .bibinfo field={'}year{'}\}): \_[Theoretical Aspects of Object-oriented Programming]\{ .bibinfo field={'}title{'}\}\_. \\chapter [Type Inference for Records in Natural Extension of ML]\{ .bibinfo field={'}chapter{'}\}, [MIT Press]\{ .bibinfo field={'}publisher{'}\}, [Cambridge, MA, USA]\{ .bibinfo field={'}address{'}\}, pp. [67--95]\{ .bibinfo field={'}pages{'}\}. \\Available at [http://dl.acm.org/citation.cfm?id=186677.186689](http://dl.acm.org/citation.cfm?id=186677.186689).},searchterm={+Didier+Theoretical+Aspects+Object+oriented+Programming+chapter+Type+Inference+Records+Natural+Extension+Press+Cambridge+Available+http+citation+http+citation++}]%
\mdSpan[class={bibitem-before}]{[\mdSpan[class={bibitem-label}]{32}]{\mdNbsp}{\mdNbsp}}\mdSpan[class={bibinfo},field={author}]{Didier R{\'{e}}my}
  (\mdSpan[class={bibinfo},field={year}]{1994}): \mdEm{\mdSpan[class={bibinfo},field={title}]{Theoretical Aspects of
  Object-oriented Programming}}.
\mdSpan[class={newblock}]{} chapter \mdSpan[class={bibinfo},field={chapter}]{Type Inference for Records in Natural
  Extension of ML}, \mdSpan[class={bibinfo},field={publisher}]{MIT Press},
  \mdSpan[class={bibinfo},field={address}]{Cambridge, MA, USA}, pp. \mdSpan[class={bibinfo},field={pages}]{67{\textendash}95}.
\mdSpan[class={newblock}]{} Available at{\mdNbsp}\mdA{http://dl.acm.org/citation.cfm?id=186677.186689}{}{http://dl.acm.org/citation.cfm?id=186677.186689}.%
\end{mdBibitem}%
\begin{mdBibitem}[class={bibitem},id=scala:effects,label={[33]\{.bibitem-label\}},elem={bibitem},cite-label={[33]\{.bibitem-label\}},caption={[Lukas Rytz]\{ .bibinfo field={'}author{'}\}, [Martin Odersky]\{ .bibinfo field={'}author{'}\} {\&} [Philipp Haller]\{ .bibinfo field={'}author{'}\} ([2012]\{ .bibinfo field={'}year{'}\}): \_[Lightweight Polymorphic Effects]\{ .bibinfo field={'}title{'}\}\_. \\In: [ [Proceedings of the 26th European Conference on Object-Oriented Programming]\{ .bibinfo field={'}booktitle{'}\}]\{font-style=oblique\}, [ECOOP{'}12]\{ .bibinfo field={'}series{'}\}, [Springer-Verlag]\{ .bibinfo field={'}publisher{'}\}, [Berlin, Heidelberg]\{ .bibinfo field={'}address{'}\}, pp. [258--282]\{ .bibinfo field={'}pages{'}\}, doi:[10.1007/978-3-642-31057-7\_13](http://dx.doi.org/10.1007/978-3-642-31057-7\_13).},searchterm={+Lukas+Rytz+Martin+Odersky+Philipp+Haller+Lightweight+Polymorphic+Effects+Proceedings+European+Conference+Object+Oriented+Programming+ECOOP+Springer+Verlag+Berlin+Heidelberg+http++}]%
\mdSpan[class={bibitem-before}]{[\mdSpan[class={bibitem-label}]{33}]{\mdNbsp}{\mdNbsp}}\mdSpan[class={bibinfo},field={author}]{Lukas Rytz}, \mdSpan[class={bibinfo},field={author}]{Martin
  Odersky} {\&} \mdSpan[class={bibinfo},field={author}]{Philipp Haller} (\mdSpan[class={bibinfo},field={year}]{2012}): \mdEm{\mdSpan[class={bibinfo},field={title}]{Lightweight
  Polymorphic Effects}}.
\mdSpan[class={newblock}]{} In: \mdSpan[font-style={oblique}]{ \mdSpan[class={bibinfo},field={booktitle}]{Proceedings of the 26th European
  Conference on Object-Oriented Programming}}, \mdSpan[class={bibinfo},field={series}]{ECOOP{'}12},
  \mdSpan[class={bibinfo},field={publisher}]{Springer-Verlag}, \mdSpan[class={bibinfo},field={address}]{Berlin, Heidelberg},
  pp. \mdSpan[class={bibinfo},field={pages}]{258{\textendash}282}, doi:\mdA{http://dx.doi.org/10.1007/978-3-642-31057-7\_13}{}{10.1007/978-3-642-31057-7\_13}.%
\end{mdBibitem}%
\begin{mdBibitem}[class={bibitem},id=sulzmann:records,label={[34]\{.bibitem-label\}},elem={bibitem},cite-label={[34]\{.bibitem-label\}},caption={[Martin Sulzmann]\{ .bibinfo field={'}author{'}\} ([1997]\{ .bibinfo field={'}year{'}\}): \_[Designing record systems]\{ .bibinfo field={'}title{'}\}\_. \\[Technical Report]\{ .bibinfo field={'}type{'}\} [YALEU/DCS/RR-1128]\{ .bibinfo field={'}number{'}\}, [Yale University]\{ .bibinfo field={'}institution{'}\}.},searchterm={+Martin+Sulzmann+Designing+record+systems+Technical+Report+YALEU+Yale+University++}]%
\mdSpan[class={bibitem-before}]{[\mdSpan[class={bibitem-label}]{34}]{\mdNbsp}{\mdNbsp}}\mdSpan[class={bibinfo},field={author}]{Martin Sulzmann}
  (\mdSpan[class={bibinfo},field={year}]{1997}): \mdEm{\mdSpan[class={bibinfo},field={title}]{Designing record systems}}.
\mdSpan[class={newblock}]{} \mdSpan[class={bibinfo},field={type}]{Technical Report} \mdSpan[class={bibinfo},field={number}]{YALEU/DCS/RR-1128},
  \mdSpan[class={bibinfo},field={institution}]{Yale University}.%
\end{mdBibitem}%
\begin{mdBibitem}[class={bibitem},id=sulzmann:recordsrev,label={[35]\{.bibitem-label\}},elem={bibitem},cite-label={[35]\{.bibitem-label\}},caption={[Martin Sulzmann]\{ .bibinfo field={'}author{'}\} ([1998]\{ .bibinfo field={'}year{'}\}): \_[Type systems for records revisited]\{ .bibinfo field={'}title{'}\}\_. \\[Unpublished]\{ .bibinfo field={'}note{'}\}.},searchterm={+Martin+Sulzmann+Type+systems+records+revisited+Unpublished++}]%
\mdSpan[class={bibitem-before}]{[\mdSpan[class={bibitem-label}]{35}]{\mdNbsp}{\mdNbsp}}\mdSpan[class={bibinfo},field={author}]{Martin Sulzmann}
  (\mdSpan[class={bibinfo},field={year}]{1998}): \mdEm{\mdSpan[class={bibinfo},field={title}]{Type systems for records
  revisited}}.
\mdSpan[class={newblock}]{} \mdSpan[class={bibinfo},field={note}]{Unpublished}.%
\end{mdBibitem}%
\begin{mdBibitem}[class={bibitem},id=talpin:effects,label={[36]\{.bibitem-label\}},elem={bibitem},cite-label={[36]\{.bibitem-label\}},caption={[Jean-Pierre Talpin]\{ .bibinfo field={'}author{'}\} {\&} [Pierre Jouvelot]\{ .bibinfo field={'}author{'}\} ([1994]\{ .bibinfo field={'}year{'}\}): \_[The type and effect discipline]\{ .bibinfo field={'}title{'}\}\_. \\[ [Inf. Comput.]\{ .bibinfo field={'}journal{'}\}]\{font-style=oblique\} [111]\{ .bibinfo field={'}volume{'}\}([2]\{ .bibinfo field={'}number{'}\}), pp. [245--296]\{ .bibinfo field={'}pages{'}\}, doi:[10.1006/inco.1994.1046](http://dx.doi.org/10.1006/inco.1994.1046).},searchterm={+Jean+Pierre+Talpin+Pierre+Jouvelot+type+effect+discipline+Comput+inco+http+inco++}]%
\mdSpan[class={bibitem-before}]{[\mdSpan[class={bibitem-label}]{36}]{\mdNbsp}{\mdNbsp}}\mdSpan[class={bibinfo},field={author}]{Jean-Pierre Talpin} {\&}
  \mdSpan[class={bibinfo},field={author}]{Pierre Jouvelot}
  (\mdSpan[class={bibinfo},field={year}]{1994}): \mdEm{\mdSpan[class={bibinfo},field={title}]{The type and effect
  discipline}}.
\mdSpan[class={newblock}]{} \mdSpan[font-style={oblique}]{ \mdSpan[class={bibinfo},field={journal}]{Inf. Comput.}}
  \mdSpan[class={bibinfo},field={volume}]{111}(\mdSpan[class={bibinfo},field={number}]{2}), pp. \mdSpan[class={bibinfo},field={pages}]{245{\textendash}296},
  doi:\mdA{http://dx.doi.org/10.1006/inco.1994.1046}{}{10.1006/inco.1994.1046}.%
\end{mdBibitem}%
\begin{mdBibitem}[class={bibitem},id=talpin:thesis,label={[37]\{.bibitem-label\}},elem={bibitem},cite-label={[37]\{.bibitem-label\}},caption={[J.P. Talpin]\{ .bibinfo field={'}author{'}\} ([1993]\{ .bibinfo field={'}year{'}\}): \_[Theoretical and practical aspects of type and effect inference]\{ .bibinfo field={'}title{'}\}\_. \\Ph.D. thesis, [Ecole des Mines de Paris and University Paris VI, Paris, France]\{ .bibinfo field={'}school{'}\}.},searchterm={+Talpin+Theoretical+practical+aspects+type+effect+inference+thesis+Ecole+Mines+Paris+University+Paris+Paris+France++}]%
\mdSpan[class={bibitem-before}]{[\mdSpan[class={bibitem-label}]{37}]{\mdNbsp}{\mdNbsp}}\mdSpan[class={bibinfo},field={author}]{J.P. Talpin} (\mdSpan[class={bibinfo},field={year}]{1993}):
  \mdEm{\mdSpan[class={bibinfo},field={title}]{Theoretical and practical aspects of type and effect
  inference}}.
\mdSpan[class={newblock}]{} Ph.D. thesis, \mdSpan[class={bibinfo},field={school}]{Ecole des Mines de Paris and
  University Paris VI, Paris, France}.%
\end{mdBibitem}%
\begin{mdBibitem}[class={bibitem},id=leijen:effects-tr,label={[38]\{.bibitem-label\}},elem={bibitem},cite-label={[38]\{.bibitem-label\}},caption={[Ross Tate]\{ .bibinfo field={'}author{'}\} {\&} [Daan Leijen]\{ .bibinfo field={'}author{'}\} ([2010]\{ .bibinfo field={'}year{'}\}): \_[Convenient Explicit Effects using Type Inference with Subeffects]\{ .bibinfo field={'}title{'}\}\_. \\[Technical Report]\{ .bibinfo field={'}type{'}\} [MSR-TR-2010-80]\{ .bibinfo field={'}number{'}\}, [Microsoft Research]\{ .bibinfo field={'}institution{'}\}.},searchterm={+Ross+Tate+Daan+Leijen+Convenient+Explicit+Effects+using+Type+Inference+with+Subeffects+Technical+Report+Microsoft+Research++}]%
\mdSpan[class={bibitem-before}]{[\mdSpan[class={bibitem-label}]{38}]{\mdNbsp}{\mdNbsp}}\mdSpan[class={bibinfo},field={author}]{Ross Tate} {\&} \mdSpan[class={bibinfo},field={author}]{Daan
  Leijen} (\mdSpan[class={bibinfo},field={year}]{2010}):
  \mdEm{\mdSpan[class={bibinfo},field={title}]{Convenient Explicit Effects using Type Inference with
  Subeffects}}.
\mdSpan[class={newblock}]{} \mdSpan[class={bibinfo},field={type}]{Technical Report} \mdSpan[class={bibinfo},field={number}]{MSR-TR-2010-80},
  \mdSpan[class={bibinfo},field={institution}]{Microsoft Research}.%
\end{mdBibitem}%
\begin{mdBibitem}[class={bibitem},id=tofte:refs,label={[39]\{.bibitem-label\}},elem={bibitem},cite-label={[39]\{.bibitem-label\}},caption={[Mads Tofte]\{ .bibinfo field={'}author{'}\} ([1990]\{ .bibinfo field={'}year{'}\}): \_[Type inference for polymorphic references]\{ .bibinfo field={'}title{'}\}\_. \\[ [Inf. Comput.]\{ .bibinfo field={'}journal{'}\}]\{font-style=oblique\} [89]\{ .bibinfo field={'}volume{'}\}([1]\{ .bibinfo field={'}number{'}\}), pp. [1--34]\{ .bibinfo field={'}pages{'}\}, doi:[10.1016/0890-5401(90)90018-D](http://dx.doi.org/10.1016/0890-5401(90)90018-D).},searchterm={+Mads+Tofte+Type+inference+polymorphic+references+Comput+http++}]%
\mdSpan[class={bibitem-before}]{[\mdSpan[class={bibitem-label}]{39}]{\mdNbsp}{\mdNbsp}}\mdSpan[class={bibinfo},field={author}]{Mads Tofte} (\mdSpan[class={bibinfo},field={year}]{1990}):
  \mdEm{\mdSpan[class={bibinfo},field={title}]{Type inference for polymorphic references}}.
\mdSpan[class={newblock}]{} \mdSpan[font-style={oblique}]{ \mdSpan[class={bibinfo},field={journal}]{Inf. Comput.}}
  \mdSpan[class={bibinfo},field={volume}]{89}(\mdSpan[class={bibinfo},field={number}]{1}), pp. \mdSpan[class={bibinfo},field={pages}]{1{\textendash}34},
  doi:\mdA{http://dx.doi.org/10.1016/0890-5401(90}{}{10.1016/0890-5401(90)90018-D}90018-D).%
\end{mdBibitem}%
\begin{mdBibitem}[class={bibitem},id=tofte:regioninf,label={[40]\{.bibitem-label\}},elem={bibitem},cite-label={[40]\{.bibitem-label\}},caption={[Mads Tofte]\{ .bibinfo field={'}author{'}\} {\&} [Lars Birkedal]\{ .bibinfo field={'}author{'}\} ([1998]\{ .bibinfo field={'}year{'}\}): \_[A region inference algorithm]\{ .bibinfo field={'}title{'}\}\_. \\[ [ACM Trans. Program. Lang. Syst.]\{ .bibinfo field={'}journal{'}\}]\{font-style=oblique\} [20]\{ .bibinfo field={'}volume{'}\}([4]\{ .bibinfo field={'}number{'}\}), pp. [724--767]\{ .bibinfo field={'}pages{'}\}, doi:[10.1145/291891.291894](http://dx.doi.org/10.1145/291891.291894).},searchterm={+Mads+Tofte+Lars+Birkedal+region+inference+algorithm+Trans+Program+Lang+Syst+http++}]%
\mdSpan[class={bibitem-before}]{[\mdSpan[class={bibitem-label}]{40}]{\mdNbsp}{\mdNbsp}}\mdSpan[class={bibinfo},field={author}]{Mads Tofte} {\&} \mdSpan[class={bibinfo},field={author}]{Lars
  Birkedal} (\mdSpan[class={bibinfo},field={year}]{1998}):
  \mdEm{\mdSpan[class={bibinfo},field={title}]{A region inference algorithm}}.
\mdSpan[class={newblock}]{} \mdSpan[font-style={oblique}]{ \mdSpan[class={bibinfo},field={journal}]{ACM Trans. Program. Lang. Syst.}}
  \mdSpan[class={bibinfo},field={volume}]{20}(\mdSpan[class={bibinfo},field={number}]{4}), pp. \mdSpan[class={bibinfo},field={pages}]{724{\textendash}767},
  doi:\mdA{http://dx.doi.org/10.1145/291891.291894}{}{10.1145/291891.291894}.%
\end{mdBibitem}%
\begin{mdBibitem}[class={bibitem},id=tolmach:monadhierarchy,label={[41]\{.bibitem-label\}},elem={bibitem},cite-label={[41]\{.bibitem-label\}},caption={[Andrew{\mdNbsp}P. Tolmach]\{ .bibinfo field={'}author{'}\} ([1998]\{ .bibinfo field={'}year{'}\}): \_[Optimizing ML Using a Hierarchy of Monadic Types]\{ .bibinfo field={'}title{'}\}\_. \\In: [ [TIC {'}98]\{ .bibinfo field={'}booktitle{'}\}]\{font-style=oblique\}, pp. [97--115]\{ .bibinfo field={'}pages{'}\}, doi:[10.1007/BFb0055514](http://dx.doi.org/10.1007/BFb0055514).},searchterm={+Andrew+Tolmach+Optimizing+Using+Hierarchy+Monadic+Types+BFb0055514+http+BFb0055514++}]%
\mdSpan[class={bibitem-before}]{[\mdSpan[class={bibitem-label}]{41}]{\mdNbsp}{\mdNbsp}}\mdSpan[class={bibinfo},field={author}]{Andrew{\mdNbsp}P. Tolmach}
  (\mdSpan[class={bibinfo},field={year}]{1998}): \mdEm{\mdSpan[class={bibinfo},field={title}]{Optimizing ML Using a
  Hierarchy of Monadic Types}}.
\mdSpan[class={newblock}]{} In: \mdSpan[font-style={oblique}]{ \mdSpan[class={bibinfo},field={booktitle}]{TIC {'}98}}, pp. \mdSpan[class={bibinfo},field={pages}]{97{\textendash}115},
  doi:\mdA{http://dx.doi.org/10.1007/BFb0055514}{}{10.1007/BFb0055514}.%
\end{mdBibitem}%
\begin{mdBibitem}[class={bibitem},id=wadler:marriage,label={[42]\{.bibitem-label\}},elem={bibitem},cite-label={[42]\{.bibitem-label\}},caption={[Philip Wadler]\{ .bibinfo field={'}author{'}\} {\&} [Peter Thiemann]\{ .bibinfo field={'}author{'}\} ([2003]\{ .bibinfo field={'}year{'}\}): \_[The marriage of effects and monads]\{ .bibinfo field={'}title{'}\}\_. \\[ [ACM Trans. Comput. Logic]\{ .bibinfo field={'}journal{'}\}]\{font-style=oblique\} [4]\{ .bibinfo field={'}volume{'}\}([1]\{ .bibinfo field={'}number{'}\}), pp. [1--32]\{ .bibinfo field={'}pages{'}\}, doi:[10.1145/601775.601776](http://dx.doi.org/10.1145/601775.601776).},searchterm={+Philip+Wadler+Peter+Thiemann+marriage+effects+monads+Trans+Comput+Logic+http++}]%
\mdSpan[class={bibitem-before}]{[\mdSpan[class={bibitem-label}]{42}]{\mdNbsp}{\mdNbsp}}\mdSpan[class={bibinfo},field={author}]{Philip Wadler} {\&}
  \mdSpan[class={bibinfo},field={author}]{Peter Thiemann}
  (\mdSpan[class={bibinfo},field={year}]{2003}): \mdEm{\mdSpan[class={bibinfo},field={title}]{The marriage of effects and
  monads}}.
\mdSpan[class={newblock}]{} \mdSpan[font-style={oblique}]{ \mdSpan[class={bibinfo},field={journal}]{ACM Trans. Comput. Logic}}
  \mdSpan[class={bibinfo},field={volume}]{4}(\mdSpan[class={bibinfo},field={number}]{1}), pp. \mdSpan[class={bibinfo},field={pages}]{1{\textendash}32},
  doi:\mdA{http://dx.doi.org/10.1145/601775.601776}{}{10.1145/601775.601776}.%
\end{mdBibitem}%
\begin{mdBibitem}[class={bibitem},id=wand:records,label={[43]\{.bibitem-label\}},elem={bibitem},cite-label={[43]\{.bibitem-label\}},caption={[Mitchell Wand]\{ .bibinfo field={'}author{'}\} ([1987]\{ .bibinfo field={'}year{'}\}): \_[Complete Type Inference for Simple Objects]\{ .bibinfo field={'}title{'}\}\_. \\In: [ [Proceedings of the 2nd. IEEE Symposium on Logic in Computer Science]\{ .bibinfo field={'}booktitle{'}\}]\{font-style=oblique\}, pp. [37--44]\{ .bibinfo field={'}pages{'}\}, doi:[10.1109/LICS.1988.5111](http://dx.doi.org/10.1109/LICS.1988.5111). \\[Corrigendum in LICS{'}88, page 132]\{ .bibinfo field={'}note{'}\}.},searchterm={+Mitchell+Wand+Complete+Type+Inference+Simple+Objects+Proceedings+IEEE+Symposium+Logic+Computer+Science+LICS+http+LICS+Corrigendum+LICS+page++}]%
\mdSpan[class={bibitem-before}]{[\mdSpan[class={bibitem-label}]{43}]{\mdNbsp}{\mdNbsp}}\mdSpan[class={bibinfo},field={author}]{Mitchell Wand}
  (\mdSpan[class={bibinfo},field={year}]{1987}): \mdEm{\mdSpan[class={bibinfo},field={title}]{Complete Type Inference for
  Simple Objects}}.
\mdSpan[class={newblock}]{} In: \mdSpan[font-style={oblique}]{ \mdSpan[class={bibinfo},field={booktitle}]{Proceedings of the 2nd. IEEE Symposium
  on Logic in Computer Science}}, pp. \mdSpan[class={bibinfo},field={pages}]{37{\textendash}44},
  doi:\mdA{http://dx.doi.org/10.1109/LICS.1988.5111}{}{10.1109/LICS.1988.5111}.
\mdSpan[class={newblock}]{} \mdSpan[class={bibinfo},field={note}]{Corrigendum in LICS{'}88, page 132}.%
\end{mdBibitem}%
\begin{mdBibitem}[class={bibitem},id=wrightfelleisen,label={[44]\{.bibitem-label\}},elem={bibitem},cite-label={[44]\{.bibitem-label\}},caption={[Andrew{\mdNbsp}K. Wright]\{ .bibinfo field={'}author{'}\} {\&} [Matthias Felleisen]\{ .bibinfo field={'}author{'}\} ([1994]\{ .bibinfo field={'}year{'}\}): \_[A syntactic approach to type soundness]\{ .bibinfo field={'}title{'}\}\_. \\[ [Inf. Comput.]\{ .bibinfo field={'}journal{'}\}]\{font-style=oblique\} [115]\{ .bibinfo field={'}volume{'}\}([1]\{ .bibinfo field={'}number{'}\}), pp. [38--94]\{ .bibinfo field={'}pages{'}\}, doi:[10.1006/inco.1994.1093](http://dx.doi.org/10.1006/inco.1994.1093).},searchterm={+Andrew+Wright+Matthias+Felleisen+syntactic+approach+type+soundness+Comput+inco+http+inco++}]%
\mdSpan[class={bibitem-before}]{[\mdSpan[class={bibitem-label}]{44}]{\mdNbsp}{\mdNbsp}}\mdSpan[class={bibinfo},field={author}]{Andrew{\mdNbsp}K. Wright} {\&}
  \mdSpan[class={bibinfo},field={author}]{Matthias Felleisen}
  (\mdSpan[class={bibinfo},field={year}]{1994}): \mdEm{\mdSpan[class={bibinfo},field={title}]{A syntactic approach to type
  soundness}}.
\mdSpan[class={newblock}]{} \mdSpan[font-style={oblique}]{ \mdSpan[class={bibinfo},field={journal}]{Inf. Comput.}}
  \mdSpan[class={bibinfo},field={volume}]{115}(\mdSpan[class={bibinfo},field={number}]{1}), pp. \mdSpan[class={bibinfo},field={pages}]{38{\textendash}94},
  doi:\mdA{http://dx.doi.org/10.1006/inco.1994.1093}{}{10.1006/inco.1994.1093}.%
\end{mdBibitem}
\end{mdBibliography}
\end{mdDiv}%
\clearpage\mdHxx[id=appendix,caption={Appendix}]{Appendix}\mdHxx[id=sec-inference,label={[A]\{.heading-label\}},toc={},caption={Type inference}]{\mdSpan[class={heading-before}]{\mdSpan[class={heading-label}]{A}.{\enspace}}Type inference}\begin{mdP}%
As a first step toward type inference, we first present in a syntax directed
version of our declarative type rules in Figure{\mdNbsp}\mdA[class={localref},target-element={figure}]{fig-styperules}{}{\mdSpan[class={figure-label}]{7}}. For this
system, the syntax tree completely determines the derivation tree.
Effectively, we removed the \mdSpan[class={rulename},font-variant={small-caps},font-size={small}]{(inst)} and \mdSpan[class={rulename},font-variant={small-caps},font-size={small}]{(gen)} rules, and
always apply instantiation in the \mdSpan[class={rulename},font-variant={small-caps},font-size={small}]{(var)} rule, and always generalize
at let-bindings. This technique is entirely standard
\mdSpan[class={citations},target-element={bibitem}]{[\mdA[class={bibref,localref},target-element={bibitem}]{hindley:types}{}{\mdSpan[class={bibitem-label}]{11}}, \mdA[class={bibref,localref},target-element={bibitem}]{jones:qualifiedtypes}{}{\mdSpan[class={bibitem-label}]{12}}, \mdA[class={bibref,localref},target-element={bibitem}]{milner:types}{}{\mdSpan[class={bibitem-label}]{26}}]} and we can show that
the syntax directed system is sound and complete with respect to the
declarative rules:%
\end{mdP}%
\begin{mdDiv}[class={theorem,block},id=th-ssound,label={[5]\{.theorem-label\}},elem={theorem}]%
\begin{mdP}%
\mdSpan[class={theorem-before}]{\mdStrong{Theorem{\mdNbsp}\mdSpan[class={theorem-label}]{5}.} }(\mdEm{Soundness of the syntax directed rules})\mdBr
When \mdSpan[class={math-inline}]{$\infers{\Gamma}{e}{\t}{\e}$} then we also have \mdSpan[class={math-inline}]{$\infer{\Gamma}{e}{\t}{\e}$}.%
\end{mdP}
\end{mdDiv}%
\begin{mdDiv}[class={theorem,block},id=th-scomplete,label={[6]\{.theorem-label\}},elem={theorem}]%
\begin{mdP}%
\mdSpan[class={theorem-before}]{\mdStrong{Theorem{\mdNbsp}\mdSpan[class={theorem-label}]{6}.} }(\mdEm{Completeness of the syntax directed rules}) \mdBr
When \mdSpan[class={math-inline}]{$\infer{\Gamma}{e}{\s}{\e}$} then we also have \mdSpan[class={math-inline}]{$\infer{\Gamma}{e}{\t}{\e}$}
where \mdSpan[class={math-inline}]{$\s$} can be instantiated to \mdSpan[class={math-inline}]{$\t$}.%
\end{mdP}
\end{mdDiv}%
\begin{mdP}%
Both proofs are by straightforward induction using standard techniques as
described for example by Jones{\mdNbsp}\mdSpan[class={citations},target-element={bibitem}]{[\mdA[class={bibref,localref},target-element={bibitem}]{jones:qualifiedtypes}{}{\mdSpan[class={bibitem-label}]{12}}]}.%
\end{mdP}%
\begin{mdDiv}[class={figure,align-center},id=fig-styperules,label={[7]\{.figure-label\}},elem={figure},toc={tof},toc-line={[7]\{.figure-label\}. Changed rules for the syntax directed system; Rule [inst]\{.rulename\} and [gen]\{.rulename\} are removed, and all other rules are equivalent to the declarative system (Figure [\#fig-typerules])},page-align={top},caption={Changed rules for the syntax directed system; Rule [inst]\{.rulename\} and [gen]\{.rulename\} are removed, and all other rules are equivalent to the declarative system (Figure [\#fig-typerules])}]%
\begin{mdTable}[class={madoko,block}]{2}{ll}

\mdTd[display={table-cell}]{\mdSpan[class={rulename},font-variant={small-caps},font-size={small}]{(var)}\mdSub{s}}&\multicolumn{1}{c}{\mdTd[text-align={center},display={table-cell}]{\mdSpan[class={code,math-inline}]{$\inference{\Gamma(\mathid{x})\mathspace{1}=\mathspace{1}\forall\overline\alpha.\,\t}{\infers{\Gamma}{\mathid{x}}{[\overline\alpha \mapsto \overline\t]\t}{\e}}$}}}\\
\mdTd[display={table-cell}]{{\mdNbsp}}&\multicolumn{1}{c}{\mdTd[text-align={center},display={table-cell}]{}}\\
\mdTd[display={table-cell}]{\mdSpan[class={rulename},font-variant={small-caps},font-size={small}]{(let)}\mdSub{s}}&\multicolumn{1}{c}{\mdTd[text-align={center},display={table-cell}]{\mdSpan[class={code,math-inline}]{$\inference{\ontop{\infers{\Gamma}{\exp_1}{\t_1}{{\langle}{\rangle}}\mathspace{1}\quad \overline\alpha \not\in \ftv\Gamma}{\infers{\Gamma,\mathid{x}:\forall\overline\alpha.\,\t_1}{\exp_2}{\t_2}{\e}}}{\infers{\Gamma}{\mathkw{let}\mathspace{1}\mathid{x}\mathspace{1}=\mathspace{1}\exp_1\mathspace{1}\mathkw{in}\mathspace{1}\exp_2}{\t_2}{\e}}$}}}\\
\end{mdTable}
\mdHr[class={figureline,madoko}]{}\mdSpan[class={figure-caption}]{\mdSpan[class={caption-before}]{\mdStrong{Figure{\mdNbsp}\mdSpan[class={figure-label}]{7}.} }Changed rules for the syntax directed system; Rule \mdSpan[class={rulename},font-variant={small-caps},font-size={small}]{(inst)} and \mdSpan[class={rulename},font-variant={small-caps},font-size={small}]{(gen)} are removed, and all other rules are equivalent to the declarative system (Figure{\mdNbsp}\mdA[class={localref},target-element={figure}]{fig-typerules}{}{\mdSpan[class={figure-label}]{3}})}%
\end{mdDiv}%
\mdHxxx[id=the-type-inference-algorithm,label={[A.1]\{.heading-label\}},toc={},caption={The type inference algorithm}]{\mdSpan[class={heading-before}]{\mdSpan[class={heading-label}]{A.1}.{\enspace}}The type inference algorithm}\begin{mdDiv}[class={figure,align-center},id=fig-inference,label={[8]\{.figure-label\}},elem={figure},toc={tof},toc-line={[8]\{.figure-label\}. Type and effect inference algorithm. Any type variables \${\textbackslash}alpha\$, \${\textbackslash}mu\$, \${\textbackslash}xi\$, and \${\textbackslash}overline{\textbackslash}alpha\$ are considered fresh.},page-align={top},caption={Type and effect inference algorithm. Any type variables \${\textbackslash}alpha\$, \${\textbackslash}mu\$, \${\textbackslash}xi\$, and \${\textbackslash}overline{\textbackslash}alpha\$ are considered fresh.}]%
\begin{mdTable}[class={madoko,block}]{2}{ll}

\mdTd[display={table-cell}]{\mdSpan[class={rulename},font-variant={small-caps},font-size={small}]{(var)}\mdSub{i}}&\multicolumn{1}{c}{\mdTd[text-align={center},display={table-cell}]{\mdSpan[class={code,math-inline}]{$\inference{\Gamma(\mathid{x})\mathspace{1}=\mathspace{1}\forall\overline\alpha.\,\t}{\inferi{\varnothing\Gamma}{\mathid{x}}{[\overline\alpha \mapsto \overline\beta]\t}{\mu}}$}}}\\
\mdTd[display={table-cell}]{}&\multicolumn{1}{c}{\mdTd[text-align={center},display={table-cell}]{}}\\
\mdTd[display={table-cell}]{\mdSpan[class={rulename},font-variant={small-caps},font-size={small}]{(lam)}\mdSub{i}}&\multicolumn{1}{c}{\mdTd[text-align={center},display={table-cell}]{\mdSpan[class={code,math-inline}]{$\inference{\inferi{\sub\Gamma,\mathid{x}:\alpha}{\exp}{\t_2}{\e_2}}{\inferi{\sub\Gamma}{\lambda \mathid{x}.\,\exp}{\tfun{\sub\alpha}{\e_2}{\t_2}}{\mu}}$}}}\\
\mdTd[display={table-cell}]{{\mdNbsp}}&\multicolumn{1}{c}{\mdTd[text-align={center},display={table-cell}]{}}\\
\mdTd[display={table-cell}]{\mdSpan[class={rulename},font-variant={small-caps},font-size={small}]{(app)}\mdSub{i}}&\multicolumn{1}{c}{\mdTd[text-align={center},display={table-cell}]{\mdSpan[class={code,math-inline}]{$\inference{\ontop{\inferi{\sub_1\Gamma}{\exp_1}{\t_1}{\e_1}\mathspace{1}\quad \inferi{\sub_2(\sub_1\Gamma)}{\exp_2}{\t_2}{\e_2}}{\unify{\sub_2\mathspace{1}\t_1}{(\tfun{\t_2}{\e_2}{\alpha})}{\sub_3}\mathspace{1}\quad \unify{\sub_3\sub_2\e_1}{\sub_3\e_2}{\sub_4}}}{\inferi{\sub_4\sub_3\sub_2\sub_1\Gamma}{\exp_1\,\exp_2}{\sub_4\sub_3\alpha}{\sub_4\sub_3\e_2}}$}}}\\
\mdTd[display={table-cell}]{{\mdNbsp}}&\multicolumn{1}{c}{\mdTd[text-align={center},display={table-cell}]{}}\\
\mdTd[display={table-cell}]{\mdSpan[class={rulename},font-variant={small-caps},font-size={small}]{(let)}\mdSub{i}}&\multicolumn{1}{c}{\mdTd[text-align={center},display={table-cell}]{\mdSpan[class={code,math-inline}]{$\inference{\ontopthree{\inferi{\sub_1\Gamma}{\exp_1}{\t_1}{\e_1}\mathspace{1}\quad \unify{\e_1}{{\langle}{\rangle}}{\sub_2}}{\s =\mathspace{1}\gen{\sub_2\sub_1\Gamma}{\sub_2\t_1}}{\inferi{\sub_3(\sub_2\sub_1\Gamma,\mathid{x}:\s)}{\exp_2}{\t}{\e}}}{\inferi{\sub_3\sub_2\sub_1\Gamma}{\keyword{\mathid{let}}\;\mathspace{1}\mathid{x}\mathspace{1}=\mathspace{1}\exp_1\;\keyword{\mathid{in}}\;\exp_2}{\t}{\e}}$}}}\\
\mdTd[display={table-cell}]{{\mdNbsp}}&\multicolumn{1}{c}{\mdTd[text-align={center},display={table-cell}]{}}\\
\mdTd[display={table-cell}]{\mdSpan[class={rulename},font-variant={small-caps},font-size={small}]{(run)}\mdSub{i}}&\multicolumn{1}{c}{\mdTd[text-align={center},display={table-cell}]{\mdSpan[class={code,math-inline}]{$\inference{\ontop{\inferi{\sub_1\Gamma}{\mathid{e}}{\t}{\e}\mathspace{1}\quad \unify{\e}{\ext{\tst{\xi}}{\mu}}{\sub_2}}{\sub_2\xi \in \textit{TypeVar}\mathspace{1}\quad \sub_2\xi \not\in \ftv{\sub_2\sub_1\Gamma,\sub_2\t,\sub_2\mu}}}{\inferi{\sub_2\sub_1\Gamma}{\run\,\mathid{e}}{\sub_2\t}{\sub_2\mu}}$}}}\\
\mdTd[display={table-cell}]{{\mdNbsp}}&\multicolumn{1}{c}{\mdTd[text-align={center},display={table-cell}]{}}\\
\mdTd[display={table-cell}]{\mdSpan[class={rulename},font-variant={small-caps},font-size={small}]{(catch)}\mdSub{i}}&\multicolumn{1}{c}{\mdTd[text-align={center},display={table-cell}]{\mdSpan[class={code,math-inline}]{$\inference{\ontopthree{\inferi{\sub_1\Gamma}{\mathid{e}_1}{\t_1}{\e_1}\mathspace{1}\quad \inferi{\sub_2(\sub_1\Gamma)}{\mathid{e}_2}{\t_2}{\e_2}}{\unify{\sub_2\e_1}{\ext{\ec{\mathid{exn}}}{\e_2}}{\sub_3}}{\unify{\sub_3\t_2}{()\mathspace{1}{\rightarrow}\mathspace{1}\sub_3\e_2\;\sub_3\sub_2\t_1}{\sub_4}}}{\infer{\sub_4\sub_3\sub_2\sub_1\Gamma}{\catch\,\mathid{e}_1\,\,\mathid{e}_2}{\sub_4\sub_3\t_2}{\sub_4\sub_3\e_2}}$}}}\\
\end{mdTable}
\mdHr[class={figureline,madoko}]{}\mdSpan[class={figure-caption}]{\mdSpan[class={caption-before}]{\mdStrong{Figure{\mdNbsp}\mdSpan[class={figure-label}]{8}.} }Type and effect inference algorithm. Any type variables \mdSpan[class={math-inline}]{$\alpha$}, \mdSpan[class={math-inline}]{$\mu$}, \mdSpan[class={math-inline}]{$\xi$}, and \mdSpan[class={math-inline}]{$\overline\alpha$} are considered fresh.}%
\end{mdDiv}%
\begin{mdDiv}[class={figure,align-center},id=fig-unify,label={[9]\{.figure-label\}},elem={figure},toc={tof},toc-line={[9]\{.figure-label\}. Unification: \${\textbackslash}unify\{{\textbackslash}t\}\{{\textbackslash}t'\}\{{\textbackslash}sub\}\$ unifies two types and returns a substitution \${\textbackslash}sub\$. It uses effect unification \${\textbackslash}runify\{{\textbackslash}e\}\{l\}\{{\textbackslash}e'\}\{{\textbackslash}sub\}\$ which takes an effect \${\textbackslash}e\$ and effect primitive \$l\$ as input, and returns effect tail \${\textbackslash}e'\$ and a substition \${\textbackslash}sub\$.},page-align={top},caption={Unification: \${\textbackslash}unify\{{\textbackslash}t\}\{{\textbackslash}t'\}\{{\textbackslash}sub\}\$ unifies two types and returns a substitution \${\textbackslash}sub\$. It uses effect unification \${\textbackslash}runify\{{\textbackslash}e\}\{l\}\{{\textbackslash}e'\}\{{\textbackslash}sub\}\$ which takes an effect \${\textbackslash}e\$ and effect primitive \$l\$ as input, and returns effect tail \${\textbackslash}e'\$ and a substition \${\textbackslash}sub\$.}]%
\begin{mdTable}[class={madoko,block}]{2}{ll}

\mdTd[display={table-cell}]{\mdSpan[class={rulename},font-variant={small-caps},font-size={small}]{(uni-var)}}&\multicolumn{1}{c}{\mdTd[text-align={center},display={table-cell}]{\mdSpan[class={code,math-inline}]{$\unify{\alpha}{\alpha}{[]}$}}}\\
\mdTd[display={table-cell}]{{\mdNbsp}}&\multicolumn{1}{c}{\mdTd[text-align={center},display={table-cell}]{}}\\
\mdTd[display={table-cell}]{\mdSpan[class={rulename},font-variant={small-caps},font-size={small}]{(uni-varl)}}&\multicolumn{1}{c}{\mdTd[text-align={center},display={table-cell}]{\mdSpan[class={code,math-inline}]{$\inference{\alpha \not\in \ftv\t}{\unify{\alpha^\mathid{k}}{\t^\mathid{k}}{[\alpha \mapsto \t]}}$}}}\\
\mdTd[display={table-cell}]{{\mdNbsp}}&\multicolumn{1}{c}{\mdTd[text-align={center},display={table-cell}]{}}\\
\mdTd[display={table-cell}]{\mdSpan[class={rulename},font-variant={small-caps},font-size={small}]{(uni-varr)}}&\multicolumn{1}{c}{\mdTd[text-align={center},display={table-cell}]{\mdSpan[class={code,math-inline}]{$\inference{\alpha \not\in \ftv\t}{\unify{\t^\mathid{k}}{\alpha^\mathid{k}}{[\alpha \mapsto \t]}}$}}}\\
\mdTd[display={table-cell}]{{\mdNbsp}}&\multicolumn{1}{c}{\mdTd[text-align={center},display={table-cell}]{}}\\
\mdTd[display={table-cell}]{\mdSpan[class={rulename},font-variant={small-caps},font-size={small}]{(uni-con)}}&\multicolumn{1}{c}{\mdTd[text-align={center},display={table-cell}]{\mdSpan[class={code,math-inline}]{$\inference{\ontop{\forall \mathid{i}\mathspace{1}\in 1..\mathid{n}.\quad \unify{\sub_{\mathid{i}-1}...\sub_1\t_\mathid{i}}{\sub_{\mathid{i}-1}...\sub_1{\mathid{t}_\mathid{i}}}{\sub_\mathid{i}}\mathspace{1}}{\mathspace{1}\kk =\mathspace{1}(\kk_1,...,\kk_\mathid{n})\mathspace{1}{\rightarrow}\mathspace{1}\kk'}}{\unify{\mathid{c}^\kk{\langle}\t_1^{\kk_1},...,\t_\mathid{n}^{\kk_\mathid{n}}{\rangle}}{\mathid{c}^\kk{\langle}\mathid{t}_1^{\kk_1},...,\mathid{t}_\mathid{n}^{\kk_\mathid{n}}{\rangle}}{\sub_\mathid{n}...\sub_1}}$}}}\\
\mdTd[display={table-cell}]{{\mdNbsp}}&\multicolumn{1}{c}{\mdTd[text-align={center},display={table-cell}]{}}\\
\mdTd[display={table-cell}]{{\mdNbsp}}&\multicolumn{1}{c}{\mdTd[text-align={center},display={table-cell}]{}}\\
\mdTd[display={table-cell}]{\mdSpan[class={rulename},font-variant={small-caps},font-size={small}]{(uni-eff)}}&\multicolumn{1}{c}{\mdTd[text-align={center},display={table-cell}]{\mdSpan[class={code,math-inline}]{$\inference{\ontop{\runify{\e_2}{\mathid{l}}{\e_3}{\sub_1}\mathspace{1}\quad \tl{\e_1}\mathspace{1}\not\in\dom{\sub_1}\mathspace{1}}{\unify{\sub_1\e_1}{\sub_1\e_3}{\sub_2}}}{\unify{\ext{\mathid{l}}{\e_1}}{\e_2}{\sub_2\sub_1}}$}}}\\
\mdTd[display={table-cell}]{{\mdNbsp}}&\multicolumn{1}{c}{\mdTd[text-align={center},display={table-cell}]{}}\\
\mdTd[display={table-cell}]{\mdSpan[class={rulename},font-variant={small-caps},font-size={small}]{(eff-head)}}&\multicolumn{1}{c}{\mdTd[text-align={center},display={table-cell}]{\mdSpan[class={code,math-inline}]{$\inference{\mathid{l}\mathspace{1}\equiv \mathid{l}'\mathspace{1}\quad \unify{\mathid{l}}{\mathid{l}'}{\sub}}{\runify{\ext{\mathid{l}'}{\e}}{\mathid{l}}{\e}{\sub}}$}}}\\
\mdTd[display={table-cell}]{{\mdNbsp}}&\multicolumn{1}{c}{\mdTd[text-align={center},display={table-cell}]{}}\\
\mdTd[display={table-cell}]{\mdSpan[class={rulename},font-variant={small-caps},font-size={small}]{(eff-swap)}}&\multicolumn{1}{c}{\mdTd[text-align={center},display={table-cell}]{\mdSpan[class={code,math-inline}]{$\inference{\mathid{l}\mathspace{1}\not\equiv \mathid{l}'\mathspace{1}\quad \runify{\e}{\mathid{l}}{\e'}{\sub}}{\runify{\ext{\mathid{l}'}{\e}}{\mathid{l}}{\ext{\mathid{l}}{\e'}}{\sub}}$}}}\\
\mdTd[display={table-cell}]{{\mdNbsp}}&\multicolumn{1}{c}{\mdTd[text-align={center},display={table-cell}]{}}\\
\mdTd[display={table-cell}]{\mdSpan[class={rulename},font-variant={small-caps},font-size={small}]{(eff-tail)}}&\multicolumn{1}{c}{\mdTd[text-align={center},display={table-cell}]{\mdSpan[class={code,math-inline}]{$\inference{\textrm{fresh }\mathspace{1}\mu'}{\runify{\mu}{\mathid{l}}{\mu'}{[\mu \mapsto \ext{\mathid{l}}{\mu'}]}}$}}}\\
\end{mdTable}
\mdHr[class={figureline,madoko}]{}\mdSpan[class={figure-caption}]{\mdSpan[class={caption-before}]{\mdStrong{Figure{\mdNbsp}\mdSpan[class={figure-label}]{9}.} }Unification: \mdSpan[class={math-inline}]{$\unify{\t}{\t'}{\sub}$} unifies two types and returns a substitution \mdSpan[class={math-inline}]{$\sub$}. It uses effect unification \mdSpan[class={math-inline}]{$\runify{\e}{l}{\e'}{\sub}$} which takes an effect \mdSpan[class={math-inline}]{$\e$} and effect primitive \mdSpan[class={math-inline}]{$l$} as input, and returns effect tail \mdSpan[class={math-inline}]{$\e'$} and a substition \mdSpan[class={math-inline}]{$\sub$}.}%
\end{mdDiv}%
\begin{mdP}%
Starting from the syntax directed rules, we can now give a the type inference
algorithm for our effect system which is shown in Figure{\mdNbsp}\mdA[class={localref},target-element={figure}]{fig-inference}{}{\mdSpan[class={figure-label}]{8}}.
Following Jones{\mdNbsp}\mdSpan[class={citations},target-element={bibitem}]{[\mdA[class={bibref,localref},target-element={bibitem}]{jones:qualifiedtypes}{}{\mdSpan[class={bibitem-label}]{12}}]} we present the algorithm as
natural inference rules of the form \mdSpan[class={math-inline}]{$\infer{\sub\Gamma}{\exp}{\t}{\e}$} where
\mdSpan[class={math-inline}]{$\sub$} is a substitution, \mdSpan[class={math-inline}]{$\Gamma$} the environment, and \mdSpan[class={math-inline}]{$\exp$}, \mdSpan[class={math-inline}]{$\t$}, and
\mdSpan[class={math-inline}]{$\e$}, the expression, its type, and its effect respectively. The rules can be
read as an attribute grammar where \mdSpan[class={math-inline}]{$\sub$}, \mdSpan[class={math-inline}]{$\t$}, and \mdSpan[class={math-inline}]{$\e$} are synthesised, and
\mdSpan[class={math-inline}]{$\Gamma$} and \mdSpan[class={math-inline}]{$\exp$} inherited. An advantage is that this highlights the
correspondence between the syntax directed rules and the inference algorithm.%
\end{mdP}%
\begin{mdP}[class={indent,para-continue}]%
The algorithm uses unification written as \mdSpan[class={math-inline}]{$\unify{\t_1}{\t_2}{\sub}$} which
unifies \mdSpan[class={math-inline}]{$\t_1$} and \mdSpan[class={math-inline}]{$\t_2$} with a most general substitution \mdSpan[class={math-inline}]{$\sub$} such that
\mdSpan[class={math-inline}]{$\sub\t_1 = \sub\t_2$}.  \%The unification algorithm is standard and effects are
unified using standard row unificiation allowing for duplicate label as
described by Leijen{\mdNbsp}\mdSpan[class={citations},target-element={bibitem}]{[\mdA[class={bibref,localref},target-element={bibitem}]{leijen:scopedlabels}{}{\mdSpan[class={bibitem-label}]{17}}]}.  The \mdSpan[class={math-inline}]{$\mathsf{gen}$} function
generalizes a type with respect to an environment and is defined as:%
\end{mdP}%
\begin{mdDiv}[class={math,para-block,input-math},elem={math}]%
\begin{mdDiv}[class={math,math-display}]%
\[\gen{\Gamma}{\t} = \forall (\ftv{\t} - \ftv{\Gamma}).\,\t \]%
\end{mdDiv}
\end{mdDiv}%
\begin{mdP}%
We can prove that the inference algorithm is sound and complete with respect
to the syntax directed rules (and by Theorem{\mdNbsp}\mdA[class={localref},target-element={theorem}]{th-ssound}{}{\mdSpan[class={theorem-label}]{5}} and
\mdA[class={localref},target-element={theorem}]{th-scomplete}{}{\mdSpan[class={theorem-label}]{6}} also sound and complete to the declarative rules):%
\end{mdP}%
\begin{mdDiv}[class={theorem,block},id=th-sound,label={[7]\{.theorem-label\}},elem={theorem}]%
\begin{mdP}%
\mdSpan[class={theorem-before}]{\mdStrong{Theorem{\mdNbsp}\mdSpan[class={theorem-label}]{7}.} }(\mdEm{Soundness})\mdBr
If \mdSpan[class={math-inline}]{$\inferi{\sub\Gamma}{\exp}{\t}{\e}$} then there exists a \mdSpan[class={math-inline}]{$\sub'$} such that 
\mdSpan[class={math-inline}]{$\infers{\sub\Gamma}{\exp}{\t'}{\e'}$} where \mdSpan[class={math-inline}]{$\sub'\t = \t'$} and \mdSpan[class={math-inline}]{$\sub'\e = \e'$}.%
\end{mdP}
\end{mdDiv}%
\begin{mdDiv}[class={theorem,block},id=th-complete,label={[8]\{.theorem-label\}},elem={theorem}]%
\begin{mdP}%
\mdSpan[class={theorem-before}]{\mdStrong{Theorem{\mdNbsp}\mdSpan[class={theorem-label}]{8}.} }(\mdEm{Completeness})\mdBr
If \mdSpan[class={math-inline}]{$\infers{\sub_1\Gamma}{\exp}{\t_1}{\e_1}$} then \mdSpan[class={math-inline}]{$\inferi{\sub_2\Gamma}{\exp}{\t_2}{\e_2}$}
and there exists a substitution \mdSpan[class={math-inline}]{$\sub$} such that \mdSpan[class={math-inline}]{$\sub_1 \approx \sub\sub_2$}, \mdSpan[class={math-inline}]{$\t_1 = \sub\t_2$} 
and \mdSpan[class={math-inline}]{$\e_1 = \sub\e_2$}.%
\end{mdP}
\end{mdDiv}%
\begin{mdP}%
Since the inference algorithm is basically just algorithm{\mdNbsp}W
\mdSpan[class={citations},target-element={bibitem}]{[\mdA[class={bibref,localref},target-element={bibitem}]{damasmilner:hm}{}{\mdSpan[class={bibitem-label}]{5}}]} together with extra unifications for effect types, the
proofs of soundness and completeness are entirely standard. The main extended
lemma is for the soundness, completeness, and termination of the  unification
algorithm which now also unifies effect types.%
\end{mdP}%
\begin{mdP}[class={indent}]%
The unification algorithm is shown in Figure{\mdNbsp}\mdA[class={localref},target-element={figure}]{fig-unify}{}{\mdSpan[class={figure-label}]{9}}. The algorithm is
an almost literal adaption of the unification algorithm for records with
scoped labels as described by Leijen{\mdNbsp}\mdSpan[class={citations},target-element={bibitem}]{[\mdA[class={bibref,localref},target-element={bibitem}]{leijen:scopedlabels}{}{\mdSpan[class={bibitem-label}]{17}}]}, and the
proofs of soundness, completeness, and termination carry over directly.%
\end{mdP}%
\begin{mdP}[class={indent,para-continue}]%
The first four rules are the standard Robinson unification rules with a slight
modification to return only kind-preserving substitutions
\mdSpan[class={citations},target-element={bibitem}]{[\mdA[class={bibref,localref},target-element={bibitem}]{gasterjones:trex}{}{\mdSpan[class={bibitem-label}]{7}}, \mdA[class={bibref,localref},target-element={bibitem}]{jones:constructorclasses}{}{\mdSpan[class={bibitem-label}]{13}}]}.  The rule \mdSpan[class={rulename},font-variant={small-caps},font-size={small}]{(uni-eff)}
unifies effect rows. The operation \mdSpan[class={math-inline}]{$\tl{\e}$} is defined as:%
\end{mdP}%
\begin{mdDiv}[class={para-block,para-block,input-mathpre},elem={pre}]%
\begin{mdDiv}[class={math-display}]%
\[\begin{mdMathprearray}
\tl{{\langle}\mathid{l}_1,...,\mathid{l}_\mathid{n}\mathspace{1}|\mathspace{1}\mu{\rangle}}\mathspace{1}&\mathspace{1}=\mathspace{1}\mu  \mathbr{}
\tl{{\langle}\mathid{l}_1,...,\mathid{l}_\mathid{n}{\rangle}}\mathspace{1}&\mathspace{1}=\mathspace{1}{\langle}{\rangle}\mathspace{2}
\end{mdMathprearray}\]%
\end{mdDiv}
\end{mdDiv}%
\begin{mdP}%
As described in detail in{\mdNbsp}\mdSpan[class={citations},target-element={bibitem}]{[\mdA[class={bibref,localref},target-element={bibitem}]{leijen:scopedlabels}{}{\mdSpan[class={bibitem-label}]{17}}]}, the check \mdSpan[class={math-inline}]{$\tl{\e_1} \not\in
\dom\sub_1$} is subtle but necessary to guarantee termination of row
unification.  The final three rules unify an effect with a specific head. In
particular,  \mdSpan[class={math-inline}]{$\runify{\e}{l}{\e'}{\sub}$} states that for a given effect row
\mdSpan[class={math-inline}]{$\e$}, we match it with a  given effect constant \mdSpan[class={math-inline}]{$l$}, and return an effect tail
\mdSpan[class={math-inline}]{$\e'$} and substitution \mdSpan[class={math-inline}]{$\sub$} such  that \mdSpan[class={code,math-inline}]{$\sub\e \equiv {\langle}\sub \mathid{l}\mathspace{1}|\mathspace{1}\sub\e'{\rangle}$}. Each
rule basically corresponds to  the equivalence rules on effect rows
(Figure{\mdNbsp}\mdA[class={localref},target-element={figure}]{fig-eqeffect}{}{\mdSpan[class={figure-label}]{2}}).%
\end{mdP}%

\end{document}